\documentclass[aps,prd,twocolumn,superscriptaddress,showpacs,floatfix]{revtex4}

\usepackage{amssymb}
\usepackage{amsmath}
\usepackage{graphicx}
\usepackage{dcolumn}
\usepackage{bm}

\RequirePackage{xspace}

\newcommand{\anu}{\ensuremath{\bar{\nu}}\xspace}
\newcommand{\nue}{\ensuremath{\nu_{e}}\xspace}
\newcommand{\anue}{\ensuremath{\anu_{e}}\xspace}
\newcommand{\numu}{\ensuremath{\nu_{\mu}}\xspace}
\newcommand{\anumu}{\ensuremath{\anu_{\mu}}\xspace}

\newcommand{\mup}{\ensuremath{\mu^+}\xspace}
\newcommand{\mum}{\ensuremath{\mu^-}\xspace}

\newcommand{\pip}{\ensuremath{\pi^+}\xspace}

\newcommand{\piz}{\ensuremath{\pi^0}\xspace}

\newcommand{\kpl}{\ensuremath{K^+}\xspace}

\newcommand{\cc}{\ensuremath{\textrm{CC}}\xspace}

\newcommand{\ccqe}{\ensuremath{\cc\textrm{QE}}\xspace}

\newcommand{\ccpip}{\ensuremath{\cc\pip}\xspace}
\newcommand{\nmccpip}{\ensuremath{\numu\cc\pip}\xspace}

\newcommand{\ccpiz}{\ensuremath{\cc\piz}\xspace}

\newcommand{\ccdis}{\ensuremath{\cc\textrm{DIS}}\xspace}

\newcommand{\ccmesonb}{\ensuremath{\cc\textrm{meson}B}\xspace}

\newcommand{\cm}    {\ensuremath{\textrm{cm}}\xspace}
\newcommand{\mev}   {\ensuremath{\textrm{MeV}}\xspace}
\newcommand{\gev}   {\ensuremath{\textrm{GeV}}\xspace}

\begin{document}


\preprint{APS/123-QED}

\title{Measurement of Neutrino-Induced Charged-Current Charged Pion Production Cross Sections on Mineral Oil at E$_{\nu}\sim 1~\gev$}

%



\newcommand{\argo}{Argonne National Laboratory; Argonne, IL 60439}
\newcommand{\bama}{University of Alabama; Tuscaloosa, AL 35487}
\newcommand{\bucknell}{Bucknell University; Lewisburg, PA 17837}
\newcommand{\cinci}{University of Cincinnati; Cincinnati, OH 45221}
\newcommand{\colorado}{University of Colorado; Boulder, CO 80309}
\newcommand{\columbia}{Columbia University; New York, NY 10027}
\newcommand{\embry}{Embry-Riddle Aeronautical University; Prescott, AZ 86301}
\newcommand{\fnal}{Fermi National Accelerator Laboratory; Batavia, IL 60510}
\newcommand{\florida}{University of Florida; Gainesville, FL 32611}
\newcommand{\indiana}{Indiana University; Bloomington, IN 47405}
\newcommand{\lanl}{Los Alamos National Laboratory; Los Alamos, NM 87545}
\newcommand{\lsu}{Louisiana State University; Baton Rouge, LA 70803}
\newcommand{\umich}{University of Michigan; Ann Arbor, MI 48109}
\newcommand{\princeton}{Princeton University; Princeton, NJ 08544}
\newcommand{\marys}{Saint Mary's University of Minnesota; Winona, MN 55987}
\newcommand{\vtech}{Virginia Polytechnic Institute \& State University; Blacksburg, VA 24061}
\newcommand{\yale}{Yale University; New Haven, CT 06520}
\newcommand{\beijing}{Institute of High Energy Physics; Beijing 100049, China}
\newcommand{\hope}{Hope College; Holland, MI 49423}
\newcommand{\iit}{Illinois Institute of Technology; Chicago, IL 60616}
\newcommand{\imsa}{Illinois Mathematics and Science Academy; Aurora IL 60506}
\newcommand{\massit}{Massachusetts Institute of Technology; Cambridge, MA 02139}
\newcommand{\namex}{Instituto de Ciencias Nucleares, Universidad Nacional Aut\'onoma de M\'exico, D.F. 04510, M\'exico}
\newcommand{\caltech}{California Institute of Technology; Pasadena, CA 91125}
\newcommand{\bu}{Boston University; Boston, MA 02215}
\newcommand{\valencia}{IFIC, Universidad de Valencia and CSIC; 46071 Valencia, Spain}
\newcommand{\ubc}{University of British Columbia, Vancouver, BC V6T 1Z1, Canada}
\newcommand{\imperial}{Imperial College; London SW7 2AZ, United Kingdom}

\affiliation{\bama}      
\affiliation{\argo}      
\affiliation{\bucknell}  
\affiliation{\cinci}     
\affiliation{\colorado}  
\affiliation{\columbia}  
\affiliation{\embry}     
\affiliation{\fnal}      
\affiliation{\florida}   
\affiliation{\indiana}   
\affiliation{\lanl}      
\affiliation{\lsu}       
\affiliation{\massit}    
\affiliation{\namex}     
\affiliation{\umich}     
\affiliation{\princeton} 
\affiliation{\marys}     
\affiliation{\vtech}     
\affiliation{\yale}      

\author{A.~A. Aguilar-Arevalo}\affiliation{\namex}
\author{C.~E.~Anderson}\affiliation{\yale}
\author{A.~O.~Bazarko}\affiliation{\princeton}
\author{S.~J.~Brice}\affiliation{\fnal}
\author{B.~C.~Brown}\affiliation{\fnal}
\author{L.~Bugel}\affiliation{\columbia}
\author{J.~Cao}\affiliation{\umich}
\author{L.~Coney}\affiliation{\columbia}
\author{J.~M.~Conrad}\affiliation{\massit}
\author{D.~C.~Cox}\affiliation{\indiana}
\author{A.~Curioni}\affiliation{\yale}
\author{R.~Dharmapalan}\affiliation{\bama}
\author{Z.~Djurcic}\affiliation{\argo}
\author{D.~A.~Finley}\affiliation{\fnal}
\author{B.~T.~Fleming}\affiliation{\yale}
\author{R.~Ford}\affiliation{\fnal}
\author{F.~G.~Garcia}\affiliation{\fnal}
\author{G.~T.~Garvey}\affiliation{\lanl}
\author{J.~Grange}\affiliation{\florida}
\author{C.~Green}\affiliation{\fnal}\affiliation{\lanl}
\author{J.~A.~Green}\affiliation{\indiana}\affiliation{\lanl}
\author{T.~L.~Hart}\affiliation{\colorado}
\author{E.~Hawker}\affiliation{\cinci}\affiliation{\lanl}
\author{R.~Imlay}\affiliation{\lsu}
\author{R.~A. ~Johnson}\affiliation{\cinci}
\author{G.~Karagiorgi}\affiliation{\massit}
\author{P.~Kasper}\affiliation{\fnal}
\author{T.~Katori}\affiliation{\indiana}\affiliation{\massit}
\author{T.~Kobilarcik}\affiliation{\fnal}
\author{I.~Kourbanis}\affiliation{\fnal}
\author{S.~Koutsoliotas}\affiliation{\bucknell}
\author{E.~M.~Laird}\affiliation{\princeton}
\author{S.~K.~Linden}\affiliation{\yale}
\author{J.~M.~Link}\affiliation{\vtech}
\author{Y.~Liu}\affiliation{\umich}
\author{Y.~Liu}\affiliation{\bama}
\author{W.~C.~Louis}\affiliation{\lanl}
\author{K.~B.~M.~Mahn}\affiliation{\columbia}
\author{W.~Marsh}\affiliation{\fnal}
\author{C.~Mauger}\affiliation{\lanl}
\author{V.~T.~McGary}\affiliation{\massit}
\author{G.~McGregor}\affiliation{\lanl}
\author{W.~Metcalf}\affiliation{\lsu}
\author{P.~D.~Meyers}\affiliation{\princeton}
\author{F.~Mills}\affiliation{\fnal}
\author{G.~B.~Mills}\affiliation{\lanl}
\author{J.~Monroe}\affiliation{\columbia}
\author{C.~D.~Moore}\affiliation{\fnal}
\author{J.~Mousseau}\affiliation{\florida}
\author{R.~H.~Nelson}\affiliation{\colorado}
\author{P.~Nienaber}\affiliation{\marys}
\author{J.~A.~Nowak}\affiliation{\lsu}
\author{B.~Osmanov}\affiliation{\florida}
\author{S.~Ouedraogo}\affiliation{\lsu}
\author{R.~B.~Patterson}\affiliation{\princeton}
\author{Z.~Pavlovic}\affiliation{\lanl}
\author{D.~Perevalov}\affiliation{\bama}
\author{C.~C.~Polly}\affiliation{\fnal}
\author{E.~Prebys}\affiliation{\fnal}
\author{J.~L.~Raaf}\affiliation{\cinci}
\author{H.~Ray}\affiliation{\florida}
\author{B.~P.~Roe}\affiliation{\umich}
\author{A.~D.~Russell}\affiliation{\fnal}
\author{V.~Sandberg}\affiliation{\lanl}
\author{R.~Schirato}\affiliation{\lanl}
\author{D.~Schmitz}\affiliation{\fnal}
\author{M.~H.~Shaevitz}\affiliation{\columbia}
\author{F.~C.~Shoemaker}\altaffiliation{deceased}\affiliation{\princeton}
\author{D.~Smith}\affiliation{\embry}
\author{M.~Soderberg}\affiliation{\yale}
\author{M.~Sorel}\altaffiliation{Present Address: \valencia}\affiliation{\columbia}
\author{P.~Spentzouris}\affiliation{\fnal}
\author{J.~Spitz}\affiliation{\yale}
\author{I.~Stancu}\affiliation{\bama}
\author{R.~J.~Stefanski}\affiliation{\fnal}
\author{M.~Sung}\affiliation{\lsu}
\author{H.~A.~Tanaka}\affiliation{\princeton}
\author{R.~Tayloe}\affiliation{\indiana}
\author{M.~Tzanov}\affiliation{\colorado}
\author{R.~Van~de~Water}\affiliation{\lanl}
\author{M.~O.~Wascko}\altaffiliation{Present Address: \imperial}\affiliation{\lsu}
\author{D.~H.~White}\affiliation{\lanl}
\author{M.~J.~Wilking}\affiliation{\colorado}
\author{H.~J.~Yang}\affiliation{\umich}
\author{G.~P.~Zeller}\affiliation{\fnal}
\author{E.~D.~Zimmerman}\affiliation{\colorado}

\collaboration{MiniBooNE Collaboration}\noaffiliation

\date{\today}

\begin{abstract}
Using a high-statistics, high-purity sample of \numu-induced charged
   current, charged pion events in mineral oil (CH$_2$), MiniBooNE
   reports a collection of interaction cross sections for this
   process. This includes measurements of the \ccpip\ cross section as
   a function of neutrino energy, as well as flux-averaged single- and
   double-differential cross sections of the energy and direction of
   both the final-state muon and pion. In addition, each of the
   single-differential cross sections are extracted as a function of
   neutrino energy to decouple the shape of the MiniBooNE energy
   spectrum from the results. In many cases, these cross sections are
   the first time such quantities have been measured on a nuclear
   target and in the 1 GeV energy range.
\end{abstract}

\maketitle

\section{Introduction}

Charged-current charged pion production (\ccpip) is a process in which
a neutrino interacts with an atomic nucleus and produces a muon, a
charged pion, and recoiling nuclear fragments.  An understanding of
\ccpip interactions is important for the next generation of
accelerator-based neutrino oscillation experiments.  The signal mode
for these experiments is the charged-current quasielastic process
(CCQE), and in the few GeV neutrino energy range where such searches
are typically conducted, the dominant charged-current background is
from \ccpip\ events.  If the pion produced in a \ccpip\ interaction is
lost, the final state will be identical to that of a
\ccqe\ event.

Further complicating these measurements is the use of nuclear targets.
The spectrum of nuclear resonance states that most often produce the
pions in \ccpip\ events is modified inside the nucleus, and
interactions that produce nucleons below the nuclear Fermi momentum
are inhibited by Pauli exclusion.  After the initial interaction takes
place, the final-state particles can interact within the nuclear
medium to absorb or produce pions, thus modifying the observed
particle composition.  When taken in concert, these effects make
difficult the extrapolation of previous measurements on hydrogen and
deuterium to heavier nuclei.

Many theoretical calculations exist that predict cross sections and
kinematics for \ccpip interactions on nuclear
targets~\cite{th1,th2,th3,th4,th5,th6,th7,th8}, but currently there are not
many neutrino-based measurements with which these models can be
evaluated.  Measurements of the \ccqe\ to \ccpip\ cross section ratio
are available from MiniBooNE~\cite{minibooneratio},
K2K~\cite{k2kratio}, and ANL~\cite{radecky}.  The only absolute cross
section measurements in the 1 GeV range were conducted decades ago on
hydrogen and deuterium bubble chambers at
ANL~\cite{campbell,barish,radecky} and BNL~\cite{kitagaki}.  The
results were based on less than 4000 events combined and differed
from one another in normalization by $\sim$20\%.

MiniBooNE has collected what is currently the world's largest sample
of \ccpip\ interactions: a total of 48 322 candidate \numu-\ccpip\
events with 90\% purity.  A reconstruction algorithm has been
developed to distinguish muons and pions based on the presence of
hadronic interactions with an 88\% success rate.  Using this kinematic
information, we report a measurement of the \ccpip\ cross section as a
function of neutrino energy, as well as the first single- and
double-differential cross sections for the final-state muon and
pion. These results provide the most complete information available on
this process as measured on a nuclear target in the 1 GeV energy
range.

\subsection{Signal Definition:  Observable \ccpip}

After the initial neutrino interaction takes place, the resulting
final-state particles must traverse the remainder of the nucleus
before they can be detected.  This results in additional interactions
with the nuclear medium that can produce or absorb pions.  Since these
intranuclear processes are not experimentally accessible, an
``observable \ccpip'' interaction has been defined in this analysis as
any event with a \mum and a \pip leaving the nucleus, regardless of
which particles were produced in the initial neutrino interaction.  In
principle, it is possible to use the Monte Carlo simulation to correct
the observed event distributions back to the initial neutrino-nucleon
interaction, however such a correction introduces a significant amount
of dependence on the chosen final-state interaction model.  To reduce
this model dependence, the measurements are reported for observable
\ccpip\ interactions.  Apart from a muon and a single pion in the
final state (no other mesons), no requirement is made on the number of
photons, nucleons, and multinucleon states.

In MiniBooNE, \ccpip\ interactions are dominantly produced either
through an intermediate resonance state or by scattering off of the
entire nucleus coherently.  In the former case, the neutrino interacts
with a single nucleon, producing a resonance state (usually a $\Delta$
at MiniBooNE energies), which then decays to a nucleon and a pion.
The results to follow are all combined measurements of both incoherent
and coherent processes.

\subsection{MiniBooNE}

The Mini-Booster Neutrino Experiment (MiniBooNE) was designed to
search for the appearance of oscillated \nue events from a high-purity
\numu\ beam.  The Booster Accelerator at the Fermi National Accelerator
Laboratory (Fermilab) provides a beam of 8 GeV kinetic energy protons,
which is directed onto a 71 cm beryllium target.  Positively charged
particles produced in the target are forward-focused by a
cylindrically symmetric horn surrounding the target.  Downstream of
the horn is a 50~m drift pipe in which the particles produced in the
target are allowed to decay to neutrinos.  These decays are dominated
by \pip$\to$\mup\numu\ with a small contribution from muon and kaon
decay channels.  The result is a neutrino beam composed of 93.6\%
\numu, 5.9\% \anumu, and a small contribution of \nue and
\anue.  At 1~\gev where the observed \ccpip\ event distribution is
peaked, the \numu\ component of the beam is 97\% of the total flux.  A
detailed description of the beamline and the neutrino flux prediction
is given in Ref.~\cite{fluxpaper}.

The MiniBooNE detector, located 541~m downstream of the target,
consists of a spherical tank 610.6~cm in radius with a 575~cm radius
main volume surrounded by a outer veto region, which is used to detect
particles entering and exiting the main volume.  The inside surface of
the main volume is lined with 1280 photomultiplier tubes (PMTs), and
an additional 240 PMTs are mounted inside the veto region.  Charged
particles are detected via the light they emit as they traverse the
818 tons of mineral oil residing in the tank.  More information on the
performance of the MiniBooNE detector, including a description of the
optical properties of the oil, is given in Ref.~\cite{detectorpaper}.

\subsection{Neutrino Interaction Simulation}

The MiniBooNE Monte Carlo simulation uses the NUANCE event
generator to simulate neutrino interactions~\cite{NUANCE}.  A detailed
description of MiniBooNE-specific modifications to NUANCE can be found
in Ref.~\cite{minibooneccqe2}, so only the most important details are
reproduced here.  NUANCE uses a relativistic Fermi gas model to
simulate the carbon nucleus.  The model is parametrized by a Fermi
momentum of $220\pm 30$ MeV/c and a binding energy of $34\pm 9$ MeV,
which are determined via electron scattering data~\cite{moniz}.

Although the details of the \ccpip\ simulation are not essential to
the extraction of the measured cross sections, comparisons of the
measurements with the default MiniBooNE prediction will be shown in
the results section.  In NUANCE, resonantly produced \ccpip\ events
are simulated using the Rein-Sehgal model~\cite{reinsehgal} with
$M_A=1.10\pm0.28~\gev/\textrm{c}^2$ determined from external
experimental data~\cite{campbell,barish,radecky,kitagaki}. The model
in NUANCE is further modified to include non-isotropic
$\Delta$-resonance decays according to Ref.~\cite{reinsehgal}. Pauli
blocking is additionally accounted for in the decay of the resonance
by requiring the momentum of the decay nucleon to be larger than the
Fermi momentum. Coherently produced \ccpip\ events are described using
Rein-Sehgal~\cite{reinsehgal2} with $M_A=1.03 \pm
0.28~\gev/\textrm{c}^2$, and with the overall cross section rescaled
by 0.65 to reproduce a prior MiniBooNE measurement of coherent pion
production in the NC channel~\cite{miniboonencpi0}.

The largest backgrounds to the \ccpip\ sample result from \ccqe\ and
\cc\ multi-$\pi$ events.  \ccqe\ interactions are simulated using
non-dipole vector form factors~\cite{bba}, a nonzero pseudoscalar form
factor~\cite{liu}, and a dipole axial-vector form factor with
$M_A=1.23 \pm 0.08~\gev/\textrm{c}^2$~\cite{minibooneccqe1} along with
an additional Pauli blocking rescaling, $\kappa=1.02 \pm 0.02$, as
measured from MiniBooNE \ccqe data~\cite{minibooneccqe1}. These values
for $M_A$ and $\kappa$ are different from those recently reported
in~\cite{minibooneccqe2} but were chosen as they were extracted using
the same, default model for resonance production in NUANCE as assumed
here. An additional 10\% normalization uncertainty is assumed to
account for the differences between the relativistic Fermi gas and
more modern nuclear models.  Multipion production processes are
modeled in NUANCE assuming $M_A=1.30 \pm 0.52~\gev/\textrm{c}^2$ such
that the sum of the exclusive CC channels reproduces CC inclusive
data.

For non-coherent scattering, NUANCE assumes neutrino interactions take
place on a single nucleon within the nucleus.  The resulting particles
(including resonances, nucleons, pions, etc.) can experience
final-state interactions as they traverse the nuclear medium.  For
example, baryonic resonances can re-interact in the nucleus producing
a pion-less final state with a probability of 20\% for $\Delta^++N$
and $\Delta^0+N$ interactions, and 10\% for $\Delta^{++}+N$ and
$\Delta^-+N$.  An uncertainty of 100\% is assumed for all four
interaction probabilities~\cite{minibooneccqe2}.  Pions can also
rescatter before exiting the target nucleus. Intranuclear pion
absorption and charge-exchange processes are assigned uncertainties of
25\% and 30\%, respectively, based on existing pion-carbon
data~\cite{ashery,jones,ransome}.  Since the signal for this analysis
is defined in terms of the particle content of the post-nuclear final
state, the measurement uncertainty is not significantly affected by
the uncertainties in intranuclear pion rescattering.

\section{Event Reconstruction}

Neutrino events are reconstructed based on the charge and time
recorded in each of the hit PMTs in the main tank volume.  For a
given set of seven initial track parameters -- energy, direction
($\theta$ and $\phi$), position ($x$, $y$, and $z$), and time -- both a
charge and a time probability distribution function (PDF) are produced
for each hit PMT.  The product of these PDFs evaluated at the
measured charge and time values give a likelihood function,
\begin{equation}\label{eq:likeraw}
\mathcal{L}({\bf x})
= \prod_{i=1}^{N_{unhit}}\mathcal{P}_i(\textrm{unhit};{\bf x})
\prod_{j=1}^{N_{hit}}\mathcal{P}_j(\textrm{hit};{\bf x})f(q_j;{\bf x})f(t_j;{\bf x}),
\end{equation}
where $N_{hit}$($N_{unhit}$) is the number of hit(unhit) PMTs in the
event and $\mathcal{P}_i(\textrm{hit};{\bf
x})$($\mathcal{P}_i(\textrm{unhit};{\bf x})$) is the probability that
PMT $i$ will be hit(unhit) for a track specified by ${\bf x}$.  The
charge and time PDFs for ${\bf x}$ ($f(q_j;{\bf x})$ and $f(t_j;{\bf
x})$, respectively) are evaluated at the measured charge, $q_j$, and
time, $t_j$, in PMT $j$.  The best set of parameters, ${\bf x}$, are
those that maximize the likelihood function.  A complete description
of the MiniBooNE extended-track reconstruction method is given in
Ref.~\cite{reconstructionnim}.

\subsection{Pion Reconstruction}

To properly reconstruct \ccpip events, a mechanism for separating
muons from charged pions is required.  Unlike the case of separating
muons and electrons, muons and pions propagate and emit Cherenkov
radiation in a very similar manner due to their similar masses.  The
main differences are due to the hadronic interactions experienced by
pions.  When such a hadronic interaction takes place, the pion
experiences an abrupt change in direction.  Since the nuclear debris
created in these interactions is generally well below Cherenkov
threshold, the only detectable prompt light is produced by the
``kinked'' pion trajectory.

To exploit these kinked pion trajectories, the straight-track fit
hypothesis has been generalized to include four new parameters that
characterize a kinked track.  The length of the upstream portion of
the track is determined by the energy lost prior to the kink point,
$\Delta E_{up}$.  The additional pion energy lost to the hadronic
system during the interaction, $\Delta E_{kink}$, is also allowed to
vary during the fit.  Finally, the independent direction of the
downstream track segment is characterized by two angles,
$\theta_{down}$ and $\phi_{down}$.

\subsubsection{Particle Identification Performance}

\begin{figure}
  \begin{center} \leavevmode
    \includegraphics[height=8cm,width=8cm,clip=true]{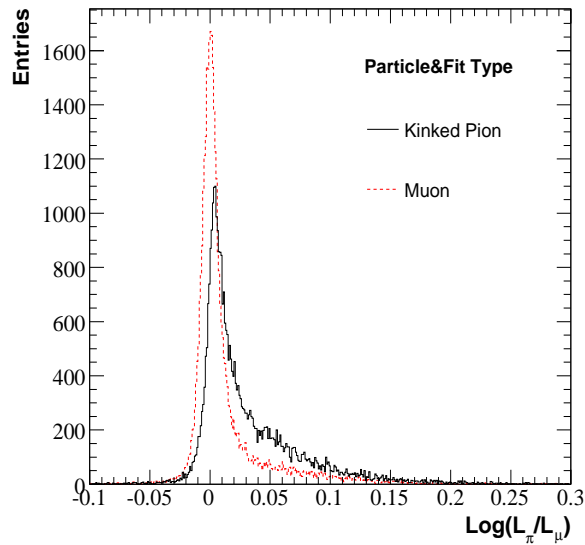}
    \caption { The straight muon and kinked pion likelihood ratios are
    shown for Monte Carlo muons (red) and pions (black) generated with
    full hadronic interactions and decays.  The particles were
    generated from a flat kinetic energy distribution ranging from 50
    to 450~MeV to approximate the true pion energy spectrum of \ccpip
    events.  There is separation in the muon and pion peaks, and a
    large excess of pion events is seen along the high side tail.  }
    \label{fig:kinklike} \end{center}
\end{figure}

The ability of the kinked pion fitter to separate muons from pions is
displayed in Fig.~\ref{fig:kinklike}.  The peak in the pion
likelihood ratio distribution is shifted relative to that of muons.
More significantly, the pion distribution has a much larger tail of
events that extend away from the muon portion of the likelihood ratio.
These are events where kinked trajectories occurred and were
successfully found by the fitter.

The $\mu/\pi$ separation provided by the kinked pion fitter is not as
clean as the $\mu$/e separation.  There is no single value of the
likelihood ratio at which a cut could be placed that would reject a
large population of muons while retaining a significant fraction of
pions.  The goal of this analysis, however, is to reconstruct events
with both a muon and a pion present, and to determine the identity of
each track.  In that case, the separation power indicated by
Fig.~\ref{fig:kinklike} is doubled due to the presence of the second
track.

\subsubsection{Kinematics Reconstruction Performance}

\begin{figure}
  \begin{center}
    \includegraphics[height=8cm,width=8cm,clip=true]{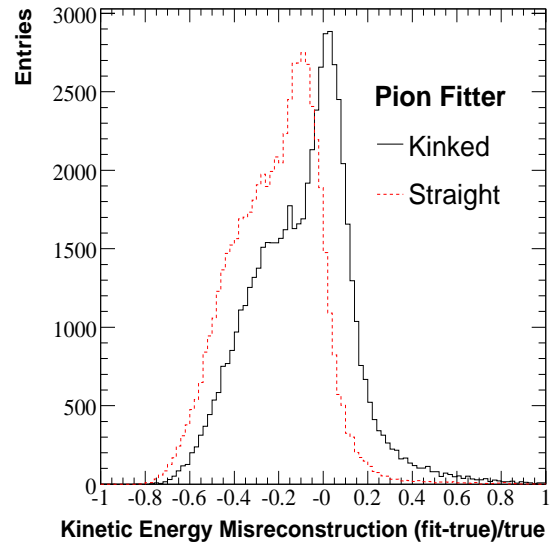}
    \caption{The fractional pion kinetic energy reconstruction bias from
	  the straight and kinked pion fits to Monte-Carlo-generated
	  single pion events is shown.  The low-energy shoulder is
	  significantly reduced in the kinked fitter, and rather than
	  being 10\% low, as is the case with the straight fitter, the
	  peak from the kinked fitter is centered at zero.  }
    \label{fig:epires1d}
  \end{center}
\end{figure}

\begin{figure}
  \begin{center}
    \includegraphics[height=8cm,width=8cm,clip=true]{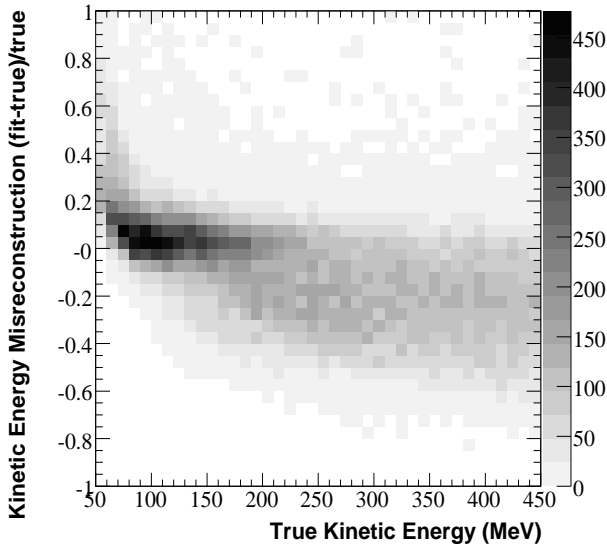}
    \caption
	{ The fractional pion kinetic energy reconstruction bias is
	  plotted versus the true energy for Monte-Carlo-generated
	  single pion events.  The low-fit-energy shoulder is
	  caused by higher energy pions.  }
    \label{fig:epiresdd}
  \end{center}
\end{figure}

\begin{figure}
  \begin{center}
    \leavevmode
    \includegraphics[height=8cm,width=8cm,clip=true]{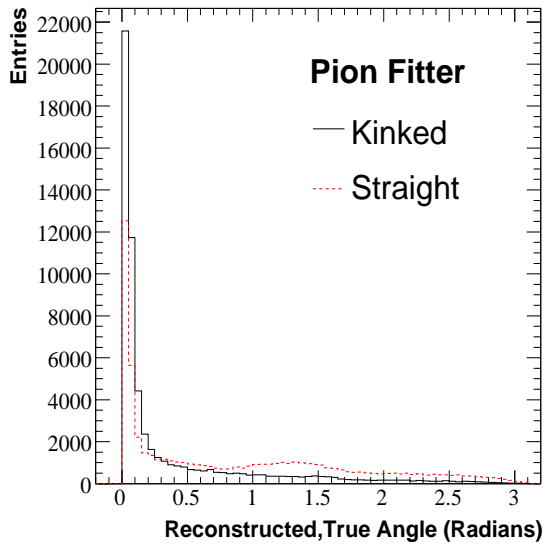}
    \caption
	{ The angle between the reconstructed and true pion directions
	  is shown for both straight and kinked pion fits to
	  Monte-Carlo-generated single pion events.  The population in
	  the lowest few bins where the properly reconstructed events
	  lie is nearly twice as large for the kinked fitter.  }
    \label{fig:piangleres}
  \end{center}
\end{figure}

Although the main motivation for developing a kinked-track fitter was
to provide a means for separating muons from charged pions, the
improved modeling of pion trajectories results in superior event
reconstruction as well.  The fractional energy reconstruction bias
(i.e. the ratio of the fit/true difference to the true value) from
both the straight and kinked pion fitters is shown in
Fig.~\ref{fig:epires1d}.  The straight pion fitter reconstructs pion
energies 10\% low, whereas the kinked fitter reconstruction bias peaks
at zero.  In addition, the ``shoulder'' just below the peak, where the
reconstructed energy underestimates the true pion energy, is reduced
by the kinked pion fitter.  The two-dimensional plot of the fractional
energy reconstruction bias versus the true energy in
Fig.~\ref{fig:epiresdd} shows that the shoulder comes from higher
energy pions that can produce multi-kink events and cause larger pion
energy losses at each kink.  The pion direction reconstruction is also
significantly improved with the kinked fitter, as shown in
Fig.~\ref{fig:piangleres}.  The event population in the first few bins
of the angle between the reconstructed and true directions are nearly
doubled in the kinked fitter relative to the straight fitter.


\subsection{\ccpip Fit}

With the ability to reconstruct charged pions, a full \ccpip\ fitter
is formed by simultaneously fitting for a straight muon and a kinked
pion track.  A \ccpip\ fit has 14 parameters: a common vertex (4
parameters), the initial energy and direction of both the muon and
pion (6 parameters), and the additional kinked-track parameters for
the pion (4 parameters).  Just as in the kinked pion fitter, the
predicted charges from all track segments (upstream pion, downstream
pion, and muon) are summed to get the total predicted charge for each
PMT\@.

\subsubsection{\ccpip Fit Seeding}

Each \ccpip\ event is assumed to have three Cherenkov rings from the
three-track segments in the event: the upstream pion segment, the
downstream pion segment, and the muon track.  Each ring is found in
succession using intermediate two- and three-track likelihood
functions.  The three tracks are then pieced together in several
different pairings to create the kinked pion track and the straight
muon track.  The pairing that produces the best likelihood is used to
seed the \ccpip\ fitter.

The first of the three rings is found by performing a one-track fit.
The results of the fit are frozen in place, and a scan for a second
track is performed over 100 equally spaced directions.  At each scan
point, a two-track likelihood function is evaluated, and the
configuration that gives the best likelihood value is used to seed a
full two-track fit.  This process is then repeated by freezing the
result of the two-track fit and scanning for a third track using a
three-track likelihood function.  The result of the scan seeds a full
three-track fit.

Once three tracks are found that characterize the Cherenkov rings in
the event, they are combined to form a straight muon and a kinked
pion.  Only pairings where the downstream pion track has a lower
energy than the upstream pion track are allowed.  This reduces the
number of possible groupings to three.  Each of these three seeds is
passed to the full \ccpip\ fitter, and the fit that produces the best
likelihood is chosen.

\subsubsection{Fit Results}

\begin{figure}
  \begin{center} \caption{The fractional muon kinetic energy
    reconstruction bias is shown for all Monte Carlo signal events, and
    for correctly paired signal events after all cuts other than the
    $m_{\pi+N}$ cut.  Most of the low-fit-energy tail is due to events
    where the pion was misidentified as the muon.}
    \includegraphics[scale=0.45,clip=true]{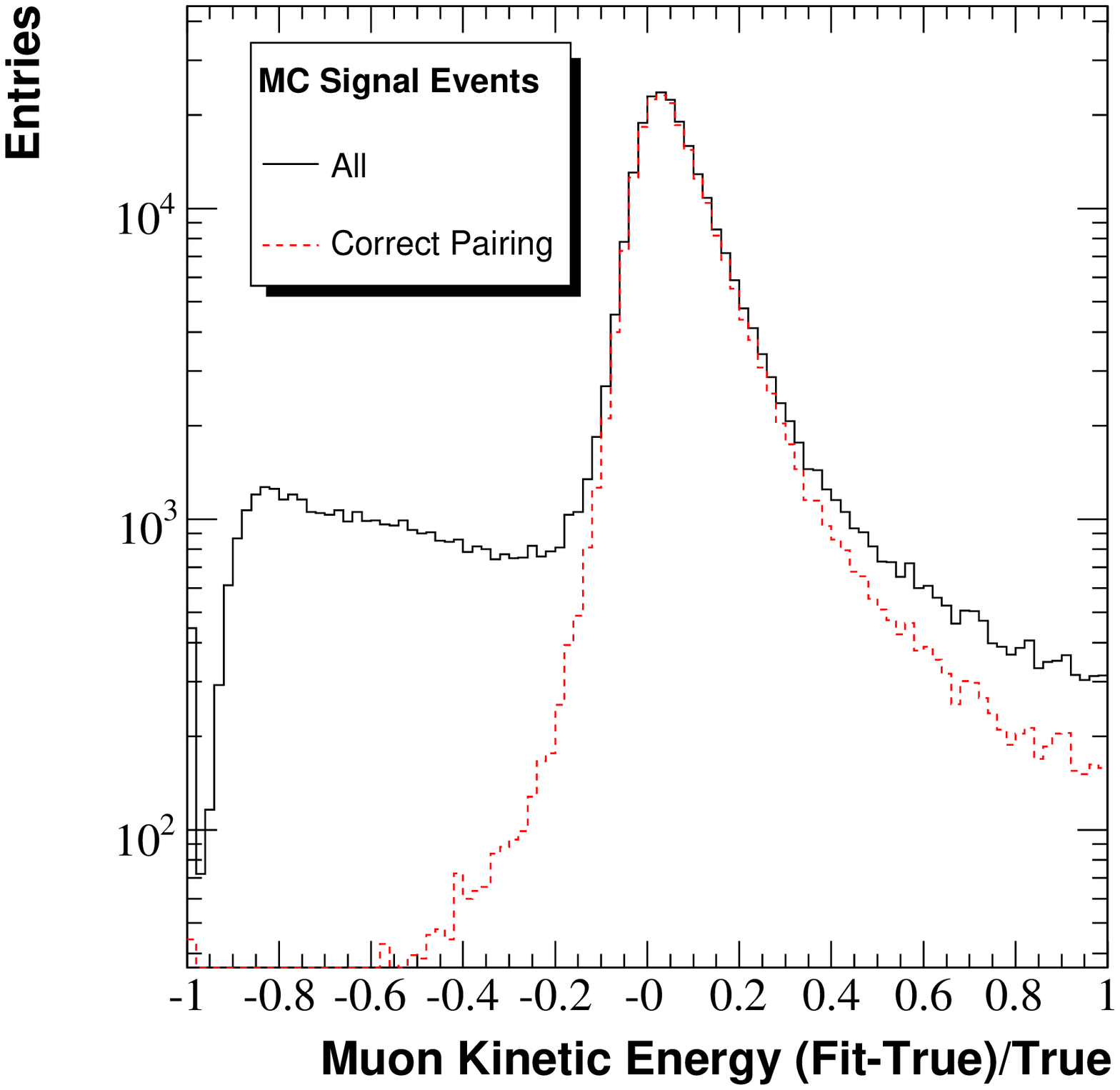}
    \label{fig:eresmu} \end{center}
\end{figure}

\begin{figure}
  \begin{center} \caption{The fractional pion kinetic energy
    reconstruction bias is shown for all Monte Carlo signal events, and
    for correctly paired signal events after all cuts other than the
    $m_{\pi+N}$ cut.  The low-reconstructed-energy shoulder from the
    pion-only fit in Fig.~\ref{fig:epires1d} is seen here as well.}
    \includegraphics[scale=0.45,clip=true]{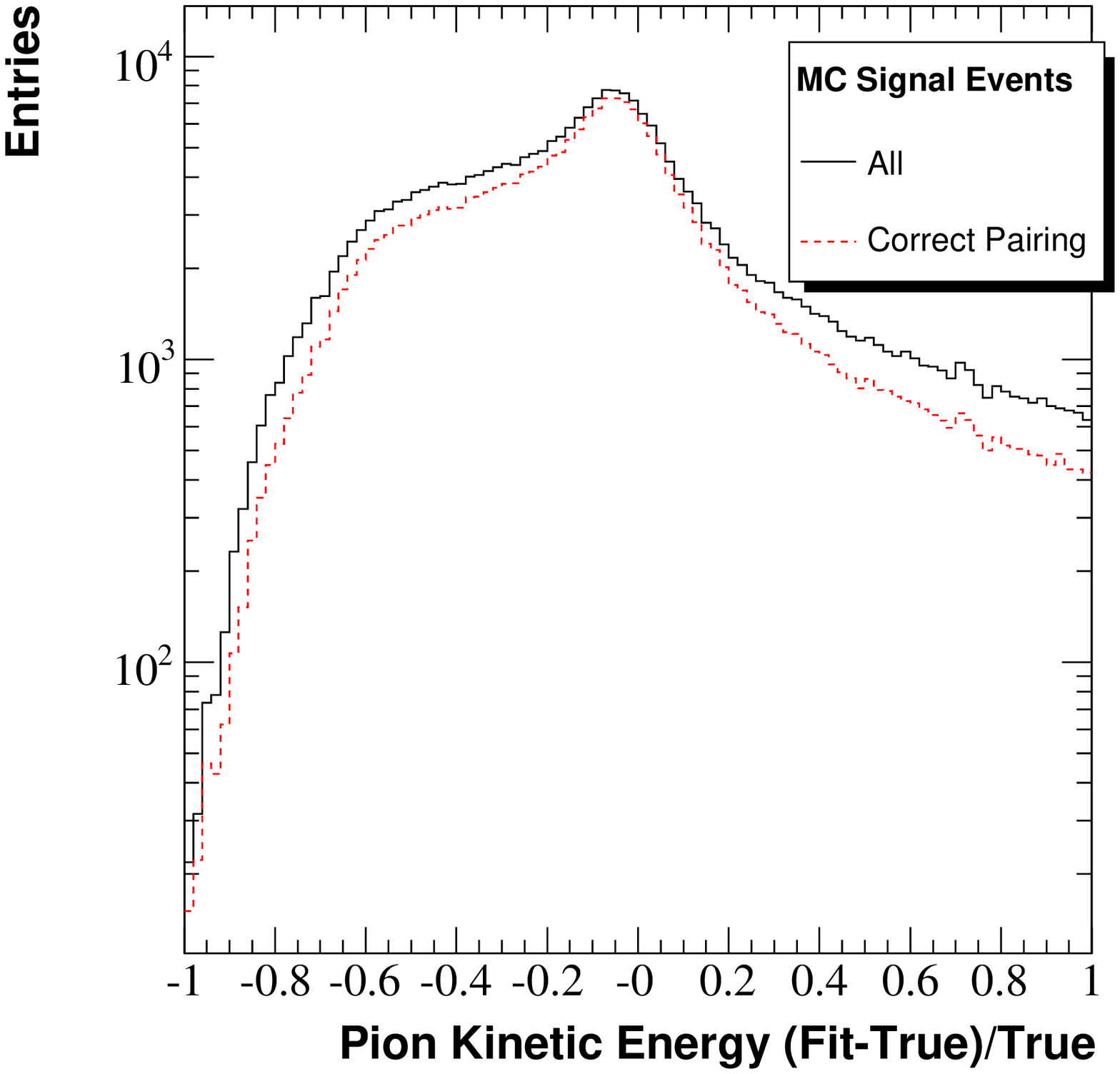}
    \label{fig:erespi} \end{center}
\end{figure}

The fractional kinetic energy reconstruction bias distributions for the
muon and pion tracks are given in Fig.s~\ref{fig:eresmu} and
\ref{fig:erespi}, respectively.  The muon kinetic energy has a small
tail at low reconstructed energy due to $\mu/\pi$ mispairing.  The
reconstructed pion kinetic energy has the same low-energy shoulder
from high energy particles seen in the pion-only fit in
Fig.~\ref{fig:epires1d}.  In addition, the fitter tends to place
about 5\% too much energy in the muon track at the expense of the pion
track.


While the pion energy fit is more accurate at low track energies, the
opposite is true for the reconstructed pion direction.  The track
direction reconstruction relies on the existence of a well-defined
Cherenkov ring from the upstream pion track segment.  At MiniBooNE
neutrino energies, 16\% of the generated pion kinetic energy spectrum
lies below 70~MeV.  This corresponds to an above-Cherenkov propagation
distance of less than 10~cm, which is often insufficient to determine
the direction.  For comparison, fewer than 1\% of muons are generated
below 70~MeV.  In addition, the pion inelastic interaction length in
mineral oil in the energy range of interest is approximately
1~m~\cite{ashery}, which means 10\% of all pions will have upstream
segments shorter than 10~cm.

\begin{figure}
  \begin{center} \leavevmode
    \includegraphics[height=8cm,width=8cm,clip=true]{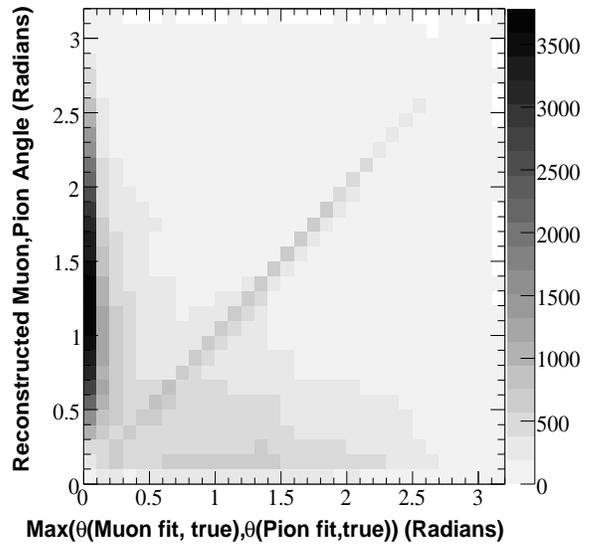}
    \caption { The reconstructed angle between the muon and pion
    directions for Monte Carlo \ccpip\ events is shown versus the
    larger (i.e. worse reconstructed) of the two reconstructed/true
    angles: $\theta(\mu_{rec},\mu_{true})$ and
    $\theta(\pi_{rec},\pi_{true})$.  The bins in the left-most columns
    represent events where both tracks have been properly
    reconstructed.  Events in which the tracks have been misidentified
    appear along the diagonal.}  \label{fig:ares} \end{center}
\end{figure}

The ability of the fitter to correctly reconstruct both the muon and
the pion directions is shown in Fig.~\ref{fig:ares}.  The
reconstructed angle between the muon and pion is plotted against the
larger of the two reconstructed/true angles.  A perfect fitter would
place all events in the lowest column where both reconstructed/true
angles are zero.  Fig.~\ref{fig:ares} shows that those bins contain
the largest event population.  The other significant event population
is along the diagonal of the plot.  These are events where the fitter
has misidentified the muon as a pion and vice versa.  In such cases,
the angle between the true and reconstructed directions of both the
muon and pion will be near the reconstructed $\mu/\pi$ angle.  This is
the first time charged pions have been tracked and their kinematics
measured in a Cherenkov detector.

\subsubsection{Neutrino Energy}\label{sec:enuresult}

\begin{figure}
  \begin{center}
    \caption{The neutrino energy reconstruction bias
    is plotted against the true neutrino energy for Monte Carlo
    generated \ccpip\ events.  The reconstructed and true values are
    well correlated over the entire energy spectrum.}
    \includegraphics[height=8cm,width=8cm,clip=true]{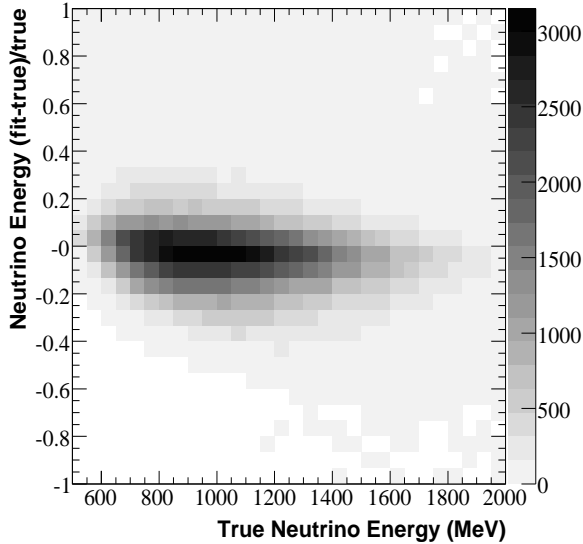}
    \label{fig:enurecvtrue}
  \end{center}
\end{figure}

With reconstructed energies and directions for both the muon and pion,
 the energy of the incident neutrino can be determined.  Assuming the
 target nucleon is at rest and the remaining, unmeasured final-state
 particle is a nucleon, the neutrino energy is specified by
\nolinebreak[4]{4-momentum} conservation,
\begin{equation}
E_{\nu}=\frac{m_{\mu}^2+m_{\pi}^2-2m_N(E_{\mu}+E_{\pi})+2p_{\mu}\cdot p_{\pi}}{2\left(E_{\mu}+E_{\pi}-\left|{\bf p}_{\mu}\right|\cos\theta_{\nu,\mu}-\left|{\bf p}_{\pi}\right|\cos\theta_{\nu,\pi}-m_N\right)},
\end{equation}
where $m_{x}$, $E_{x}$, $p_{x}$, and $\left|{\bf p}_{x}\right|$ are
the mass, energy, 4-momentum, and 3-momentum magnitude of particle $x$
in the detector frame, and $\theta_{\nu,\mu}$($\theta_{\nu,\pi}$) is
the angle between the directions of the muon(pion) and the neutrino.
The neutrino direction is determined by the event vertex location and
the mean neutrino emission point from the beam Monte Carlo prediction,
although the large distance between the beam and the detector means
this angle is never larger than one degree.  The comparison between
reconstructed and true neutrino energy is given in
Fig.~\ref{fig:enurecvtrue}.  The resolution is 13.5\% over most of the
sensitive range, with a slight increase at the highest energies.

The decreased pion angular resolution at lower pion energies has
little impact on the neutrino energy reconstruction since the neutrino
energy calculation becomes less sensitive to the reconstructed pion
direction as the pion energy is reduced.  In addition, events with
misidentified tracks that are otherwise well-reconstructed will
produce nearly the same neutrino energy, since muons and pions have
similar masses.

\subsubsection{Invariant mass of the hadronic system}

By making the aforementioned assumptions required to calculate the
neutrino energy, the kinematics of the interaction are fully
specified.  Previous attempts to measure \ccpip\ interactions by
reconstructing only the muon required the additional assumption that
the recoiling particle was an on-shell $\Delta$ baryon~\cite{morgan}.
Since the width of the $\Delta$ resonance is about 10\% of its mass,
this assumption results in an irreducible contribution to the neutrino
energy resolution.  By measuring the pion kinematics, the $\Delta$
mass constraint can be removed.

\begin{figure}
  \begin{center}

  \caption{The reconstructed $\pi+N$ mass distribution is shown for
    both the data and the Monte Carlo simulation with full systematic
    uncertainties.  The MC distribution has been normalized to the
    data.  The signal and background components of the Monte Carlo
    distribution are also shown.  At MiniBooNE energies, the majority
    of \ccpip\ events come from decays of the $\Delta(1232)$
    resonance.}
    \includegraphics[scale=0.52,clip=true]{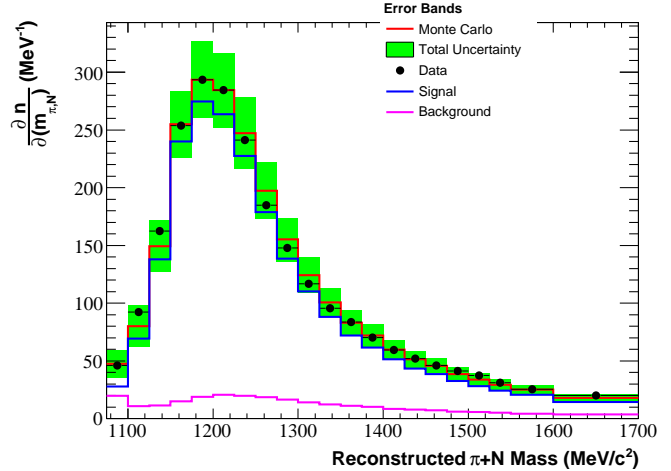}
    \label{fig:dmass} \end{center}
\end{figure}

\begin{figure}
  \begin{center}
    \includegraphics[scale=0.42,clip=true]{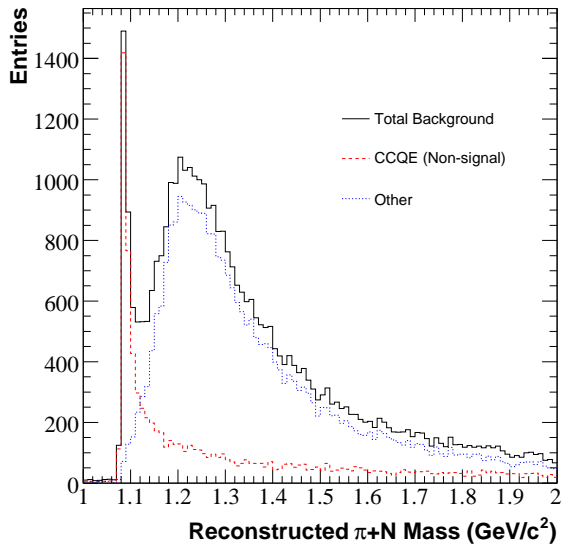}
    \caption
	{ The reconstructed Monte Carlo \pip+$N$ background
	  distribution is divided into \ccqe\ background events, and
	  all other backgrounds.  Since the \ccqe\ events do not
	  contain a pion, they are peaked near threshold
	  ($m_{\pi}+m_N$).  }
    \label{fig:dmassbkg}
  \end{center}
\end{figure}

The absence of a $\Delta$ mass constraint also means that the $\pip+N$
invariant mass, which is dominated by the $\Delta$ resonance, can be
measured.  Fig.~\ref{fig:dmass} shows the reconstructed $\pip+N$
mass, and a breakdown of the background composition is given in
Fig.~\ref{fig:dmassbkg}.  The \ccqe\ background features a sharp
peak near threshold.  \ccqe\ interactions typically do not produce a
pion, and the fitter correctly assigns very little kinetic energy to
the hadronic system in these events.


\subsubsection{Momentum Transfer}

\begin{figure}
  \begin{center}
    \caption{The $Q^2$ (fit-true)/true distribution for Monte Carlo
    events is plotted versus the true $Q^2$.  Each column of true
    $Q^2$ has been normalized to unity.  The reconstruction is able to
    determine the true $Q^2$ over the full range of the measurement.}
    \includegraphics[scale=0.45,clip=true]{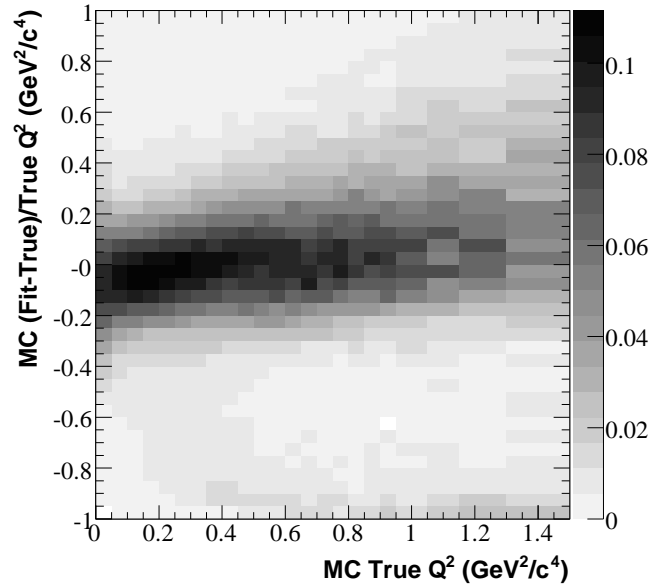}
    \label{fig:qsqresvtrue}
  \end{center}
\end{figure}

The final variable measured in this analysis is the 4-momentum
transfer, $q$, from the leptonic current to the hadronic portion of
the decay, which is characterized by its relativistic invariant,
\begin{equation}
Q^2\equiv-(p_{\mu}-p_{\nu})^2.
\end{equation}
Since $Q^2$ is a property of the exchanged $W$ boson, it is completely
specified by the change in the leptonic current.  However, this also
means that, unlike the neutrino energy calculation, the reconstructed
$Q^2$ distribution is quite sensitive to $\mu/\pi$ misidentification.
Fig.~\ref{fig:qsqresvtrue} shows the fractional error in the
reconstructed $Q^2$ distribution, normalized in columns of true $Q^2$.
Most of the columns peak near zero, but at high $Q^2$, a second
population of events appears in which the fit underestimates the true
$Q^2$.  These events are dominated by a high energy muon that has been
misidentified as a pion, giving the impression that most of the
neutrino momentum was transferred to the hadronic system.  After all
analysis cuts, the $Q^2$ resolution is 18\% below 0.3 GeV$^2/c^4$ and
20\% below 1~\gev$^2/c^4$.

\section{Event Selection}\label{sec:evtsel}

MiniBooNE events are recorded in a 19.2~$\mu s$ time window beginning
4.6~$\mu s$ prior to the arrival of the 1.6~$\mu s$ beam pulse.  The
event time window is further subdivided into groups of PMT hits called
subevents.  The hit times in each subevent must have no more than two
gaps longer than 10 ns and no gaps longer than 20 ns.

The first subevent in each event consists of the final-state particles
produced in the neutrino interaction.  Muons and pions can stop in the
detector and decay to produce Michel electrons.  These Michel
electrons result in additional subevents that provide a simple and
powerful tool for separating neutrino event types.  \ccqe\ events,
which contain a muon in the final state, most often produce two
subevents.  Since \ccpip\ events are more likely to produce two Michel
electrons from the decays of the final-state muon and pion, events are
required to have three subevents.

\begin{figure}
  \begin{center}
    \leavevmode
    \includegraphics[scale=0.45,clip=true]{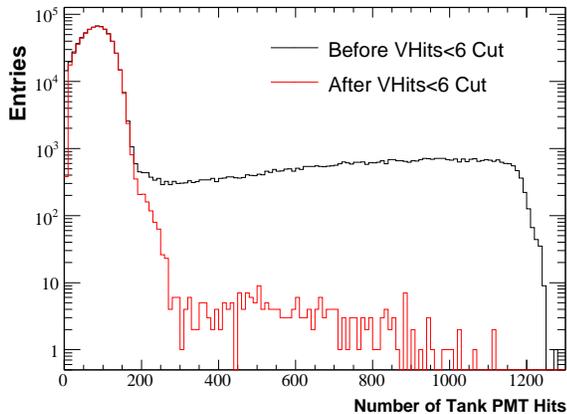}
    \caption
	{ The tank hits distribution in data is shown for the second
	  subevent before and after requiring fewer than 6 hits in the
	  veto.  The Michel electron peak is mostly unaffected, while
	  the large tail from entering comic rays is mostly removed.
	  }
    \label{fig:thitsvhits}
  \end{center}
\end{figure}

To remove backgrounds from cosmic rays, fewer than six hits are
allowed in the veto region in all three subevents.  The effect of this
cut on the second subevent is shown in Fig.~\ref{fig:thitsvhits}.
If a cosmic muon enters the tank before the beginning of the event
time window, it can stop and decay within the event time window to
simulate a neutrino interaction.  These events are removed by
requiring a minimum number of PMT hits in the main tank in the first
subevent.  If this cut is placed at 200 hits, more than 99.9\% of beam
unrelated backgrounds are removed~\cite{detectorpaper}.  For the
present analysis, the tank hits requirement has been relaxed to 175
hits since it is unusual for a Michel electron event to produce three
subevents.  The second and third subevents are required to have
between 20 and 200 hits in the main tank to accept subevents from
Michel electrons.

\begin{figure}
  \begin{center}
    \centering
    \caption{The distance between the tank wall and event vertex along
    the muon trajectory is shown for both data and Monte Carlo events.
    All other cuts have been applied and the error bars include only
    data statistics.  The cut on this distribution removes all events
    below 150 cm.}
    \label{fig:rtowallmu}
    \includegraphics[scale=0.45,clip=true]{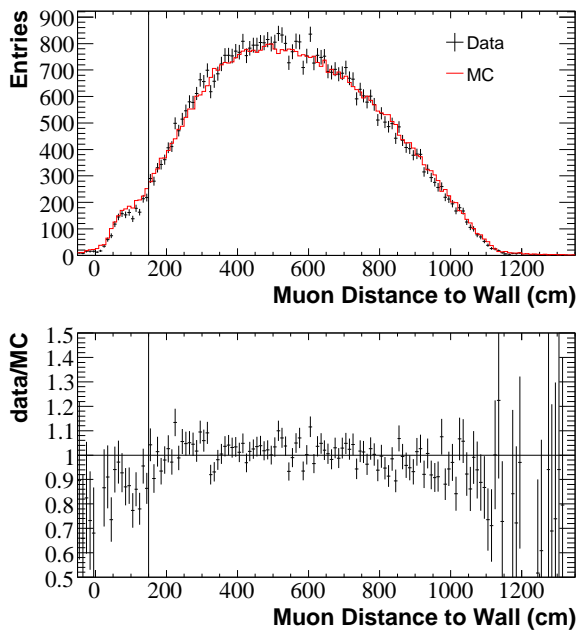}
  \end{center}
\end{figure}

\begin{figure}
  \begin{center}
    \centering
     \caption{The distance between the tank wall and event vertex
     along the pion trajectory is shown for both data and Monte Carlo
     events.  All other cuts have been applied and the error bars
     include only data statistics.  The cut on this distribution
     removes all events below 150 cm.}
    \label{fig:rtowallpi}
    \includegraphics[scale=0.45,clip=true]{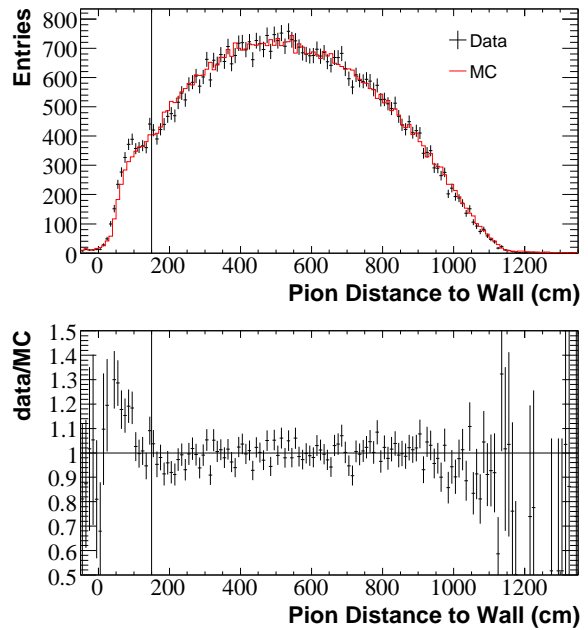}
  \end{center}
\end{figure}

To remove events that occur close to the edge of the detector, the
muon and pion tracks are required to travel at least 150~cm before
reaching the wall.  Particle trajectories that begin near the edge of
the tank and are directed toward the tank wall are poorly
reconstructed since they are detected by a small number of PMTs.
Conversely, events that occur just inside the tank wall but consist of
tracks that all point toward the interior of the tank are well
reconstructed.  An illustration of this cut is given in
Fig.s~\ref{fig:rtowallmu} and \ref{fig:rtowallpi}.

\begin{figure}
  \begin{center} \leavevmode
    \includegraphics[scale=0.42,clip=true]{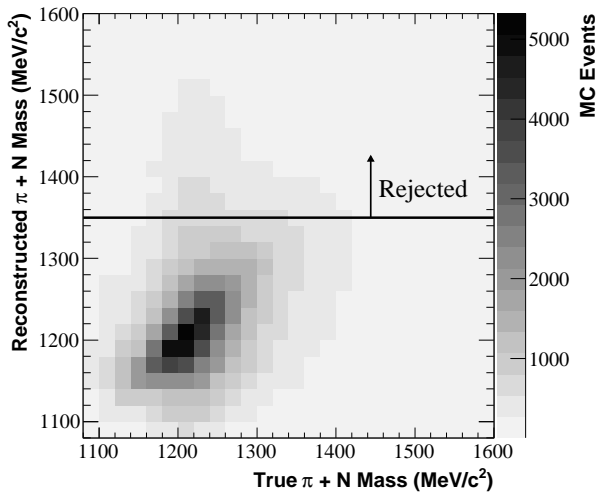}
    \caption { The Monte Carlo $m_{pi+N}$ distribution shows a
    correlation between the reconstructed and true distributions at
    low mass.  At high reconstructed mass, the distribution is
    dominated by events with a high energy muon misidentified as a
    pion.  A cut is placed at 1350~\mev/c$^2$ to remove these events.
    } \label{fig:dmasscut} \end{center}
\end{figure}

\begin{figure}
  \begin{center} \leavevmode
    \includegraphics[scale=0.45,clip=true]{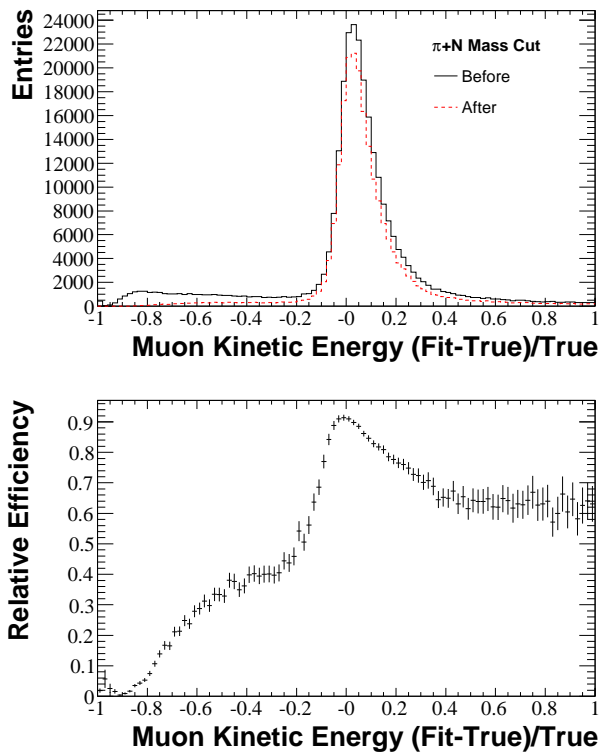}
    \caption { The fractional muon kinetic energy reconstruction bias
    is shown for Monte Carlo \ccpip\ events before and after the
    $m_{\pi+N}$ cut with all other cuts applied.  The lower plot shows
    the fraction of events that pass the cut in each bin.  The tail at
    low-fit-energy is significantly reduced.  }
    \label{fig:muonkedmasscut} \end{center}
\end{figure}

A final cut on the $\pi+N$ mass ($m_{\pi+N}$) is used to eliminated
events where the final-state particles are misidentified.  The fitter
misreconstructs the muon as a pion, and vice versa, 21.4\% of the
time.  In cases where a high energy muon is misreconstructed as a
pion, the reconstructed kinematics give a spuriously high $m_{\pi+N}$.
The relationship between the generated and reconstructed $m_{\pi+N}$
is shown in Fig.~\ref{fig:dmasscut}.  Beyond reconstructed masses of
1350~\mev/c$^2$, the population of misreconstructed events begins to
dominate, so a cut is implemented to remove these events.
Fig.~\ref{fig:muonkedmasscut} shows the improvement in the
reconstructed muon kinetic energy resulting from the $m_{\pi+N}$ cut.
The properly reconstructed events are mostly retained while the tail
at low reconstructed energy is greatly reduced.

A well-matched event is defined in terms of the angles between true
and reconstructed tracks.  There are four such angles: $\theta$(true
$\mu$, fit $\mu$), $\theta$(true $\pi$, fit $\pi$), $\theta$(true
$\mu$, fit $\pi$), $\theta$(true $\pi$, fit $\mu$).  If the minimum of
these four angles is between a true track and its corresponding
reconstructed track, the event is said to be well-matched.  This means
that an event containing two properly reconstructed tracks is
well-matched if the fitter correctly identifies the muon and pion
tracks.  Events where the direction of one of the tracks is
misreconstructed (e.g. one track is below Cherenkov threshold) are
still considered to be well-matched if the angle between the measured
track and its corresponding true track is small.  The fraction of
well-matched events increases from 78.6\% to 88.0\% with the
introduction of the $m_{\pi+N}<1350~\mev/\textrm{c}^2$ cut.

After all cuts, 48,322 events are seen in the data with an overall
signal efficiency of 12.7\%, and a purity of 90.0\%.  The background
contributions are labeled according to the particles produced in the
initial neutrino interaction (i.e. prior to any final-state effects),
rather than the final state emerging from the nucleus.  The largest
backgrounds (listed by percentage of the total sample) are from
CC multi-$\pi$ events (3.1\%), \ccqe\ events (2.7\%), and \ccpip\ events
(1.3\%) in which the pion content of the final state is altered via
nuclear effects and are therefore not considered signal events.  The
complete lists of both signal and background compositions are given in
Tables~\ref{tab:sigrates} and \ref{tab:bkgrates}, respectively.

\begin{table*}
\caption[Signal efficiency, purity, and composition]{The signal
efficiency and purity are shown after each successive analysis cut.
The efficiency is given relative to all interactions that occur within
the detector, including the outer veto region; if restricted to events
produced within 100 cm of the tank center, the effiency rises to 27\%.
A full list of efficiency and purity by NUANCE event type is given in
Ref.~\cite{wilkingthesis}.
\label{tab:sigrates}}
\begin{tabular}{llcc}\hline\hline
Cut \# & \multicolumn{1}{c}{Description} & Effic.(\%) & Purity (\%) \cr\hline
1 & no cuts & 100 & 18.7 \cr
2 & 1st subevent, tank hits $> 175$ and veto hits $< 6$ & 47.5 & 23.6 \cr
3 & number of subevents = 3 & 23.2 & 60.5 \cr
4 & 2nd and 3rd subevents: tank hits $< 200$ and veto hits $< 6$ & 19.6 & 86.3 \cr
5 & muon and pion distance to wall $> 150$ cm & 16.9 & 87.2 \cr
6 & $\pi+N$ mass $< 1350~\mev/\textrm{c}^2$ & 12.7 & 90.0 \cr\hline\hline
\end{tabular}
\end{table*}

\begin{table*}
\caption[Background acceptance and composition]{The composition of
the background after all analysis cuts is given in terms of the NUANCE
interaction mode (i.e. before final-state interactions).  The total
background is 10\% of the final event sample.  The background
acceptance and contamination after each cut is given in
Ref.~\cite{wilkingthesis}.
\label{tab:bkgrates}}
\begin{tabular}{ccc}\hline\hline
NUANCE Interaction & Initial State Description & Background Fraction (\%) \cr\hline
CC multi-$\pi$ & \multicolumn{1}{l}{1 muon, $>$1 pion, and 1 nucleon} & 31.4\cr
\ccqe      & \multicolumn{1}{l}{1 muon and 1 proton} & 26.8 \cr
\ccpip     & \multicolumn{1}{l}{1 muon, 1 \pip, and $\le$1 nucleon} & 13.4 \cr
\ccmesonb  & \multicolumn{1}{l}{1 muon, 1 non-$\pi$ meson, and 1 baryon} & 8.5 \cr
\ccdis     & \multicolumn{1}{l}{1 muon and multiple hadrons} & 6.6 \cr
\ccpiz     & \multicolumn{1}{l}{1 muon, 1 \piz, and 1 proton} & 5.7 \cr
\anu       & \multicolumn{1}{l}{all \anu\ interactions} & 1.1 \cr
other      &  & 6.5 \cr\hline
\end{tabular}
\end{table*}

\section{Cross Section Measurements}\label{sec:xsecmeasurements}

The observable \ccpip\ analysis includes measurements of the cross
section in terms of several kinematic variables.  The integrated cross
section has been measured as a function of neutrino energy.  The other
one-dimensional measurements are differential cross sections as a
function of the muon and pion energies and $Q^2$.  Since these
one-dimensional measurements are necessarily averaged over the full
neutrino energy spectrum, a corresponding two-dimensional measurement
of each variable is made in bins of neutrino energy.  In addition, the
energy and direction are measured together for both the muon and pion
in two double-differential cross section measurements.

\subsection{Data Unfolding}\label{sec:unfolding}

Because of biases and imperfect resolution in the event
reconstruction, the event distributions measured in the data do not
perfectly reflect the underlying true distributions.  For example, as
shown in Fig.~\ref{fig:epiresdd}, the pion energy reconstructed by the
\ccpip\ fitter is systematically high at low true energy, and falls
below the true energy as the energy increases.  Since such biases are
modeled in the Monte Carlo simulation, it is possible to ``unfold''
these bin-migration effects.

\begin{figure}
  \begin{center}
    \leavevmode
    \includegraphics[scale=0.4,clip=true]{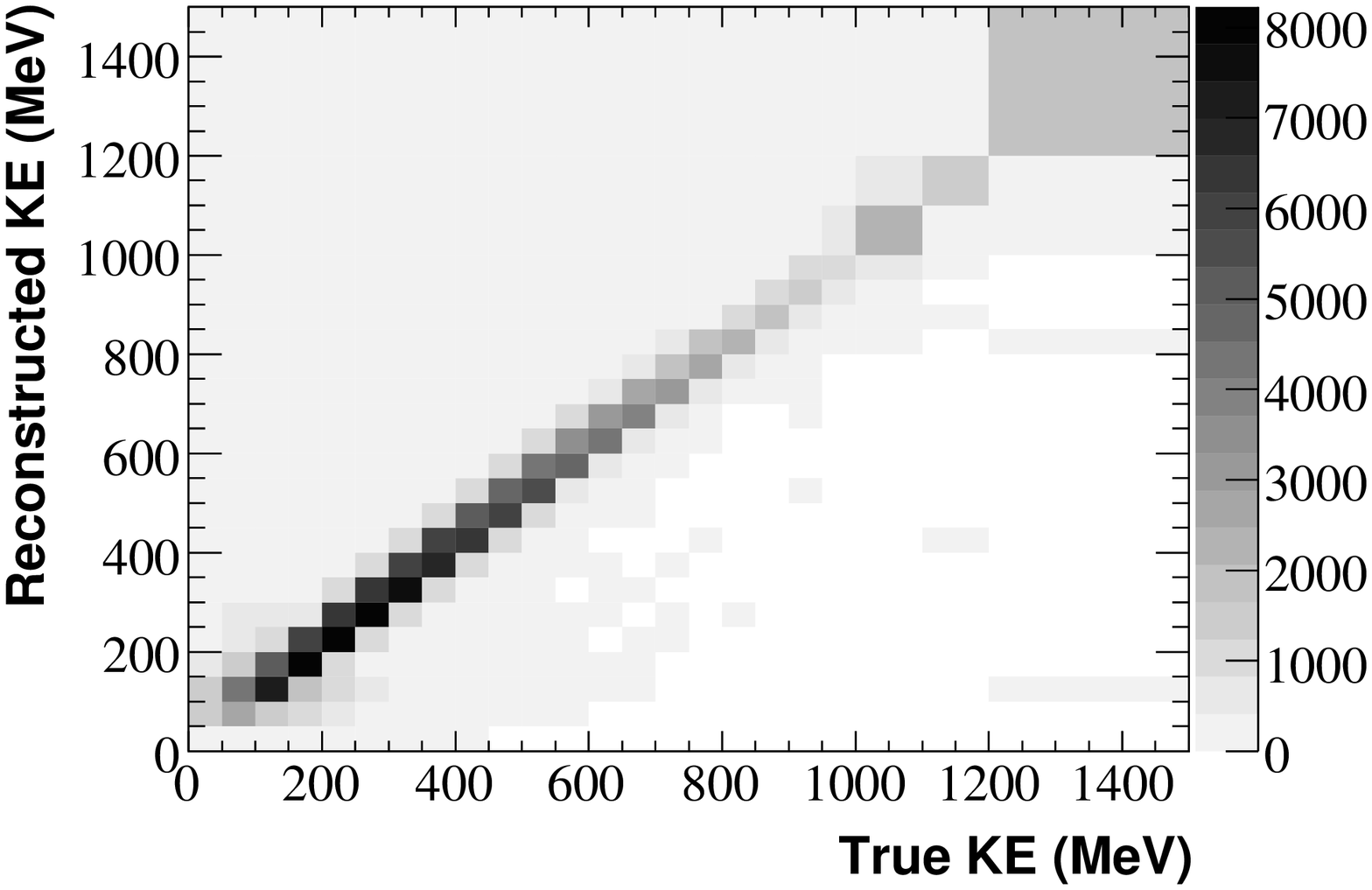}
    \includegraphics[scale=0.4,clip=true]{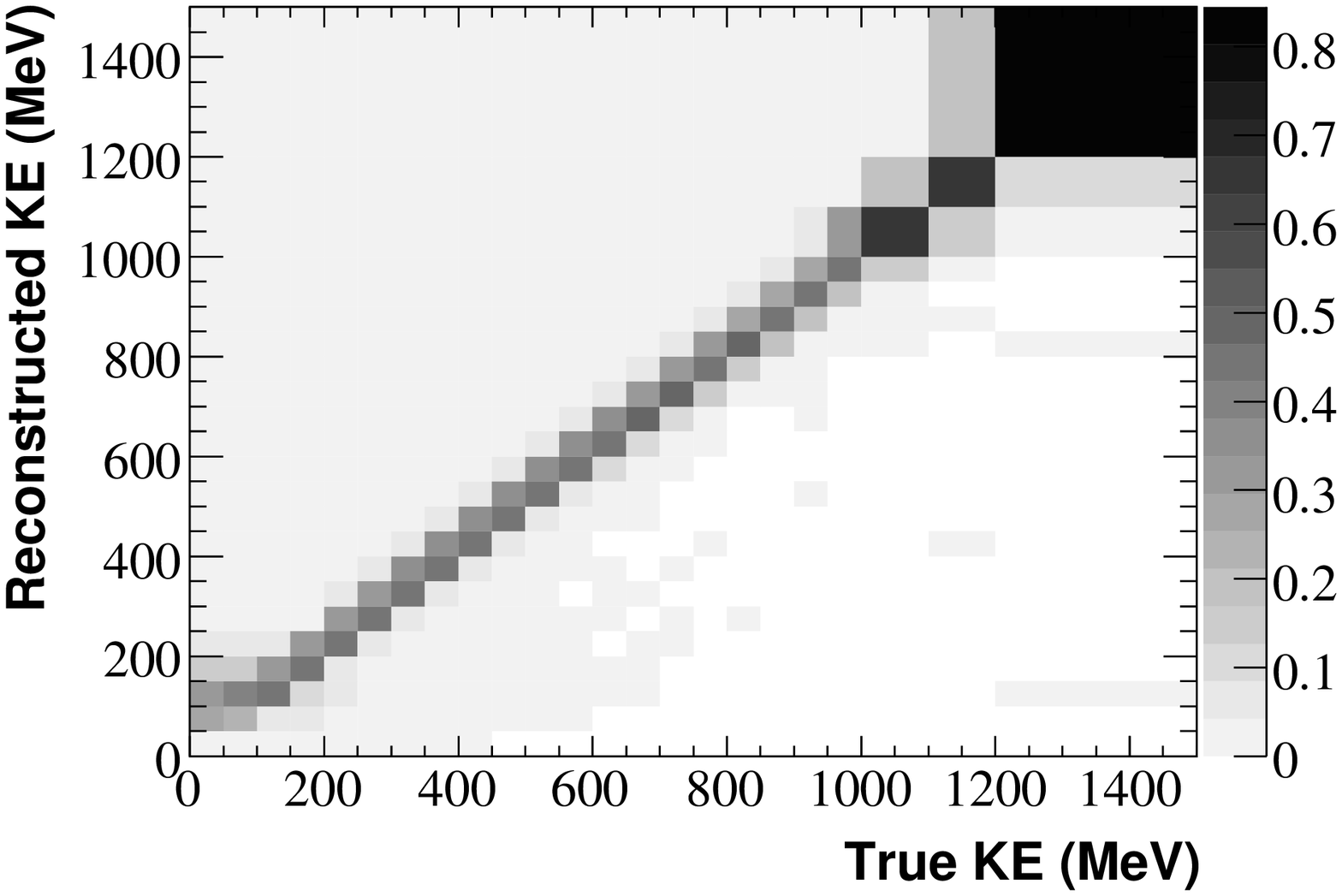}
    \includegraphics[scale=0.4,clip=true]{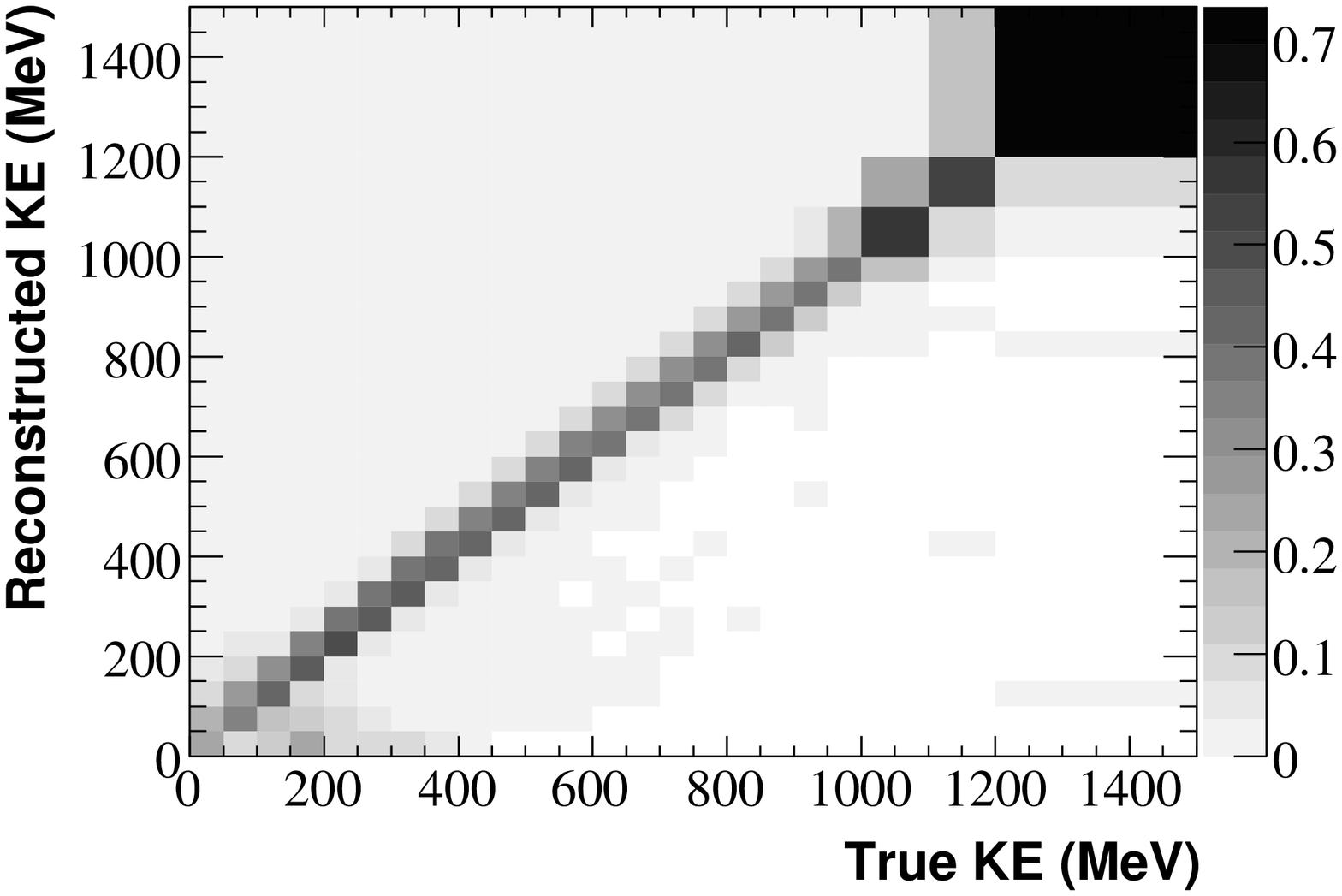}
    \caption
      { The reconstructed-versus-true (top), bin-migration (middle),
      and unfolding (bottom) matrices are shown for muon kinetic
      energy.  The bin-migration matrix is formed by normalizing the
      true columns of the Monte Carlo reconstructed-versus-true matrix
      to unity, while the unfolding matrix is formed by normalizing
      the reconstructed rows.  }
    \label{fig:foldunfold}
  \end{center}
\end{figure}

The Monte Carlo bin-migration matrix for a given variable, $v$, is
constructed by forming a two-dimensional histogram of the
reconstructed value of $v$ versus the true value, and normalizing each
true column to unity as illustrated in Fig.~\ref{fig:foldunfold}.
Each element, $F_{ji}$, in the bin-migration matrix represents the
probability that an event generated with a value of $v$ in bin $i$
will be reconstructed in bin $j$.  By definition,
\begin{equation}\label{eq:fold}
N^{int}_j = \sum_iF_{ji}T_i,
\end{equation}
where $N^{int}_j$ is the reconstructed distribution and $T_i$ is the true
distribution.

To produce an unfolding matrix, the reconstructed-versus-true
histogram is, instead, normalized in reconstructed rows.  This
produces a matrix, $M_{ij}$, that performs the inverse operation of
Eq.~\ref{eq:fold}.  This method, proposed by D'Agostini, avoids the
problem of incorporating large statistical variance in the unfolding
matrix, which is often an issue with inverting the bin-migration
matrix~\cite{unfoldmeth}.  The unfolding matrix can be extended to
two-dimensional distributions in a straight forward way by arbitrarily
ordering each of the two-dimensional bins and repeating the same
process used in the one-dimensional case.

With the introduction of the unfolding matrix, $M_{ij}$, the full expression for the differential cross section can be written as
\begin{equation}\label{eq:diffxsecexp}
\frac{\partial\sigma}{\partial v}(v_i) = \frac{\sum_jM_{ij}(D_j-B_j)}{\epsilon_i\Delta v_iN_{targ}\Phi}.
\end{equation}
where $v_i$ is the variable to be measured in true bin $i$, $D_j$ is
the measured data distribution in reconstructed bin $j$, $B_j$ is the
predicted background distribution, $\epsilon_i$ is the efficiency,
$\Delta v_i$ is the width of bin $i$, $N_{targ}$ is the number of
target molecules in the fiducial volume, and $\Phi$ is the integrated
flux in units of neutrinos per unit area, as described in
Sec.~\ref{sec:fluxfac}.

\subsection{Unfolding Bias}\label{sec:unfoldingbias}

Although the use of $M_{ij}$ rather than $F_{ij}^{-1}$ avoids the
statistical variance issues involved with matrix inversion, it does
introduce some bias.  In general, unfolding procedures often require
the introduction of some amount of bias in order to reduce the
statistical variances associated with matrix inversion such that the
overall uncertainty is reduced~\cite{unfoldmeth}.  Since the
bin-migration matrix, $F_{ij}$, is normalized in columns of the true
distribution, $F_{ij}$ and $F_{ij}^{-1}$ are fully independent of the
true Monte Carlo distribution, and are therefore unbiased
transformations.  The $M_{ij}$ matrix is normalized in reconstructed
rows, which means any change to the shape of the true distribution
slice within a reconstructed bin (i.e. changes to the Bayesian prior
probabilities as described in Ref.~\cite{unfoldmeth}) will result in
the reconstructed events in that bin being assigned to the true bins
in different proportions.

To quantify the size of the unfolding bias, an iterative technique is
used.  The background-subtracted, unfolded data provide an inferred
true distribution as described in Section~\ref{sec:unfolding}.  Each
Monte Carlo event is then assigned a weight given by the binned ratio
of inferred true data to the true Monte Carlo simulation.  Using these
weights, a new reconstructed-versus-true histogram is created from
which a new $M_{ij}$ unfolding matrix is formed.  The data
distribution is unfolded again using the new unfolding matrix and the
processes is repeated.

Successive iterations of the inferred data distribution have two
distinct features in both the one- and two-dimensional cases.  The
first is that the largest excursion relative to the uniterated
inferred distribution is in the first iteration.  The other is that
each successive iteration oscillates about an intermediate preferred
value, which is a convolution of the true underlying distribution and
any systematic biases in the unfolding matrix.  The amplitude of these
oscillations decreases as the number of iterations increases.  The
size of the largest systematic variation, the first iteration, is
taken as the systematic uncertainty.

\subsection{Efficiency Correction}

After the data are unfolded, the inferred true data distribution is
corrected for events lost due to data selection cuts and detector
inefficiency.  The numerator of the efficiency correction is the true
distribution of all Monte Carlo events that pass the cuts.  The
denominator is the generated Monte Carlo distribution, formed before
any cuts are imposed.  The ratio of these two distributions gives the
fraction of events in a particular bin that survive the analysis cuts,
\begin{equation}\label{eq:eff}
\epsilon_i = \frac{N_i^{true~after~cuts}}{N_i^{generated}}.
\end{equation}
The efficiency is insensitive to changes in the
underlying physical parameters used to produce the generated
distribution.  If any portion of the generated distribution is
incorrectly enhanced, a proportional effect should be seen in the true
distribution and thus cancel in the efficiency.

Monte Carlo events are generated out to a radius of 610.6~\cm\ to
include all neutrino interactions in the main tank, the veto region,
and the tank wall.  Since the measurement being performed is a
neutrino cross section in mineral oil, all other materials must be
excluded in forming the generated Monte Carlo distribution.  To avoid
the PMTs and, in particular, the material voids inside the PMTs, the
efficiency denominator is formed from a subset of these events
generated within a radius of 550~\cm.

Nearly all of the events generated outside of 575~\cm\ are removed by
the veto hits cut; however, there will be a population of events
generated between 550~\cm\ and 575~\cm\ that pass all cuts,
particularly in the upstream portion of the tank.  The extra
contribution from events in the 550-575~\cm\ shell are in good
agreement in the reconstructed data and Monte Carlo vertex
distributions.

The number of interaction targets in the cross section formula,
$N_{targ}$, corresponds to the definition of the generated volume
used in the efficiency calculation.  To extract the number of targets
from the volume, the only experiment-dependent quantity needed is the
oil density, which has been measured to be
$0.845\pm0.001~\textrm{g}/\textrm{cm}^3$~\cite{detectorpaper}.  Since
the cross section only depends on the relative amount of each atomic
species, the interaction target is chosen to be an average single unit
on the hydrocarbon chain, CH$_{2.08}$.

\subsection{Flux Factor}\label{sec:fluxfac}

The flux factor, $\Phi$, in Eq.~\ref{eq:diffxsecexp} takes the
form of either a distribution in neutrino energy or a single value,
depending on the type of cross section measurement being performed.
For measurements binned in $E_{\nu}$, the flux factor is the number of
incident neutrinos per unit area in each measured bin.  These
measurements are flux-averaged over the width of each bin.  In the
differential and double-differential cross section measurements, the
flux factor is the fully integrated neutrino flux, and the cross
section is flux-averaged over the entire neutrino energy spectrum.
The total integrated \numu\ flux factor for the MiniBooNE experiment,
normalized to protons on target (POT), is $5.19\times 10^{-19}$
($\numu$/cm$^2$/POT).

Flux-averaged differential cross section measurements implicitly
contain the shape of the neutrino energy spectrum, which must also be
reported to fully specify the results.  To mitigate the dependence on
the experiment-dependent shape of the energy spectrum, each of the
differential cross section measurements has also been performed in
bins of neutrino energy.

\subsection{Systematic Uncertainties}\label{sec:xsecsyst}

The systematic uncertainties are grouped by error sources.  Each
source is a set of correlated uncertainties from a particular stage of
the simulation.  The parameters of each error source are related by an
error matrix from which a set of correlated parameter variations,
called a ``multisim,'' can be drawn.  Each new set of parameters
produces a systematically varied version of any reconstructed
distribution.  The spread in the reconstructed multisim distributions
is used to calculate the total systematic uncertainty.

There are two distinct types of systematic variations.  Some
systematics, such as the flux and cross section uncertainties, only
affect the probability with which an event will occur.  For this type
of uncertainty, a systematically varied distribution can be produced
by reweighting the central value Monte Carlo distribution.  Each event
is multiplied by the ratio of the event probability calculated with
the systematically varied set of parameters to the central value event
probability.

\begin{figure}
  \begin{center}
    \leavevmode
    \includegraphics[scale=0.45,clip=true]{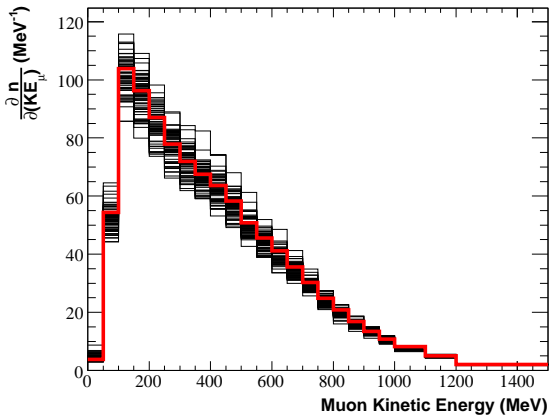}
    \caption
	{ The reconstructed muon kinetic energy is plotted in each of
	  the 67 optical model multisims.  The central value Monte
	  Carlo distribution (red) is overlayed for comparison.  }
    \label{fig:mukeommultis}.
  \end{center}
\end{figure}

The other type of systematic variation affects the measured properties
of the event after it is produced, such as the amount of light
generated as a function of wavelength and the propagation of the light
through the oil.  In general, these variations cannot be accomplished
via reweighting.  Instead, these errors are determined using 67
data-sized Monte Carlo simulations that are generated using parameter
draws from the optical model error matrix.  A plot of these optical
model multisims in muon kinetic energy is shown in
Fig.~\ref{fig:mukeommultis}.

\subsubsection{Error Matrices}\label{sec:errormatrices}

The uncertainties in the measured cross sections are described by an
error matrix that characterizes the correlated uncertainties in the
measured values in each bin.  For each error source, an error matrix,
$E_{ij}^{source}$, is calculated from the bin population differences
in the multisims compared to the central value,
\begin{equation}
E_{ij}^{source} = \frac{1}{N}\sum_{m=1}^N\left(n_{m,i}-n_{CV,i}\right)\left(n_{m,j}-n_{CV,j}\right),
\end{equation}
where $N$ is the number of multisims, $n_{m,i}$ is the number of
events in bin $i$ of multisim $m$, and $n_{CV,i}$ is the number of
events in bin $i$ in the central value Monte Carlo simulation.  Once
an error matrix has been calculated for each source, the total error
matrix is given by summing each component matrix.

Since the statistical fluctuations in a particular bin are unrelated
to the fluctuations in any other bin, statistical error matrices are
always diagonal.  By design, these uncertainties are built into the
optical model error matrix since each optical model multisim was
constructed to have the same number of events as the data.
Unfortunately, this also has the effect of adding statistical
fluctuations to the off-diagonal terms.  As more optical model
multisims are incorporated into the calculation of the error matrix,
the size of these spurious fluctuations is decreased.  The
fluctuations are also smaller for bins with significant event
populations.  For this reason, cross section results will only be
reported for bins with at least 25 unfolded data events.  In the
one-dimensional cross section measurements, the event populations are
large enough that this effect is negligible.  The reweighting
multisims do not suffer from this effect.

To evaluate the systematic uncertainties in the cross section, the
full cross section calculation procedure outlined in
Eq.~\ref{eq:diffxsecexp} is performed in each multisim.  The
multisim distributions replace the corresponding central value
distributions in the calculation.  The reconstructed data distribution
remains the same, but multisim distributions are used for the
unfolding matrix, the background prediction, and the signal
efficiency.  The formula for the differential cross section with
multisim dependent quantities explicitly identified is
\begin{equation}
\frac{\partial\sigma^m}{\partial v}(v_i) = \frac{\sum_jM^m_{ij}(D_j-B^m_j)}{\epsilon^m_i\Delta v_iN_{targ}\Phi^m},
\end{equation}
where $m$ is the multisim index.

The results presented in the remainder of this report will list only
the diagonal errors on each bin.  The full error matrices for each
distribution are quite large, and in the case of the two-dimensional
measurements, they can contain over one million elements.

\subsubsection{Flux Uncertainties}

\begin{figure}
  \begin{center}
    \leavevmode
    \includegraphics[scale=0.45,clip=true]{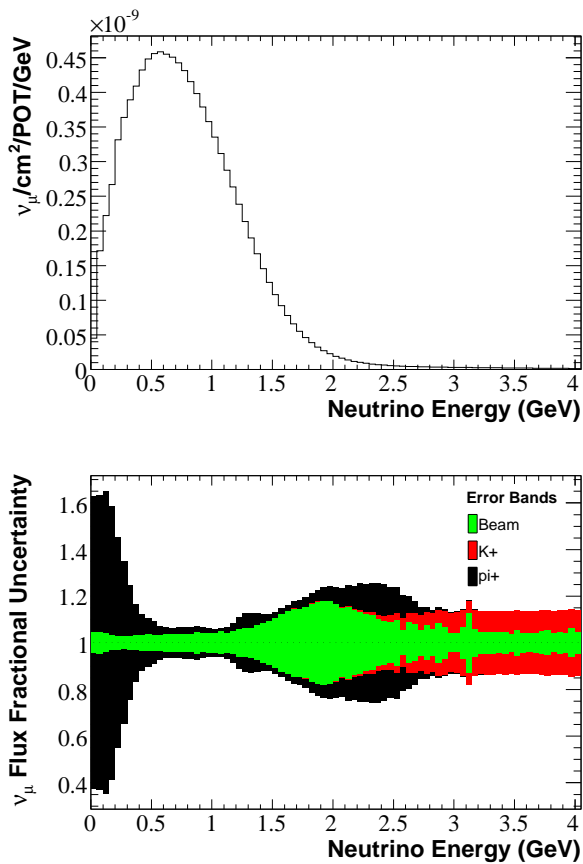}
    \caption
    { The \numu\ energy spectrum is shown (top) along with the fractional flux errors (bottom).  The fractional errors are cumulative such that each successive error band includes a quadrature sum of the previous errors such that the outer band gives the total error.  A numerical description of the energy spectrum is given in Table~\ref{tab:numuflux}}
    \label{fig:fluxsyst}
  \end{center}
\end{figure}

The predicted neutrino spectrum is modified to correspond to the
systematically varied flux parameters in each multisim.  The diagonal
uncertainties of the flux variations are shown in
Fig.~\ref{fig:fluxsyst}.

The largest of the flux errors is the uncertainty in the beam
\pip~production in proton-beryllium interactions.  The differential
cross section for these interactions is parametrized using an
empirical model from Sanford and Wang~\cite{sw}.  The parameters are
determined in a simultaneous fit to data from the HARP and E910
experiments~\cite{harp,e910}.  The flux prediction is not tuned in
any way to data from MiniBooNE.  As described in
Ref.~\cite{fluxpaper}, the shape of the Sanford-Wang parametrization
is not fully compatible with the data, but it is still used in the
Monte Carlo simulation to provide physical constraints such as driving
the cross section to zero at vanishing pion momentum.  To determine
the uncertainties in the pion production, the HARP data are fit with a
spline function, and the resulting fit parameters are systematically
varied according to the error matrix returned by the fit.  The
covariance of these spline function variations with respect to the
Sanford-Wang central value are used to form the beam \pip\ production
error matrix as described in Ref.~\cite{minibooneccqe2}.  This
produces an uncertainty given by the HARP data errors in regions where
the Sanford-Wang parametrization agrees with the spline function, and
increases the errors in regions where they differ.

The portion of the flux relevant to observable \ccpip\ interactions
occurs at neutrino energies larger than about 400~\mev.  For the peak
neutrino energies (0.5-1~\gev), the beam \pip\ flux uncertainties are
around 8\%.  At higher neutrino energies, the beam
\pip\ uncertainty grows to as large as 25\%.  The beam \pip\
fluctuations also exhibit very distinct features in shape.  The
residual effect of the wiggling behavior of the spline fit to the HARP
data is apparent.  In particular, the low-energy flux exhibits very
large systematic excursions since there are no HARP data to constrain
the fits in that region.

The other two flux related error sources have a smaller effect on the
total uncertainty.  The ``Beam'' uncertainties, which contain all
systematic effects involving the proton beam and horn, are generally
around the 5\% level below 1~\gev\ and then expand at higher neutrino
energies.  The main contributor at high energies is the horn current
skin depth uncertainty, which causes a $\sim$15\% effect between 1.5
and 2.5~\gev.  The \kpl\ production uncertainties are mostly
irrelevant for this analysis.  \kpl\ mesons become the dominant source
of \numu\ production at 2.3~\gev, and the uncertainties become
dominant at neutrino energies greater than 2.5~\gev, where the flux is
very small.

For the measurements not involving neutrino energy, the cross section
calculation is only affected by the uncertainty in the integrated
flux.  The size of these variations for each of the flux error sources
is given in Table~\ref{tab:intflux}.

\begin{table}[h!]
\caption[Uncertainties in the integrated flux from each of the flux
error sources]{The uncertainties in the integrated flux are given for
each of the flux error sources.\label{tab:intflux}}
\begin{center}
\begin{tabular}{|c|c|}\hline
Error Source & Integrated Flux Uncertainty \cr\hline
\pip & 10.4\% \cr\hline
Beam &  4.1\% \cr\hline
\kpl &  0.4\% \cr\hline
\end{tabular}
\end{center}
\end{table}

\subsection{Results}\label{sec:xsecresults}

\begin{figure}
  \begin{center}
    \leavevmode
    \includegraphics[scale=0.45,clip=true]{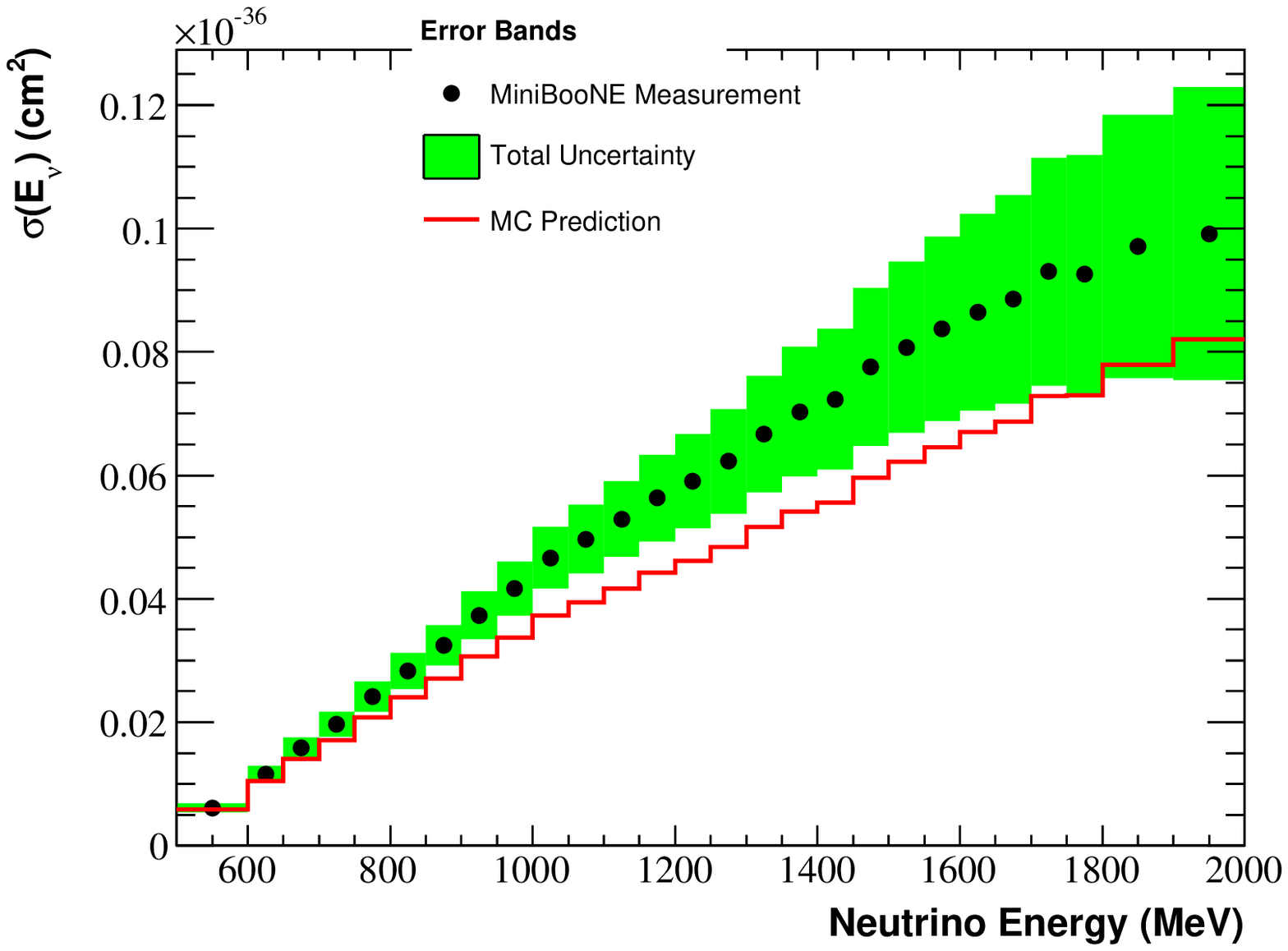}
    \includegraphics[scale=0.45,clip=true]{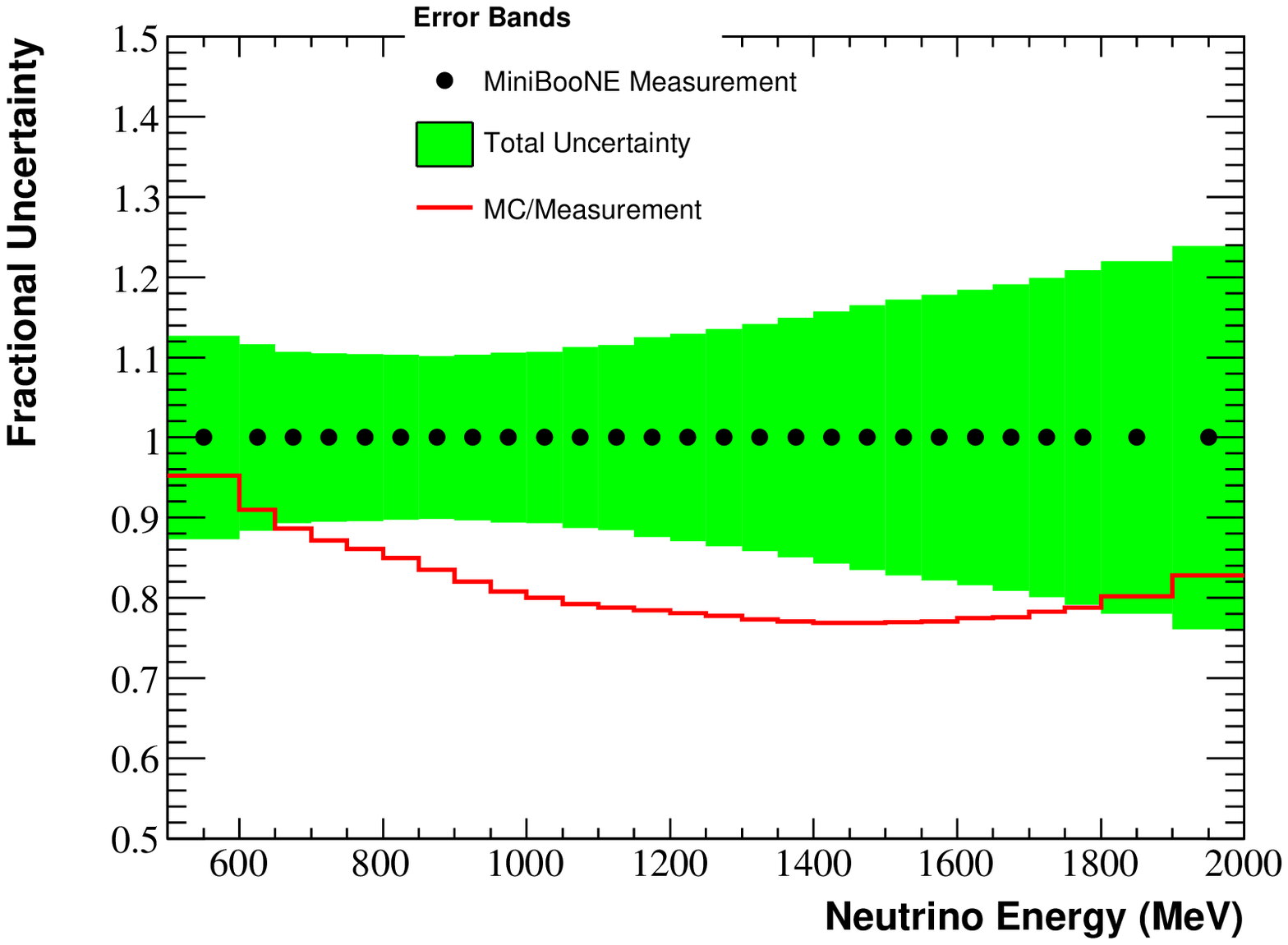}
    \caption{The $\sigma(E_{\nu})$ measurement is shown with
    cumulative systematic errors.  The absolutely normalized Monte
    Carlo prediction is shown for comparison.  The bottom plot shows
    the fractional uncertainties and the ratio of the Monte Carlo
    prediction to the measurement.  }
    \label{fig:enuxsec0}
  \end{center}
\end{figure}

\begin{figure}
  \begin{center}
    \leavevmode
    \includegraphics[scale=0.45,clip=true]{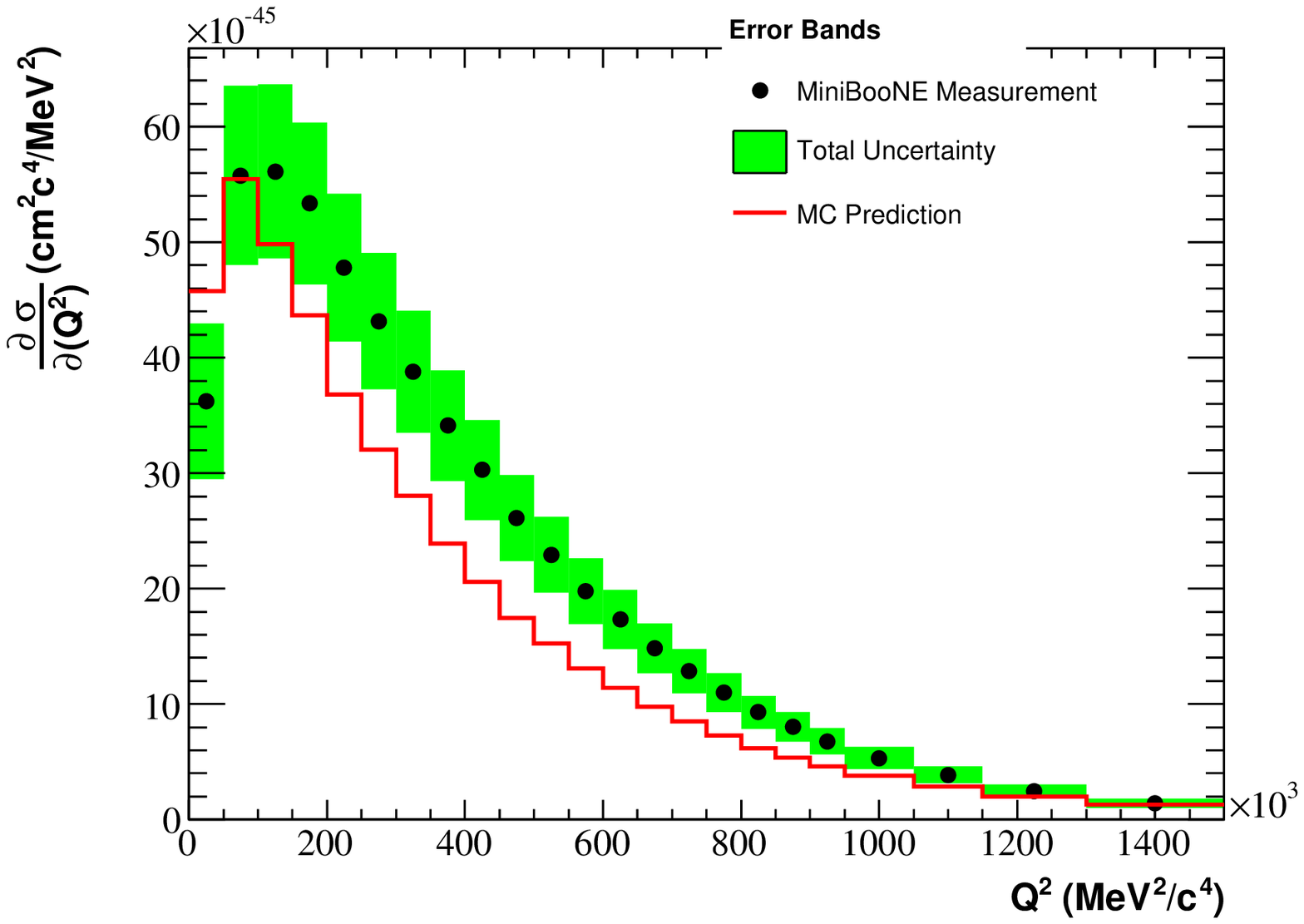}
    \includegraphics[scale=0.45,clip=true]{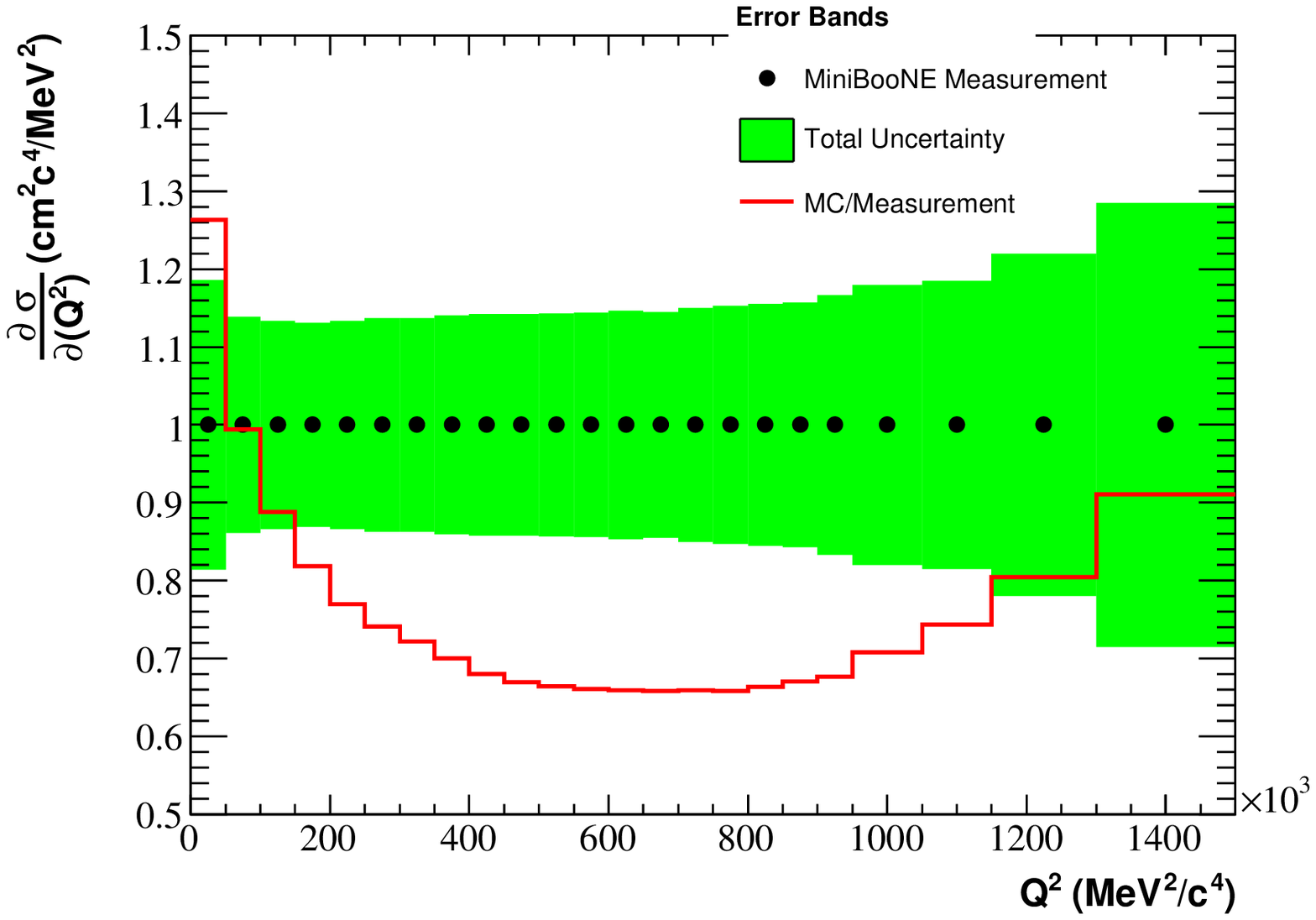}
    \caption{The $\partial\sigma/\partial (Q^2)$ measurement is
    shown with cumulative systematic errors.  The absolutely
    normalized Monte Carlo prediction is shown for comparison.  The
    bottom plot shows the fractional uncertainties and the ratio of
    the Monte Carlo prediction to the measurement.  }
    \label{fig:qsqxsec0}
  \end{center}
\end{figure}

\begin{figure}
  \begin{center}
    \leavevmode
    \includegraphics[scale=0.45,clip=true]{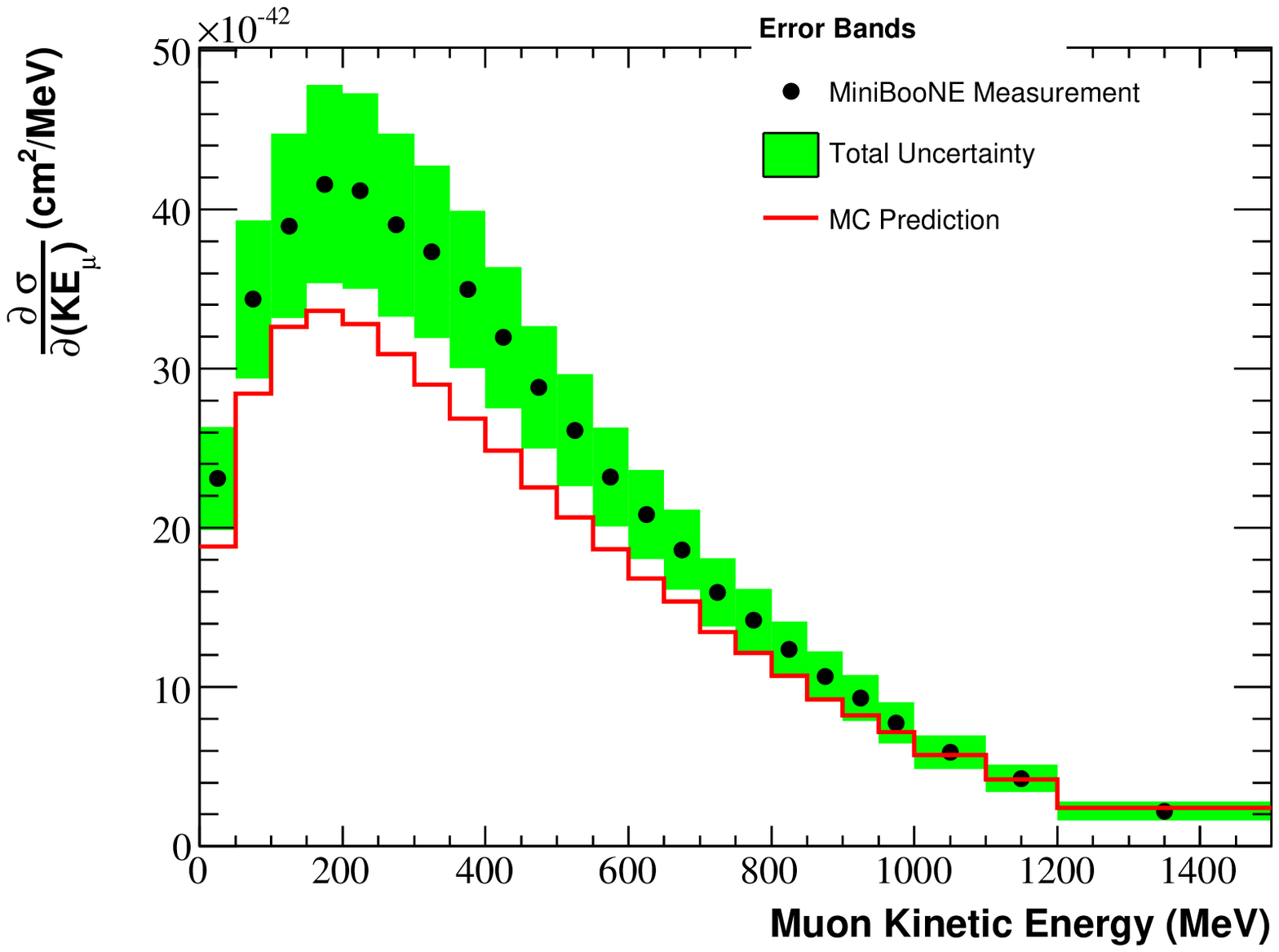}
    \includegraphics[scale=0.45,clip=true]{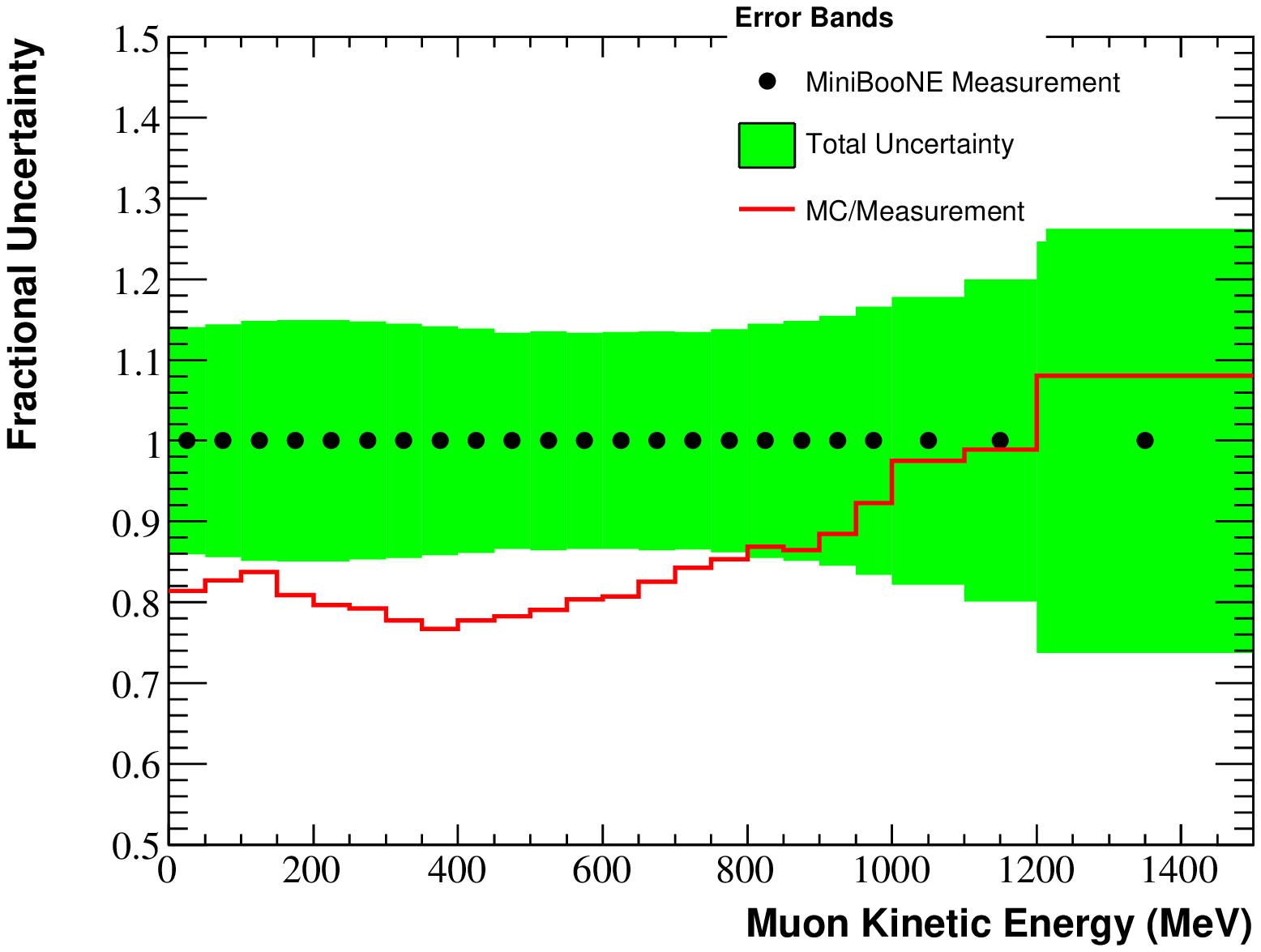}
    \caption{The $\partial\sigma/\partial (KE_{\mu})$ measurement is
    shown with cumulative systematic errors.  The absolutely
    normalized Monte Carlo prediction is shown for comparison.  The
    bottom plot shows the fractional uncertainties and the ratio of
    the Monte Carlo prediction to the measurement.  }
    \label{fig:mukexsec0}
  \end{center}
\end{figure}

\begin{figure}
  \begin{center}
    \leavevmode
    \includegraphics[scale=0.45,clip=true]{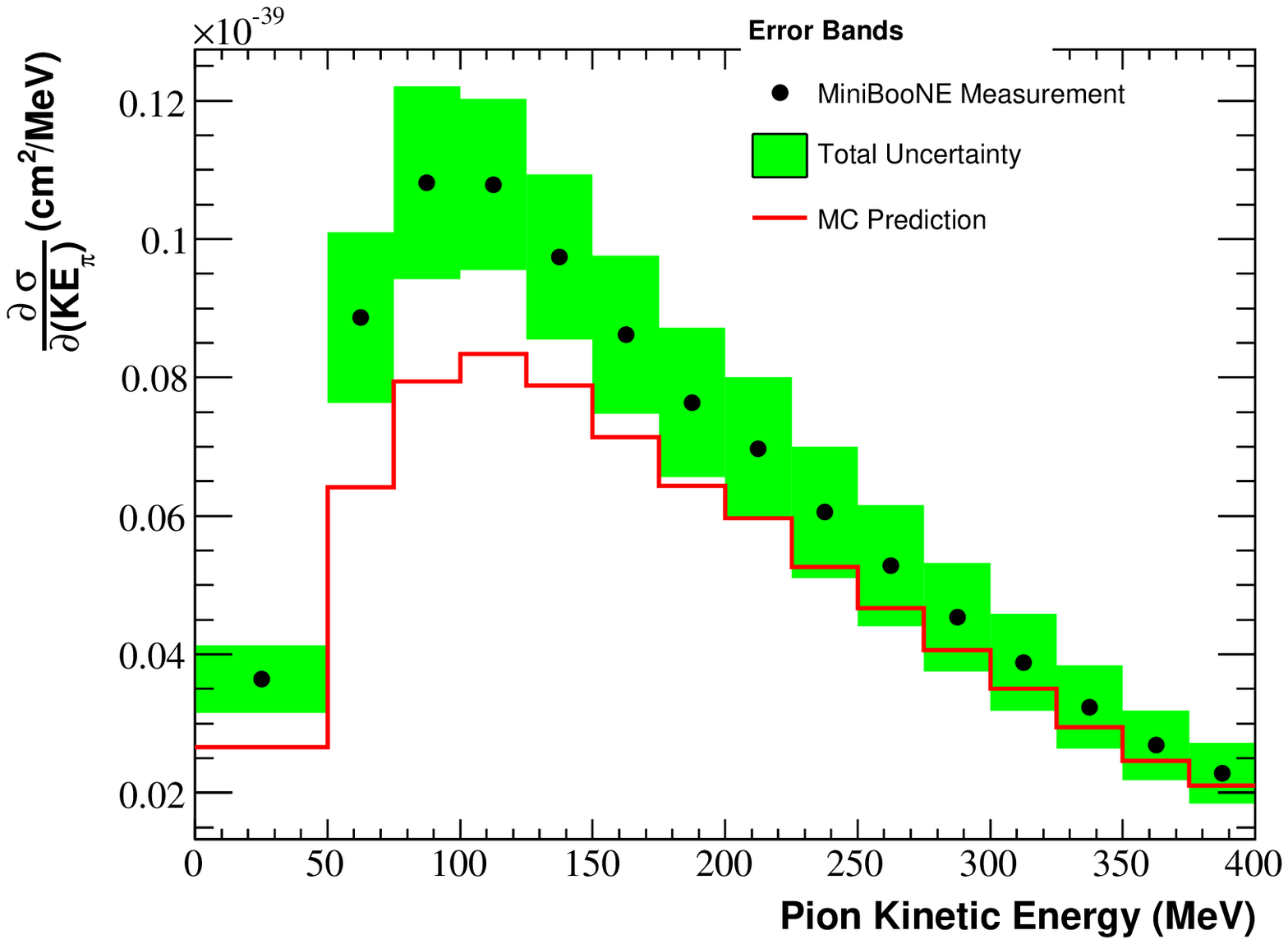}
    \includegraphics[scale=0.45,clip=true]{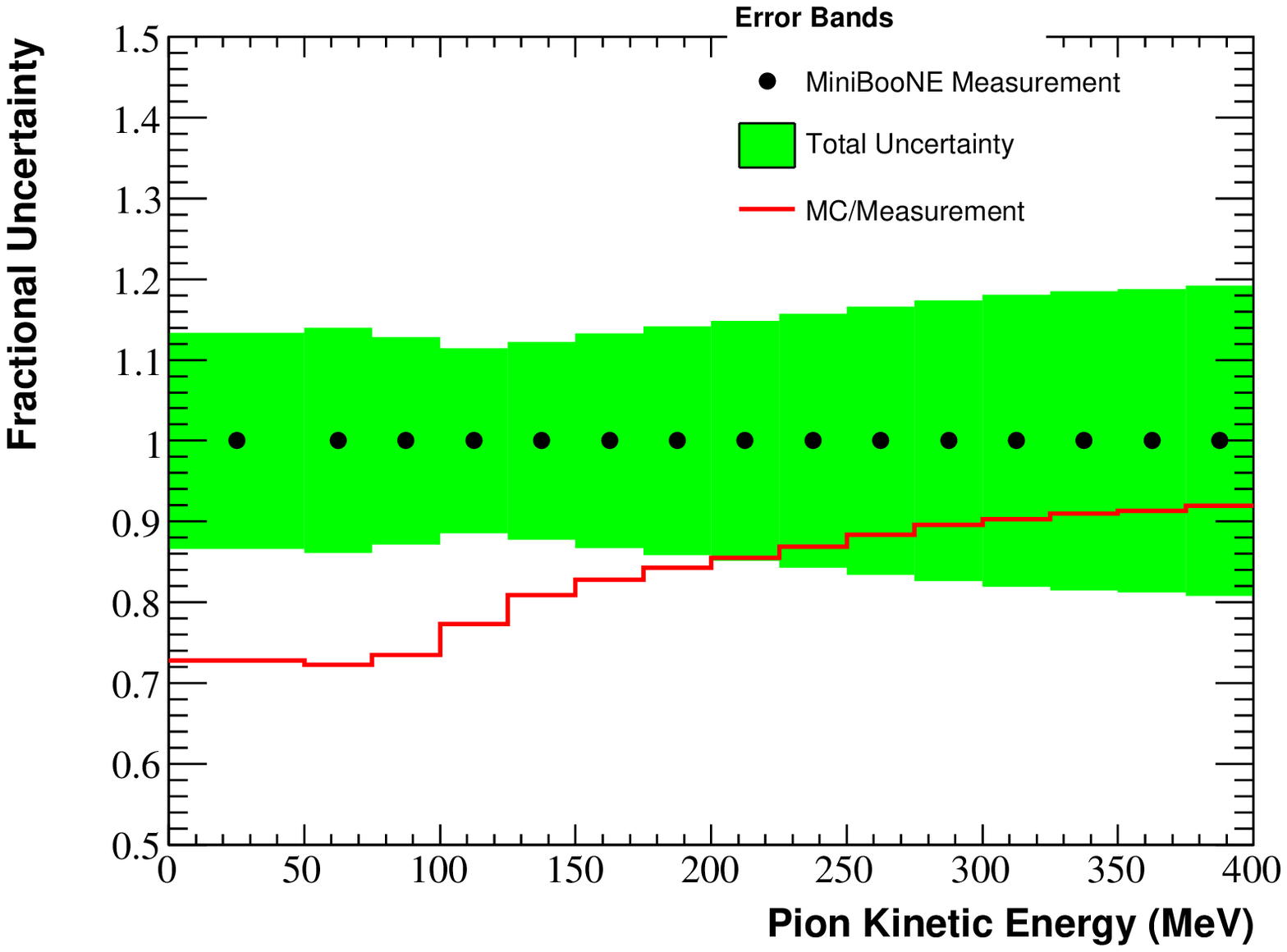}
    \caption{The $\partial\sigma/\partial (KE_{\pi})$ measurement is
    shown with cumulative systematic errors.  The absolutely
    normalized Monte Carlo prediction is shown for comparison.  The
    bottom plot shows the fractional uncertainties and the ratio of
    the Monte Carlo prediction to the measurement.  }
    \label{fig:pikexsec0}
  \end{center}
\end{figure}

\begin{figure}
  \begin{center} \centering
    \includegraphics[scale=0.4,clip=true]{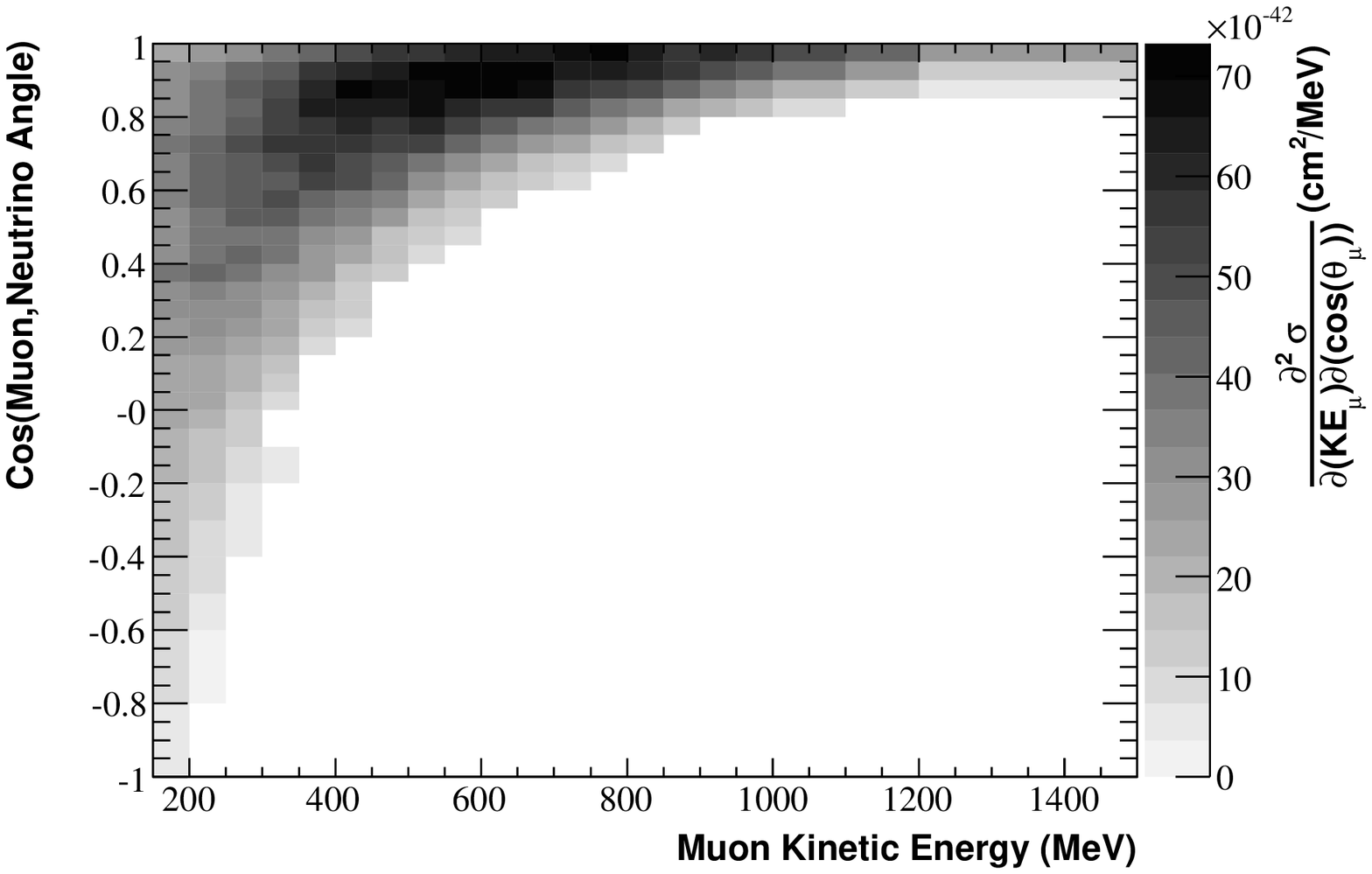}
    \includegraphics[scale=0.4,clip=true]{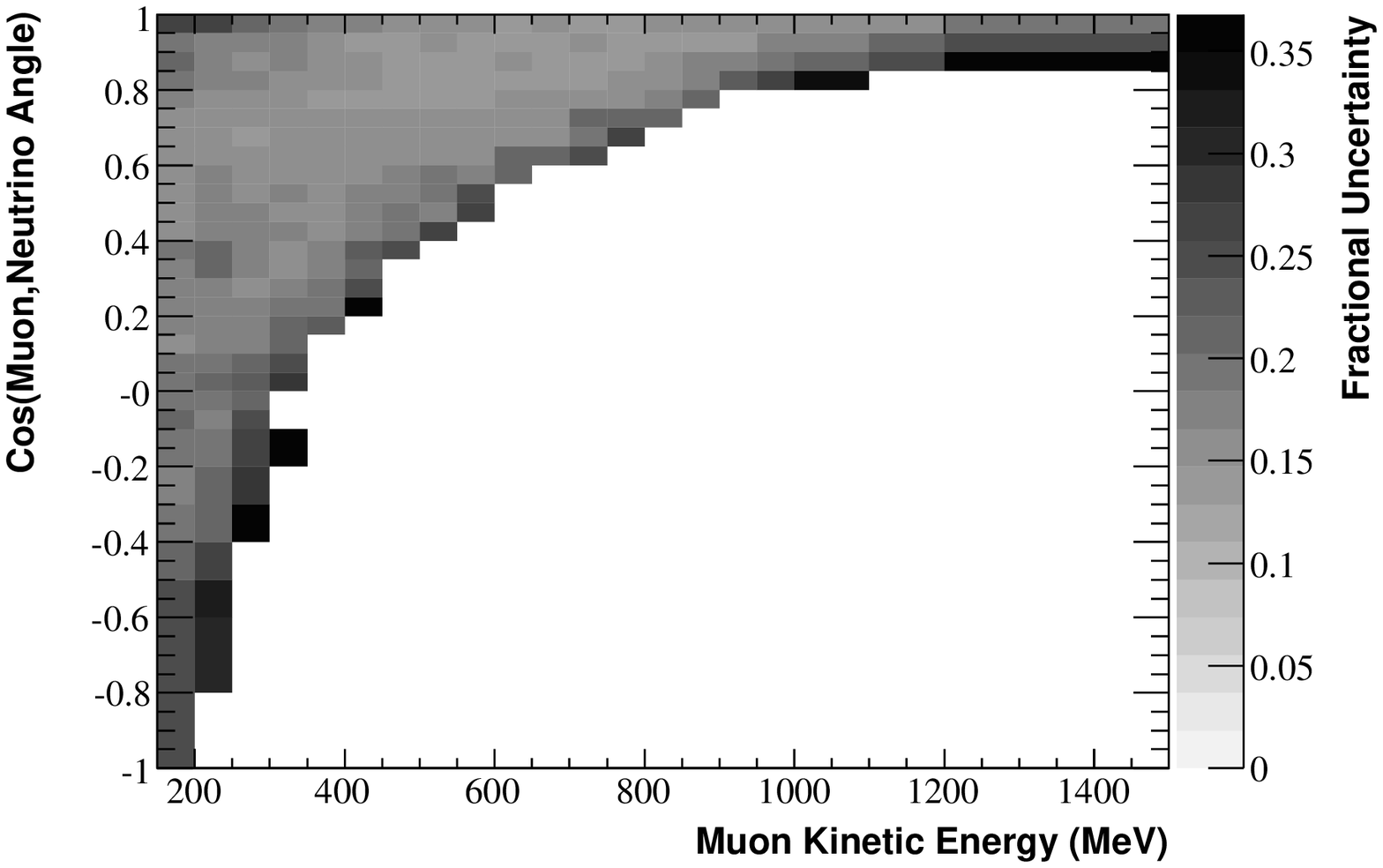}
    \includegraphics[scale=0.4,clip=true]{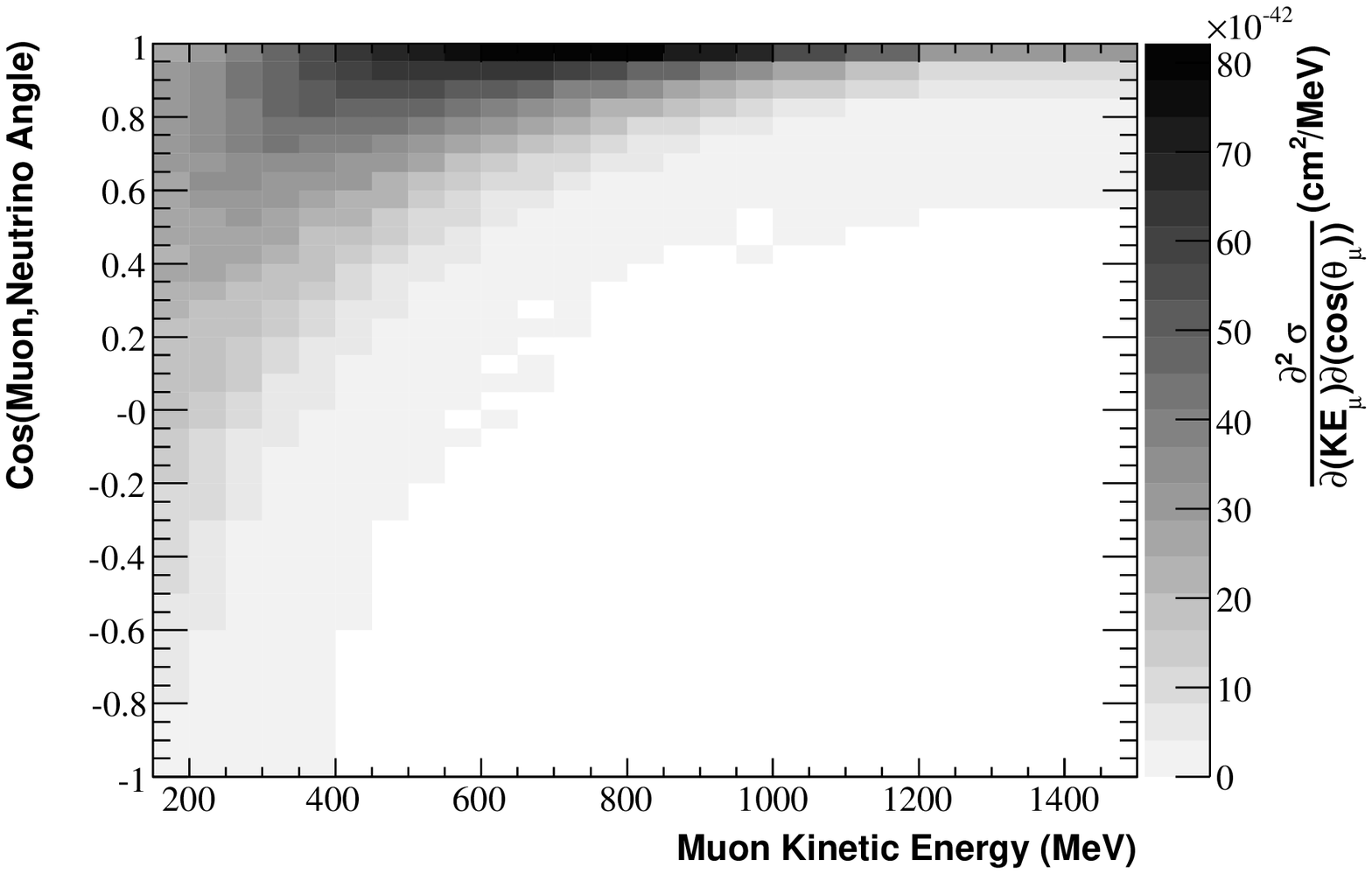}
    \caption{The measured
    $\partial^2\sigma/\partial(cos(\theta_{\mu,\nu}))\partial(KE_{\mu})$
    values are shown (top) along with the total fractional
    uncertainties (middle).  Empty bins indicate regions where no
    measurement has been made.  The Monte Carlo predicted cross
    section is shown for comparison (bottom).}
    \label{fig:muctvkexsec0} \end{center}
\end{figure}

\begin{figure}
  \begin{center} \centering
    \includegraphics[scale=0.4,clip=true]{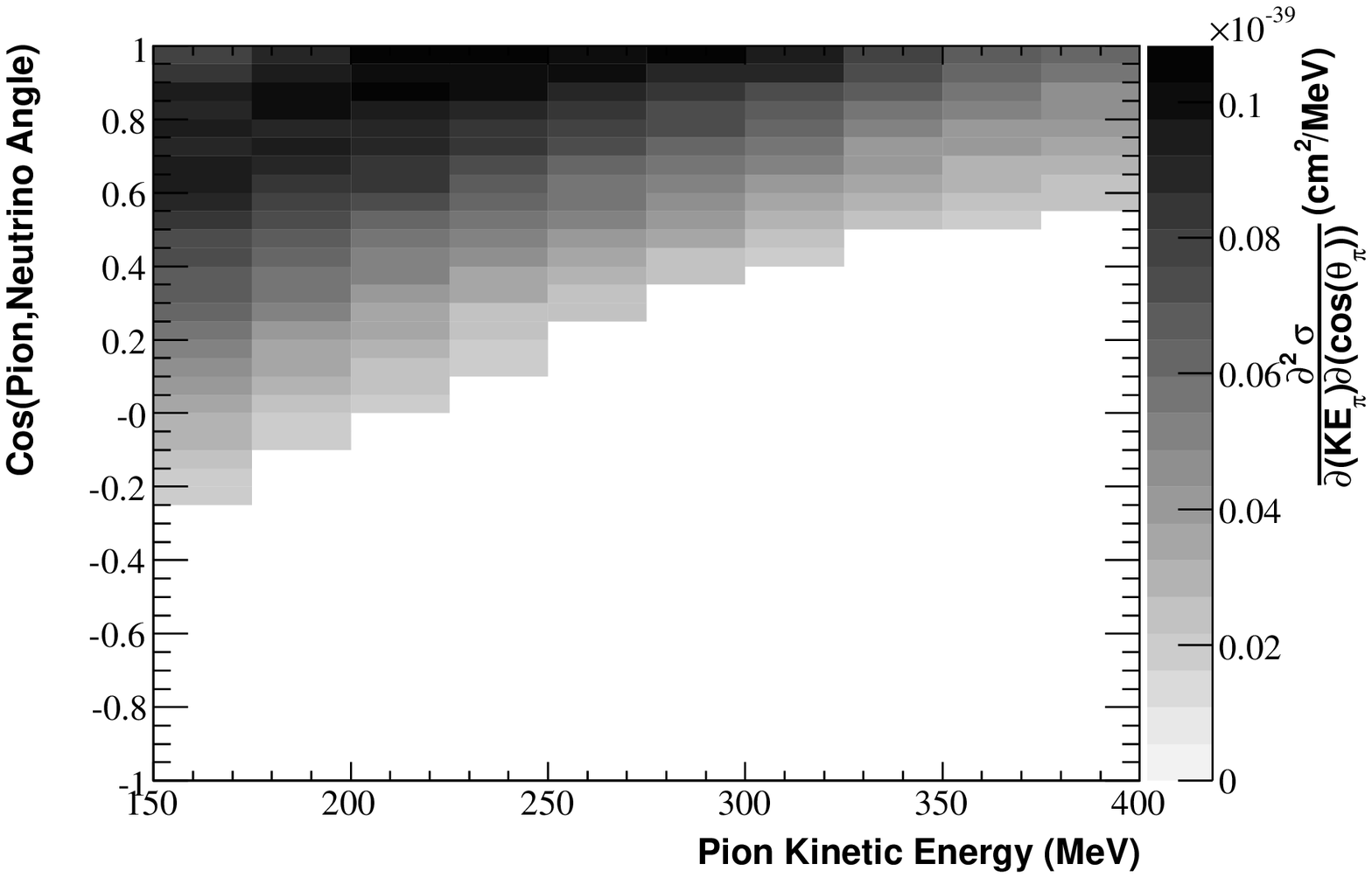}
    \includegraphics[scale=0.4,clip=true]{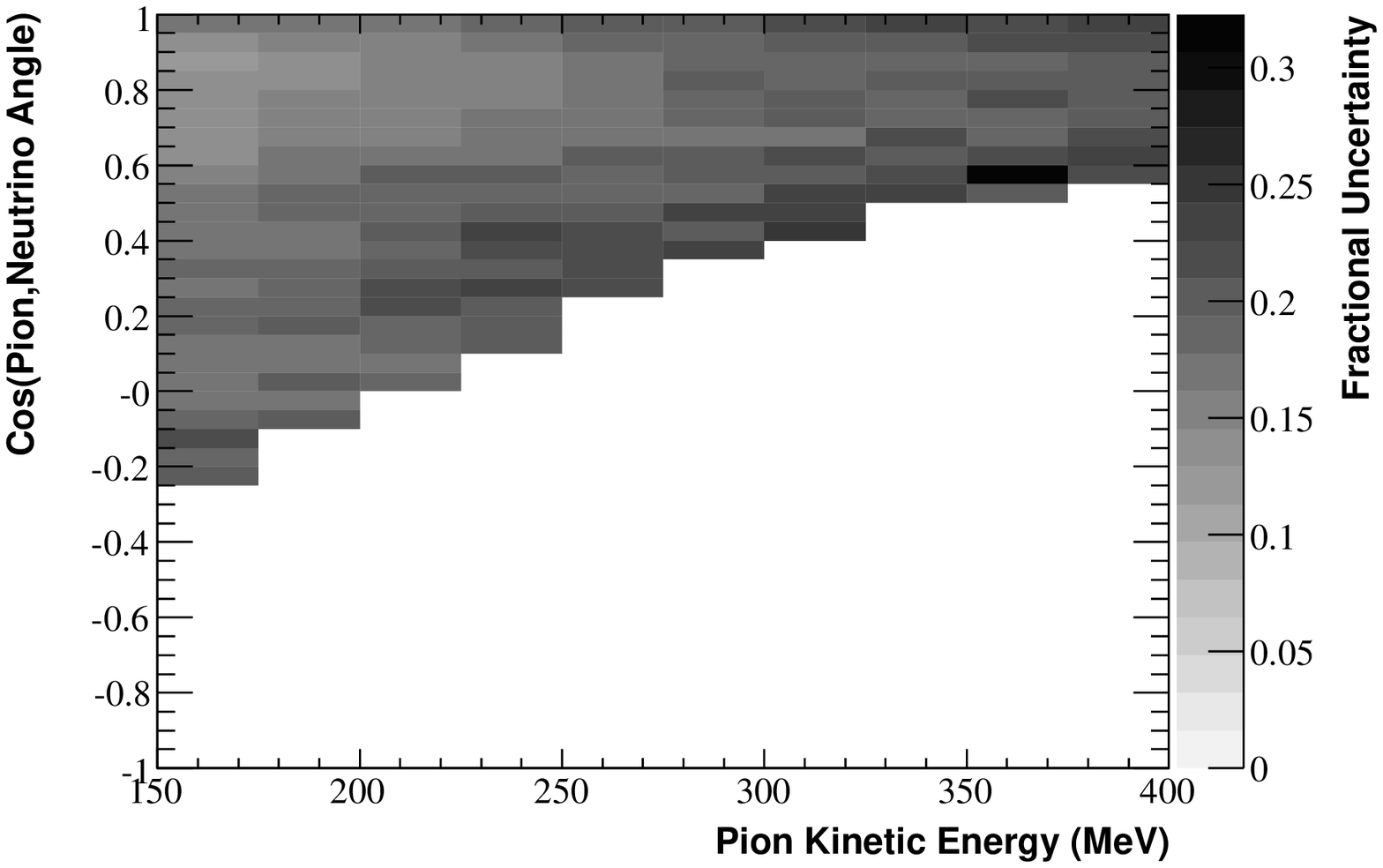}
    \includegraphics[scale=0.4,clip=true]{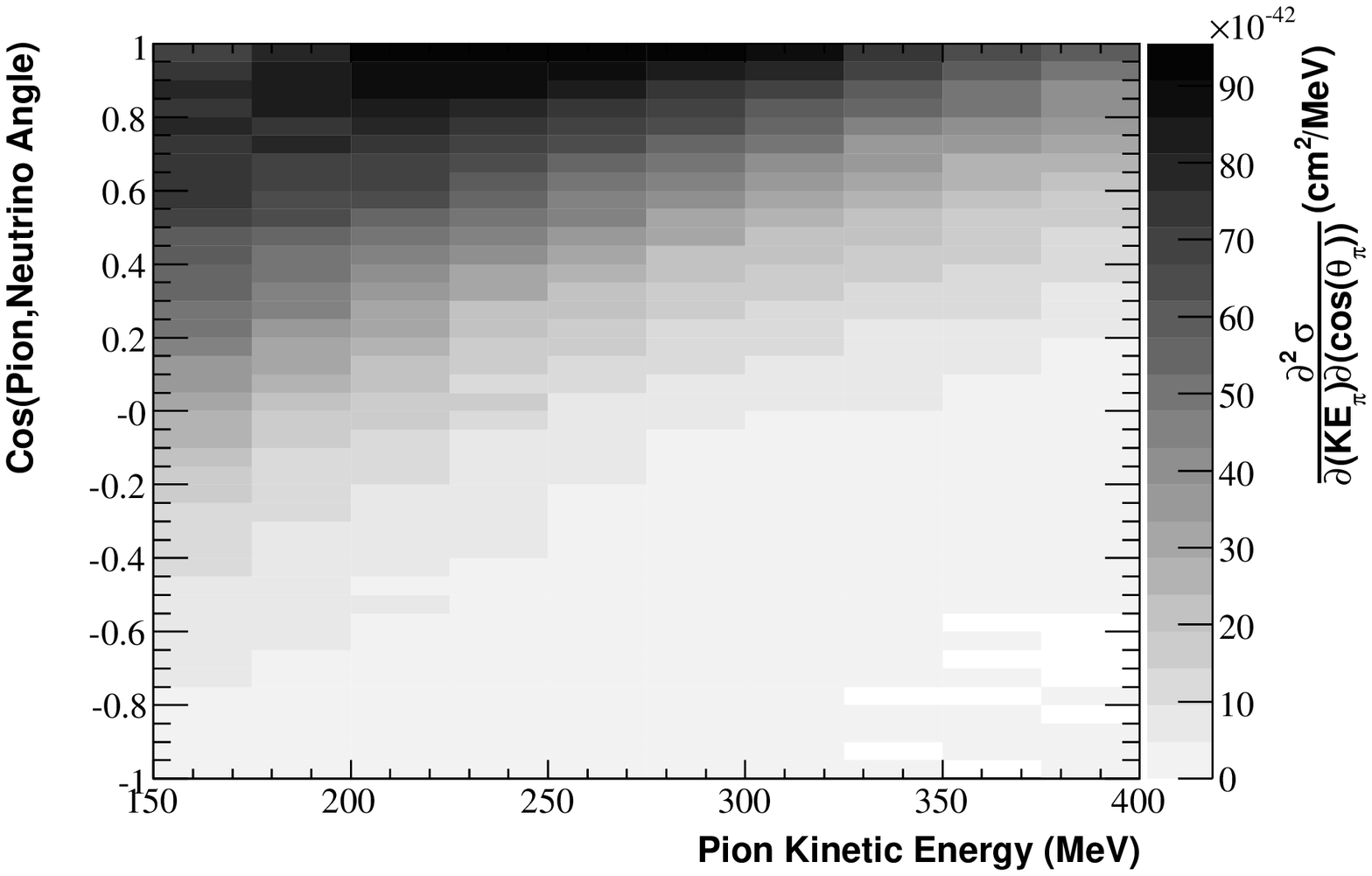}
    \caption{The measured
    $\partial^2\sigma/\partial(cos(\theta_{\pi,\nu}))\partial(KE_{\pi})$
    values are shown (top) along with the total fractional
    uncertainties (middle).  Empty bins indicate regions where no
    measurement has been made.  The Monte Carlo predicted cross
    section is shown for comparison (bottom).}
    \label{fig:pictvkexsec0} \end{center}
\end{figure}

\begin{figure}
  \begin{center}
    \includegraphics[scale=0.4,clip=true]{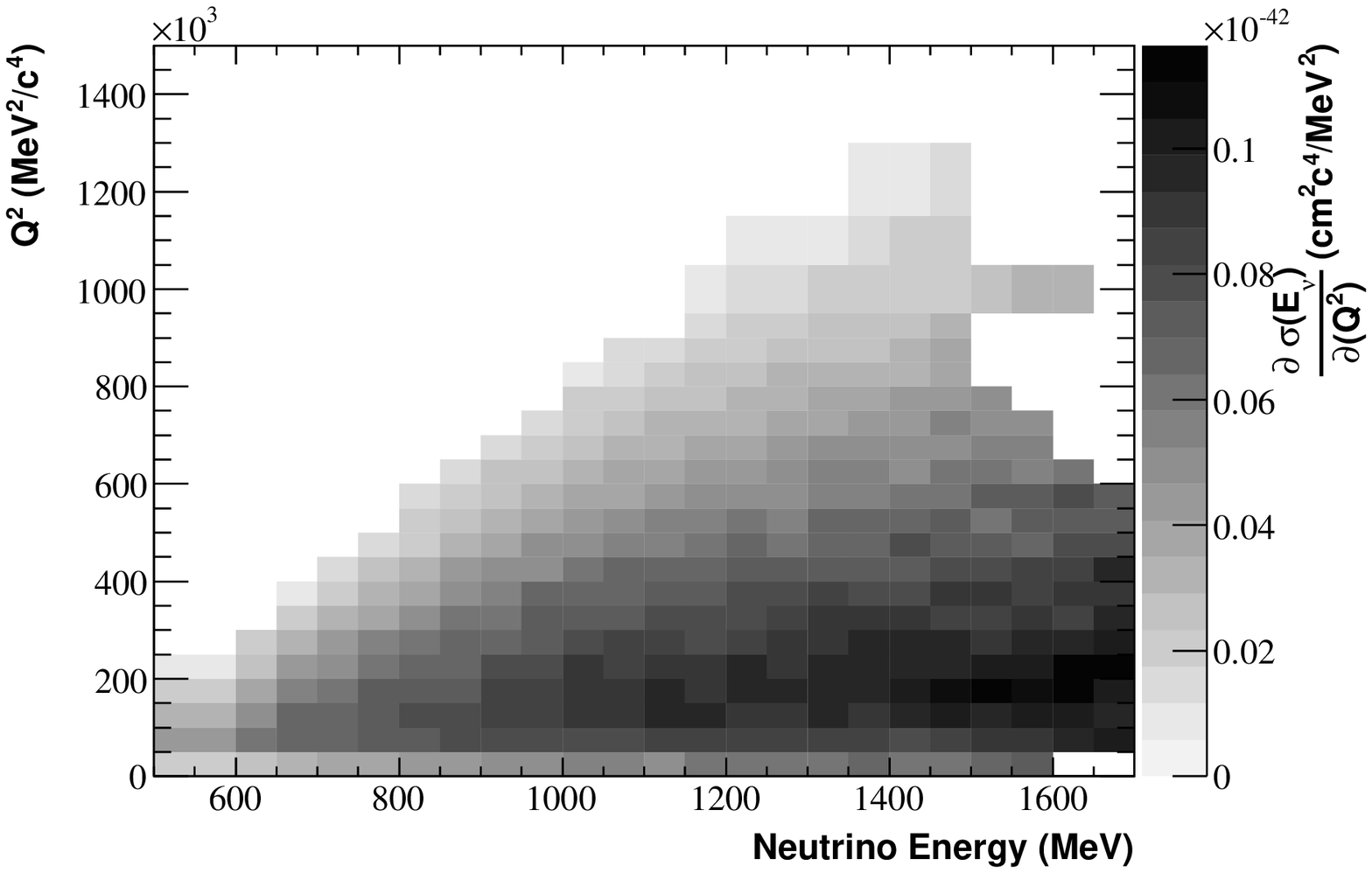}
    \includegraphics[scale=0.4,clip=true]{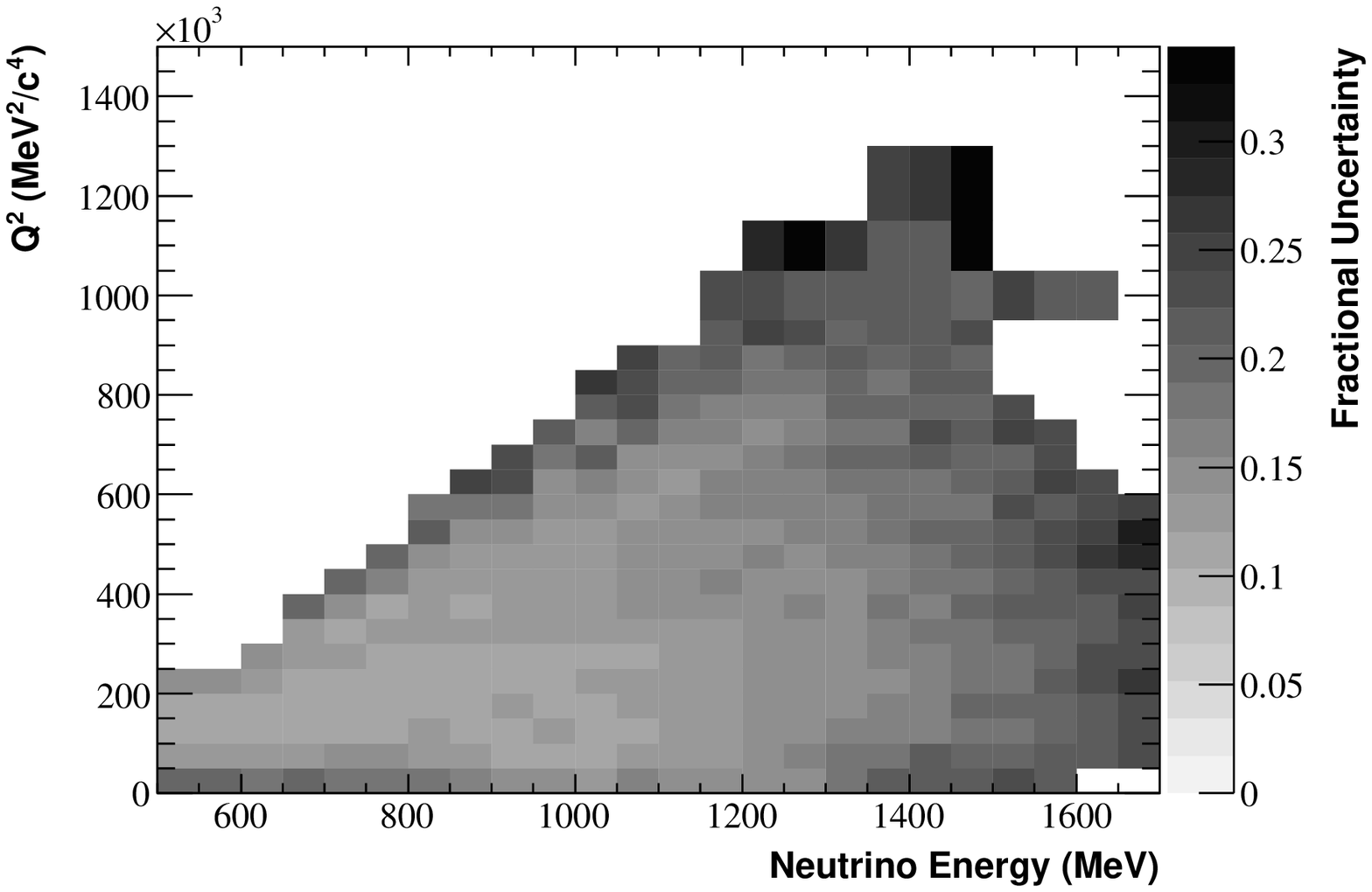}
    \includegraphics[scale=0.4,clip=true]{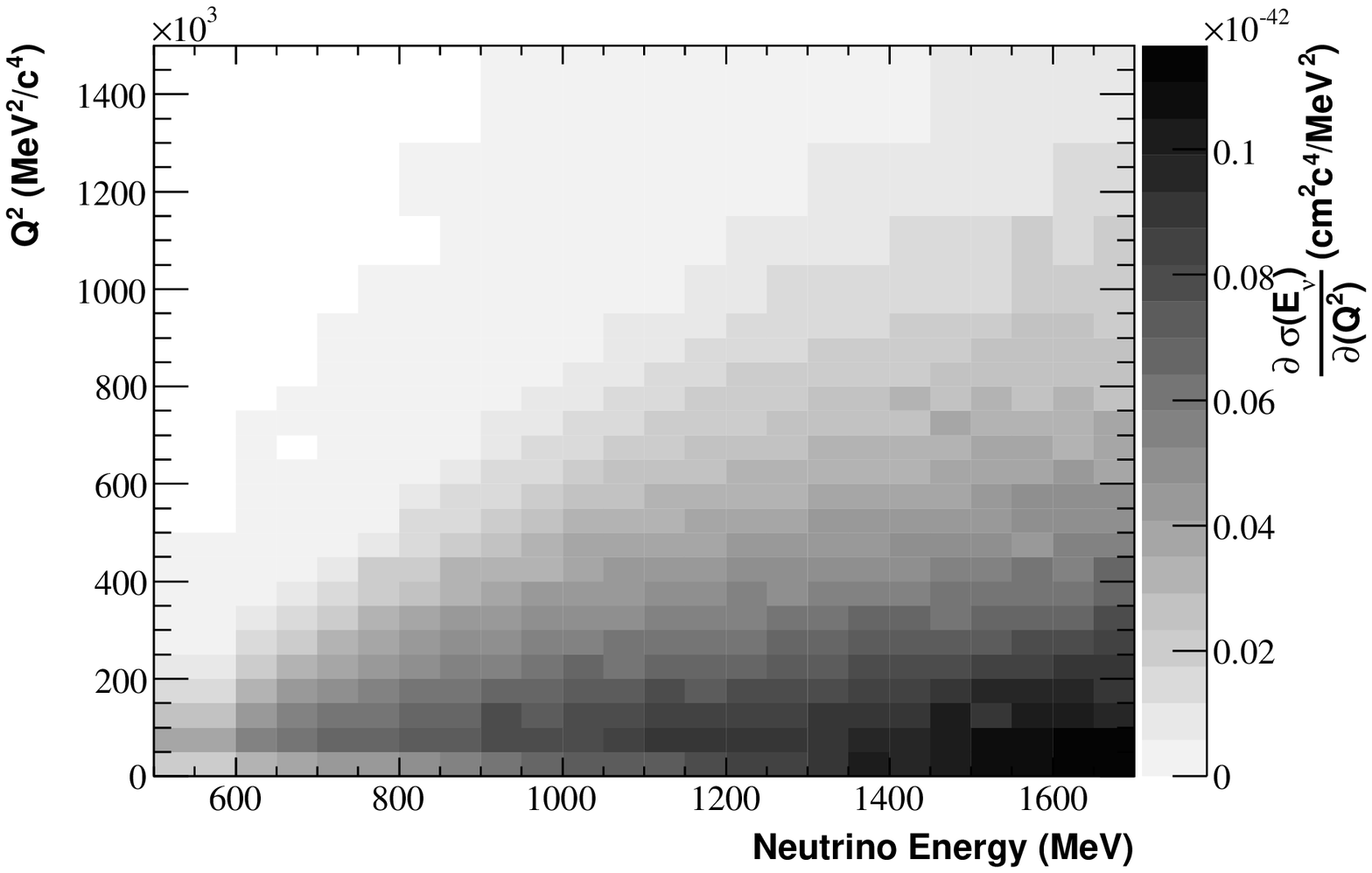}
    \caption{The measured $\partial\sigma(E_{\nu})/\partial(Q^2)$
    values are shown (top) along with the total fractional
    uncertainties (middle).  Empty bins indicate regions where no
    measurement has been made.  The Monte Carlo predicted cross
    section is shown for comparison (bottom).}
    \label{fig:qsqvenuxsec0} \end{center}
\end{figure}

\begin{figure}
  \begin{center} \centering
    \includegraphics[scale=0.4,clip=true]{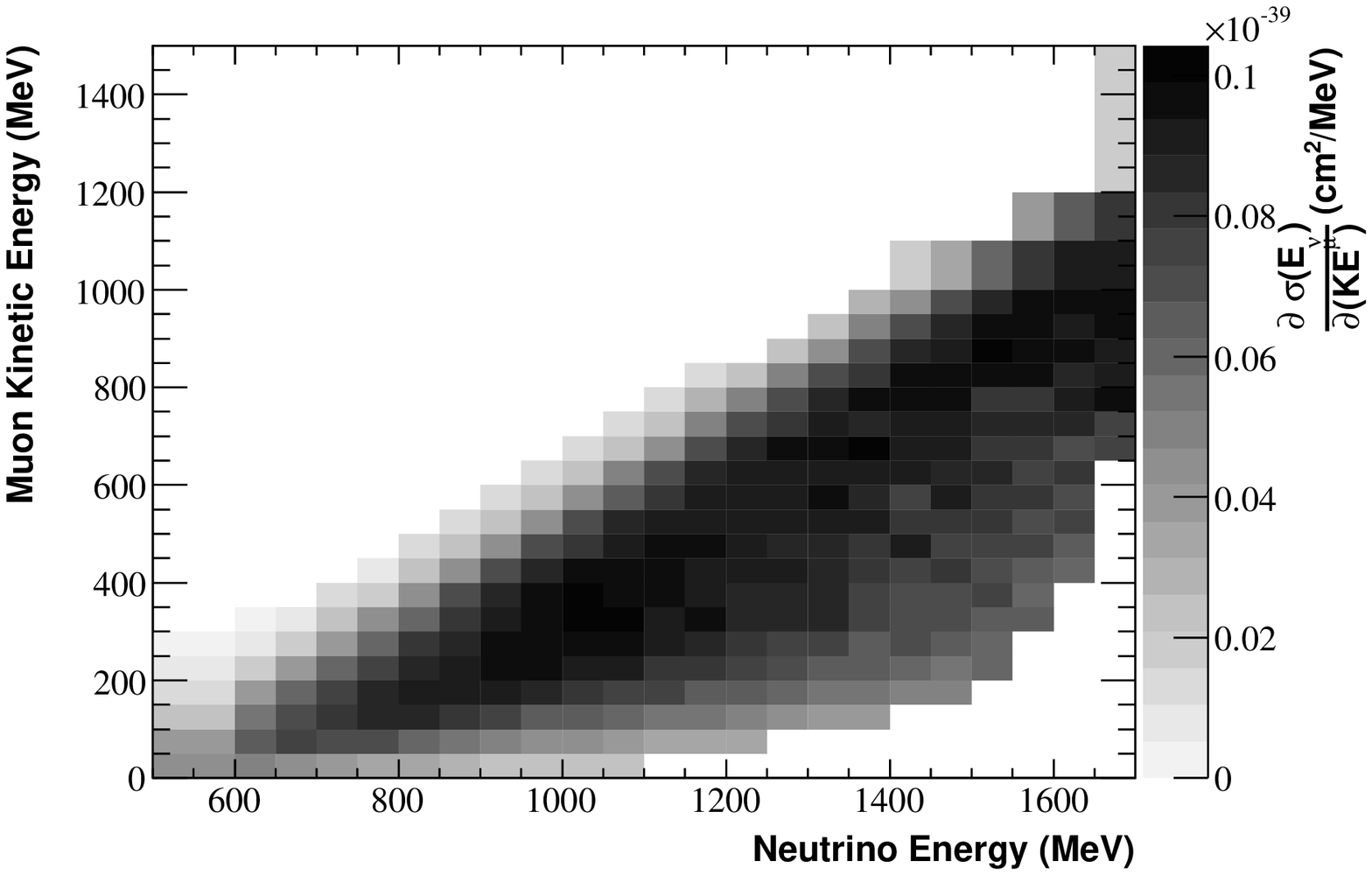}
    \includegraphics[scale=0.4,clip=true]{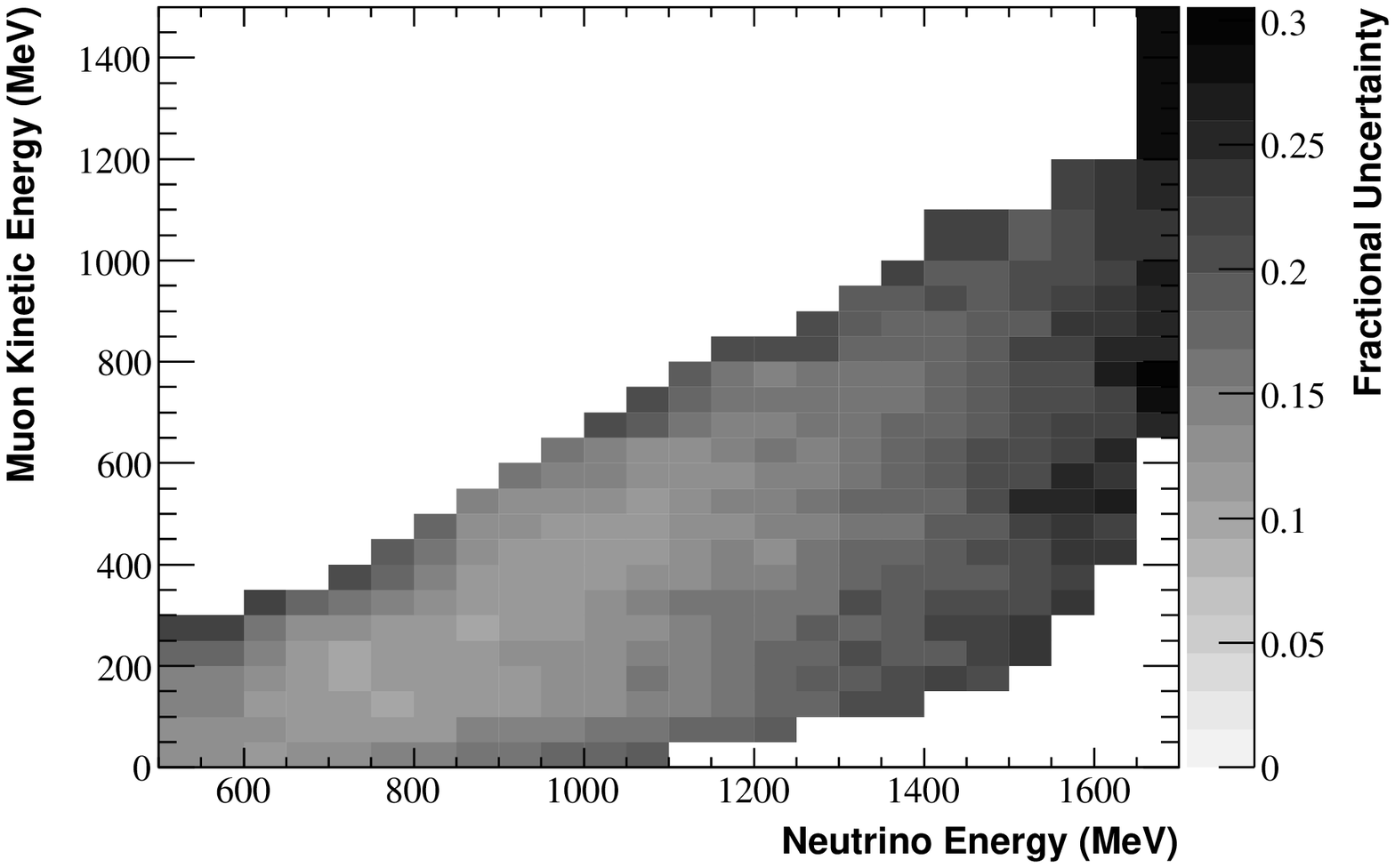}
    \includegraphics[scale=0.4,clip=true]{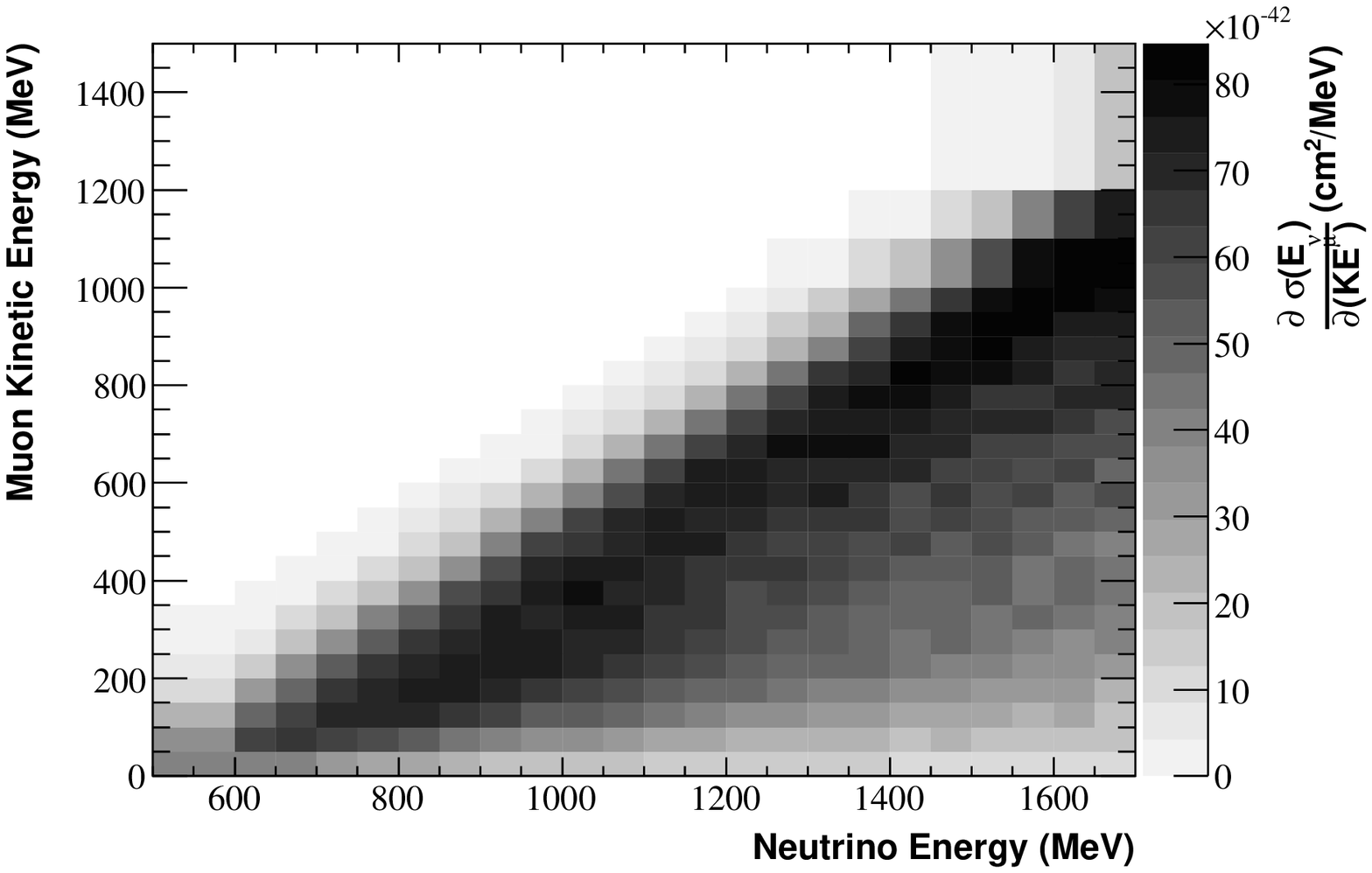}
    \caption{The measured $\partial\sigma(E_{\nu})/\partial(KE_{\mu})$
    values are shown (top) along with the total fractional
    uncertainties (middle).  Empty bins indicate regions where no
    measurement has been made.  The Monte Carlo predicted cross
    section is shown for comparison (bottom).}
    \label{fig:mukevenuxsec0} \end{center}
\end{figure}

\begin{figure}
  \begin{center} \centering
    \includegraphics[scale=0.4,clip=true]{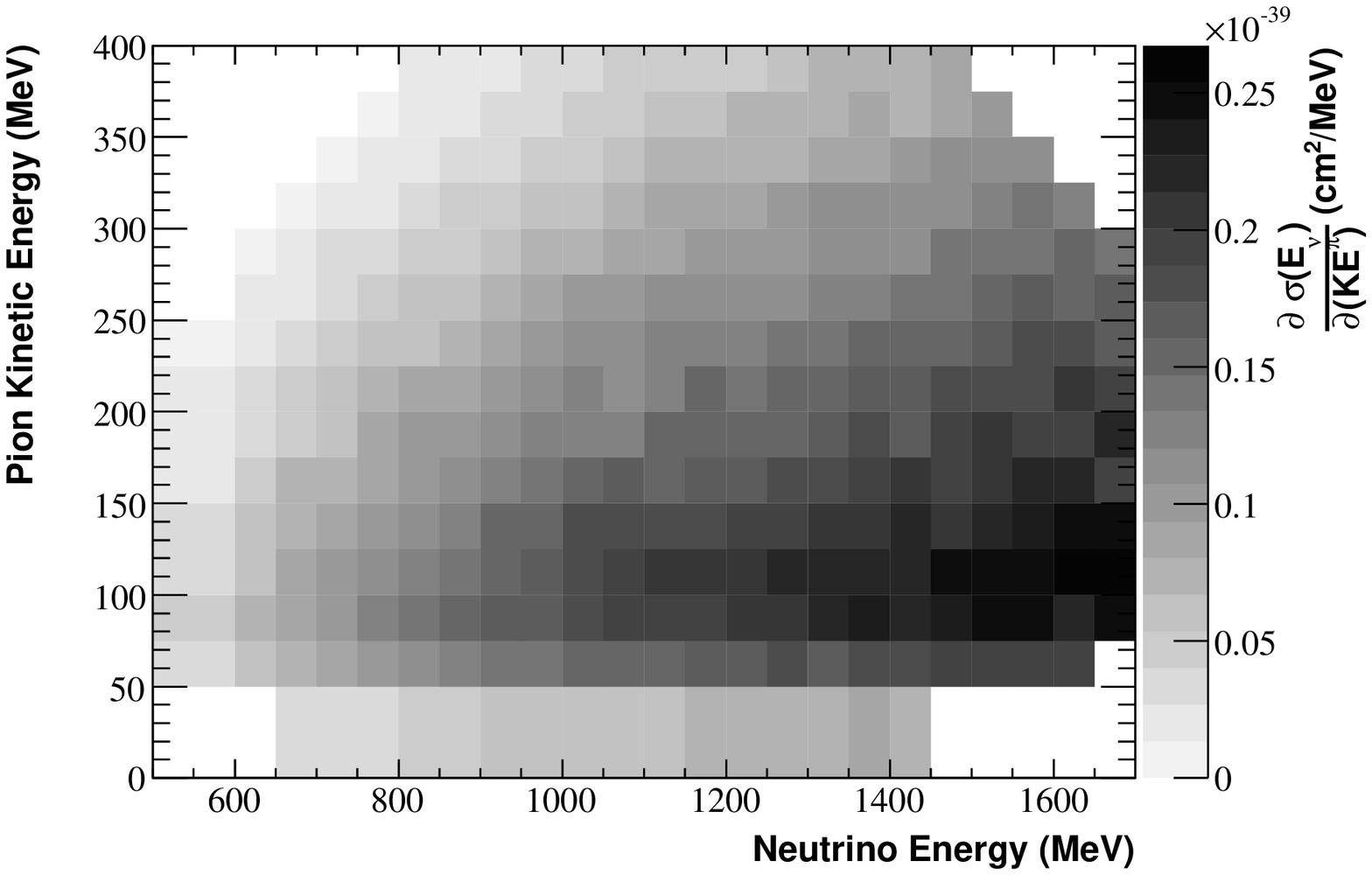}
    \includegraphics[scale=0.4,clip=true]{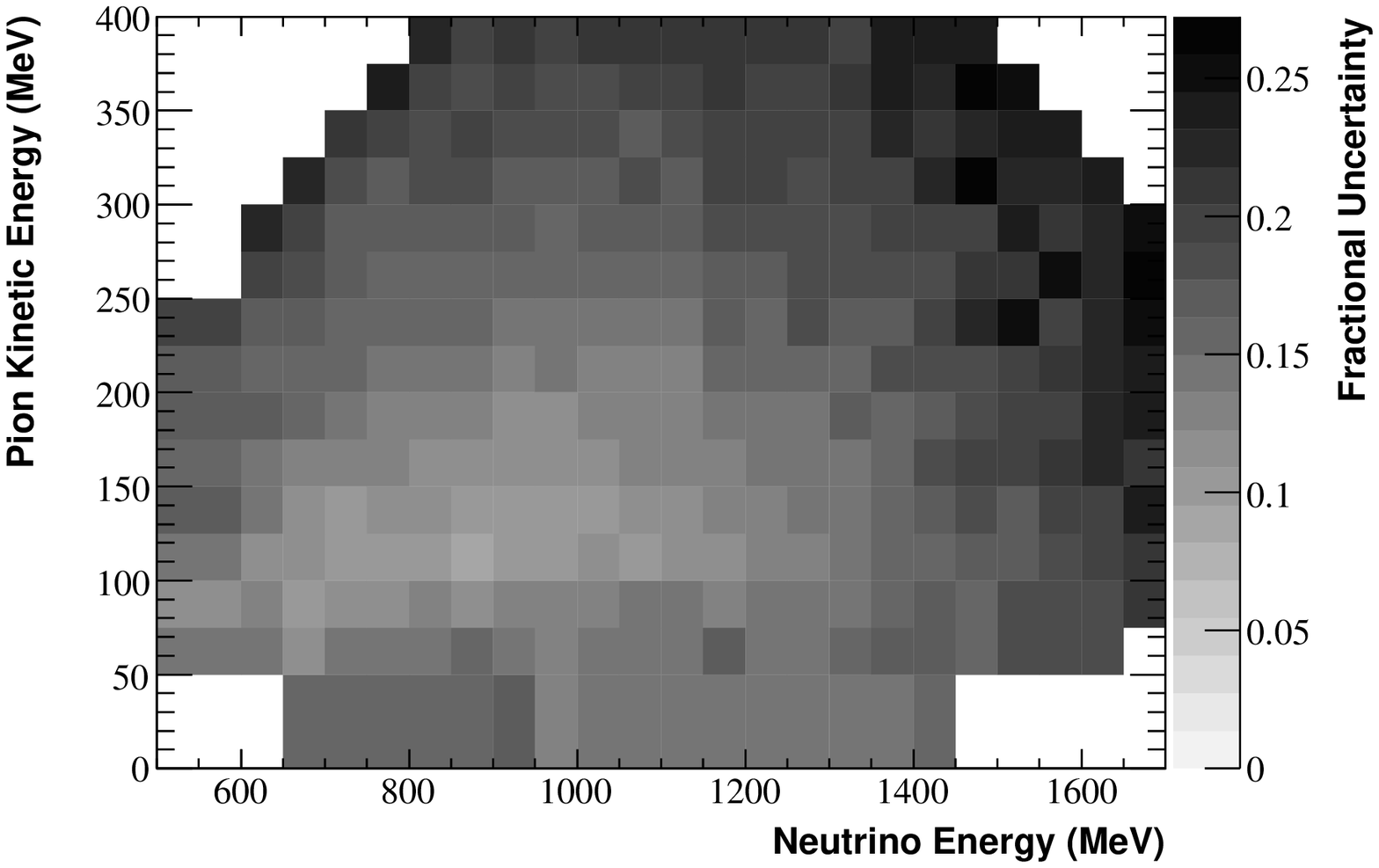}
    \includegraphics[scale=0.4,clip=true]{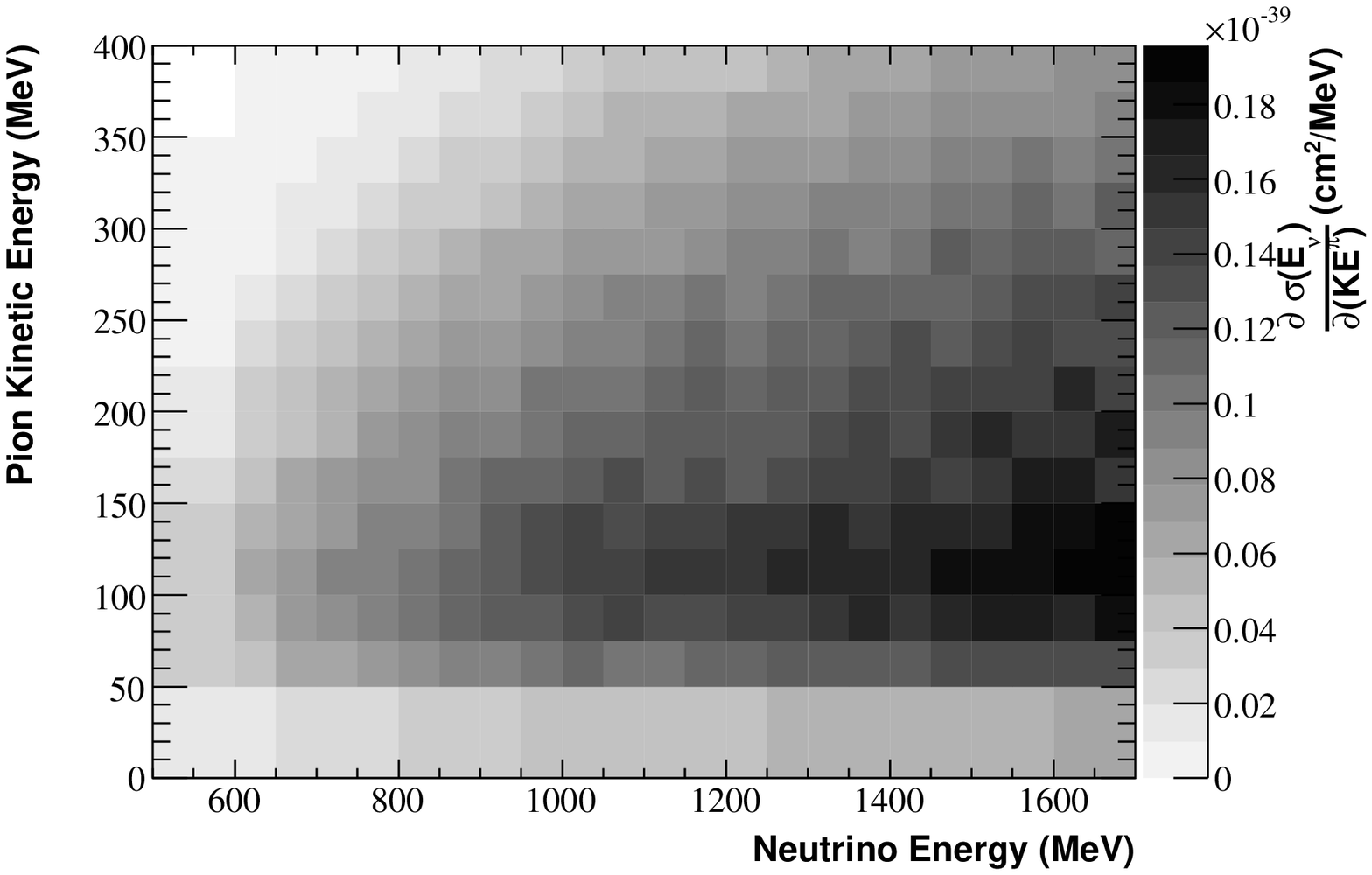}
    \caption{The measured $\partial\sigma(E_{\nu})/\partial(KE_{\pi})$
    values are shown (top) along with the total fractional
    uncertainties (middle).  Empty bins indicate regions where no
    measurement has been made.  The Monte Carlo predicted cross
    section is shown for comparison (bottom).}
    \label{fig:pikevenuxsec0} \end{center}
\end{figure}

The observable \nmccpip\ cross section on a CH$_{2.08}$ target has
been measured in a variety of forms: a total cross section as a
function of neutrino energy (Fig.~\ref{fig:enuxsec0}), a
differential cross section in $Q^2$ (Fig.~\ref{fig:qsqxsec0}),
differential cross sections in the kinetic energy of the muon
(Fig.~\ref{fig:mukexsec0}) and the pion
(Fig.~\ref{fig:pikexsec0}), double-differential cross sections in
the muon kinetic energy and angle (Fig.~\ref{fig:muctvkexsec0}) and
the pion kinetic energy and angle (Fig.~\ref{fig:pictvkexsec0}), and
two-dimensional measurements of each of the differential cross
sections in bins of neutrino energy to provide results independent of
the MiniBooNE energy spectrum (Fig.s~\ref{fig:qsqvenuxsec0},
\ref{fig:mukevenuxsec0}, and \ref{fig:pikevenuxsec0}).  This is the
first time model-independent differential cross sections have been
provided for the muon and pion kinematics in these interactions.

The binning for each of the one-dimensional distributions has been
chosen such that the true Monte Carlo prediction in each bin exceeds
250 events after all cuts.  The one-dimensional bin sizes are used for
the two-dimensional measurements as well to retain sufficient
precision in the most interesting regions of phase space.  This
results in several bins with very small numbers of predicted events.
The data-sized optical model multisims produce unreliable
uncertainties in bins with small event populations, and therefore
results will only be reported for bins that contain at least 25
inferred true data events.

\begin{table}[h!]
\caption{The uncertainties in the total, flux-averaged cross section are given for the dominant error sources.\label{tab:fracerrors}}
\begin{center}
\begin{tabular}{|c|c|}\hline
Error Source & Cross Section Uncertainty \cr\hline
Beam \pip Production & 9.2\% \cr\hline
$\nu$ Cross Sections & 8.2\% \cr\hline
Proton Beam and Horn & 4.3\% \cr\hline
Optical Model & 1.5\% \cr\hline
Other & $<$3\% \cr\hline
\end{tabular}
\end{center}
\end{table}

The uncertainties from the most significant error sources in the total
cross section, averaged over the neutrino energy spectrum, are shown
in Table~\ref{tab:fracerrors}.  In each of the one-dimensional
differential cross section measurements, the two largest sources of
uncertainty are the beam \pip\ production and the neutrino interaction
cross sections.  The \pip\ production uncertainties in the
flux-averaged results are dominated by the large uncertainties at low
neutrino energy.  Since the low-energy region has relatively little
impact on the measurements binned in neutrino energy, the beam \pip\
production uncertainties are is significantly lower and generally
remain below 10\% except at the highest neutrino energies.

The largest effects in the cross section uncertainties are
pion absorption and charge-exchange interactions that take place after
the pion has left the target nucleus.  If the pion is absorbed, it
will not produce a Michel electron and the event will fail the
three-subevent requirement; therefore, pion absorption and
charge-exchange interactions will directly affect the cut
efficiencies.  A 50\% uncertainty is assigned to the pion
charge-exchange cross section and a 35\% uncertainty is assigned to
pion absorption based on the agreement between the GCALOR Monte Carlo
simulation~\cite{gcalor} and external
data~\cite{ashery,jones,ransome}.  The remainder of the cross section
uncertainty is due to variations in the interaction cross sections of
each background process.

Since the cross section measurements in Q$^2$ and neutrino energy
include contributions from the incident neutrino, they must be
unfolded back to the initial neutrino interaction, and are therefore
dependent on the modeling of nuclear effects.  In particular,
additional uncertainties in the kinematics of the target nucleons are
absorbed in these results.  Conversely, the measurements in the muon
and pion kinematic variables are properties of only the final,
post-nuclear state, and are therefore largely insensitive to nuclear
model uncertainties.

Finally, for most of the measured phase space, the uncertainty due to
unfolding bias is negligible; however, it becomes significant at low
$Q^2$ in both the one- and two-dimensional measurements.  This
particular region has two features that generally make unfolding
difficult.  The first is that the shape is rapidly changing, which
strongly affects bin migration.  Also, this is a region where the
shapes in the data and Monte Carlo simulation significantly disagree,
which increases the probability that the shape of the true $Q^2$
distributions within each reconstructed bin are incorrect.  Despite
these features, the unfolding uncertainty is still not the dominant
systematic effect, and is of comparable size only in the few bins at
low $Q^2$ which are susceptible to these effects.

\section{Conclusion}
\label{sec:conclusion}

Results have been presented for the observable \ccpip\ cross section
as a function of several fundamental kinematic variables.  Of these
results, the cross sections measured as a function of the neutrino
energy are the least experiment dependent, since the predicted
neutrino flux has been accounted for separately in each bin.  Previous
measurements of the neutrino energy cross section below 2~\gev\ have
uncertainties larger than 20\%, upon which the present results provide
a significant improvement.

The present measurement is on average 23\% higher than the NUANCE
prediction for the observable cross section, which is compatible
with some previous results, most notably Ref.~\cite{kitagaki}.
However, a direct comparison to past data is difficult since previous
cross section measurements in this energy range were conducted on
hydrogen or deuterium.  Since MiniBooNE employs a nuclear target, it
is unclear if the source of discrepancy seen in the current results
lies in the single nucleon cross section, or whether it is due to
nuclear effects.

The remaining kinematic \ccpip\ cross sections have not been reported
previously.  Since the single- and double-differential cross section
measurements are necessarily averaged over the shape of the neutrino
flux prediction given in Fig.~\ref{fig:fluxsyst}, each measurement
of a final-state kinematic quantity has also been measured in bins of
neutrino energy to remove the dependence on the MiniBooNE energy
spectrum.  The integrated one-dimensional measurements have also been
included due to the familiarity of many in the community with
flux-averaged results.  This is the most complete set of information
that has ever been available for \ccpip\ on nuclear targets.

\clearpage

\appendix

\section{Cross Section Tables}\label{sec:tables}

Tables~\ref{tab:enu}-\ref{tab:upictvke}
give the numerical values of each measured cross section and Table~\ref{tab:numuflux} gives the MiniBooNE \numu\ flux prediction.

\begin{table}[h!]
\begin{center}
\caption{The $\sigma(E_{\nu})$ results from Fig.~\ref{fig:enuxsec0} are given with the total uncertainty.  Each row is a bin of neutrino energy (MeV) labeled according to its low edge.\label{tab:enu}}
\begin{tabular}{cc}\hline\hline
Bin (MeV) & Result $(10^{-39} \cm^2)$ \cr\hline
 500 & 6.1 $\pm$ 0.8 \cr
 600 & 11.5 $\pm$ 1.3 \cr
 650 & 15.8 $\pm$ 1.7 \cr
 700 & 19.6 $\pm$ 2.1 \cr
 750 & 24.1 $\pm$ 2.5 \cr
 800 & 28.3 $\pm$ 2.9 \cr
 850 & 32.5 $\pm$ 3.3 \cr
 900 & 37.3 $\pm$ 3.9 \cr
 950 & 41.6 $\pm$ 4.4 \cr
1000 & 46.6 $\pm$ 5.0 \cr
1050 & 49.7 $\pm$ 5.6 \cr
1100 & 52.9 $\pm$ 6.1 \cr
1150 & 56.3 $\pm$ 7.0 \cr
1200 & 59.1 $\pm$ 7.6 \cr
1250 & 62.3 $\pm$ 8.4 \cr
1300 & 66.7 $\pm$ 9.4 \cr
1350 & 70.3 $\pm$ 10.5 \cr
1400 & 72.3 $\pm$ 11.4 \cr
1450 & 77.6 $\pm$ 12.8 \cr
1500 & 80.8 $\pm$ 13.9 \cr
1550 & 83.7 $\pm$ 14.9 \cr
1600 & 86.4 $\pm$ 15.9 \cr
1650 & 88.5 $\pm$ 16.9 \cr
1700 & 93.0 $\pm$ 18.5 \cr
1750 & 92.6 $\pm$ 19.3 \cr
1800 & 97.1 $\pm$ 21.4 \cr
1900 & 99.2 $\pm$ 23.7 \cr
 \hline
\end{tabular}
\end{center}
\end{table}

\begin{table}[h!]
\begin{center}
\caption{The $\partial\sigma/\partial(KE_{\pi})$ results from Fig.~\ref{fig:pikexsec0} are given with the total uncertainty.  Each row is a bin of pion kinetic energy (MeV) labeled according to its low edge.\label{tab:pike}}
\begin{tabular}{cc}\hline\hline
Bin (MeV) & Result $(10^{-41} \cm^2/\mev)$ \cr\hline
   0 & 3.6 $\pm$ 0.5 \cr
  50 & 8.9 $\pm$ 1.2 \cr
  75 & 10.8 $\pm$ 1.4 \cr
 100 & 10.8 $\pm$ 1.2 \cr
 125 & 9.7 $\pm$ 1.2 \cr
 150 & 8.6 $\pm$ 1.1 \cr
 175 & 7.6 $\pm$ 1.1 \cr
 200 & 7.0 $\pm$ 1.0 \cr
 225 & 6.1 $\pm$ 1.0 \cr
 250 & 5.3 $\pm$ 0.9 \cr
 275 & 4.5 $\pm$ 0.8 \cr
 300 & 3.9 $\pm$ 0.7 \cr
 325 & 3.2 $\pm$ 0.6 \cr
 350 & 2.7 $\pm$ 0.5 \cr
 375 & 2.3 $\pm$ 0.4 \cr
 \hline
\end{tabular}
\end{center}
\end{table}

\begin{table}[h!]
\begin{center}
\caption{The $\partial\sigma/\partial(Q^2)$ results from Fig.~\ref{fig:qsqxsec0} are given with the total uncertainty.  Each row is a bin of $Q^2$ (GeV$^2$/c$^4$) labeled according to its low edge.\label{tab:qsq}}
\begin{tabular}{cc}\hline\hline
Bin (GeV$^2$/c$^4$) & Result $(10^{-45} \cm^2 c^4/\mev^2)$ \cr\hline
0.00 & 36.2 $\pm$ 6.7 \cr
0.05 & 55.8 $\pm$ 7.8 \cr
0.10 & 56.1 $\pm$ 7.5 \cr
0.15 & 53.4 $\pm$ 7.0 \cr
0.20 & 47.8 $\pm$ 6.4 \cr
0.25 & 43.2 $\pm$ 5.9 \cr
0.30 & 38.8 $\pm$ 5.3 \cr
0.35 & 34.1 $\pm$ 4.8 \cr
0.40 & 30.3 $\pm$ 4.3 \cr
0.45 & 26.1 $\pm$ 3.7 \cr
0.50 & 22.9 $\pm$ 3.3 \cr
0.55 & 19.8 $\pm$ 2.9 \cr
0.60 & 17.3 $\pm$ 2.5 \cr
0.65 & 14.8 $\pm$ 2.1 \cr
0.70 & 12.9 $\pm$ 1.9 \cr
0.75 & 11.0 $\pm$ 1.7 \cr
0.80 & 9.3 $\pm$ 1.4 \cr
0.85 & 8.0 $\pm$ 1.3 \cr
0.90 & 6.8 $\pm$ 1.1 \cr
0.95 & 5.3 $\pm$ 1.0 \cr
1.05 & 3.9 $\pm$ 0.7 \cr
1.15 & 2.5 $\pm$ 0.5 \cr
1.30 & 1.4 $\pm$ 0.4 \cr
 \hline
\end{tabular}
\end{center}
\end{table}

\begin{table}[h!]
\begin{center}
\caption{The $\partial\sigma/\partial(KE_{\mu})$ results from Fig.~\ref{fig:mukexsec0} are given with the total uncertainty.  Each row is a bin of muon kinetic energy (MeV) labeled according to its low edge.\label{tab:muke}}
\begin{tabular}{cc}\hline\hline
Bin (MeV) & Result $(10^{-42} \cm^2/\mev)$ \cr\hline
   0 & 23.1 $\pm$ 3.2 \cr
  50 & 34.4 $\pm$ 5.0 \cr
 100 & 39.0 $\pm$ 5.8 \cr
 150 & 41.6 $\pm$ 6.2 \cr
 200 & 41.2 $\pm$ 6.1 \cr
 250 & 39.0 $\pm$ 5.7 \cr
 300 & 37.3 $\pm$ 5.4 \cr
 350 & 35.0 $\pm$ 4.9 \cr
 400 & 32.0 $\pm$ 4.4 \cr
 450 & 28.8 $\pm$ 3.8 \cr
 500 & 26.1 $\pm$ 3.5 \cr
 550 & 23.2 $\pm$ 3.1 \cr
 600 & 20.8 $\pm$ 2.8 \cr
 650 & 18.6 $\pm$ 2.5 \cr
 700 & 15.9 $\pm$ 2.1 \cr
 750 & 14.2 $\pm$ 2.0 \cr
 800 & 12.3 $\pm$ 1.8 \cr
 850 & 10.7 $\pm$ 1.6 \cr
 900 & 9.3 $\pm$ 1.4 \cr
 950 & 7.7 $\pm$ 1.3 \cr
1000 & 5.9 $\pm$ 1.0 \cr
1100 & 4.3 $\pm$ 0.8 \cr
1200 & 2.2 $\pm$ 0.6 \cr
 \hline
\end{tabular}
\end{center}
\end{table}

\begin{table*}[h!]
\begin{center}
\caption{The $\partial\sigma(E_{\nu})/\partial(Q^2)$ results from Fig.~\ref{fig:qsqvenuxsec0} are shown ($10^{-44}\cm^2 c^4/\mev^2$).  Each bin is labeled according to its low edge.  The columns are bins of neutrino energy (MeV) and the rows are bins of $Q^2$ (GeV$^2$/c$^4$).  Empty bins indicate regions where no measurement has been made.\label{tab:qsqvenu}}
\begin{tabular}{l|ccccccccccccccccccccccc}\hline\hline
Bin &  500 &  600 &  650 &  700 &  750 &  800 &  850 &  900 &  950 & 1000 & 1050 & 1100 & 1150 & 1200 & 1250 & 1300 & 1350 & 1400 & 1450 & 1500 & 1550 & 1600 & 1650 \cr\hline
1.15 &  &  &  &  &  &  &  &  &  &  &  &  &  &  &  &  & 1.1 & 1.0 & 1.2 &  &  &  &  \cr
1.05 &  &  &  &  &  &  &  &  &  &  &  &  &  & 0.9 & 1.1 & 1.2 & 1.5 & 1.8 & 1.8 &  &  &  &  \cr
0.95 &  &  &  &  &  &  &  &  &  &  &  &  & 1.1 & 1.4 & 1.7 & 2.0 & 2.2 & 2.3 & 2.3 & 2.6 & 3.1 & 3.1 &  \cr
0.90 &  &  &  &  &  &  &  &  &  &  &  &  & 1.6 & 2.0 & 2.0 & 2.3 & 2.4 & 2.8 & 3.4 &  &  &  &  \cr
0.85 &  &  &  &  &  &  &  &  &  &  & 1.2 & 1.5 & 1.8 & 2.2 & 2.6 & 2.8 & 2.8 & 3.1 & 3.5 &  &  &  &  \cr
0.80 &  &  &  &  &  &  &  &  &  & 1.1 & 1.6 & 2.0 & 2.6 & 2.7 & 2.9 & 3.4 & 3.4 & 3.4 & 3.8 &  &  &  &  \cr
0.75 &  &  &  &  &  &  &  &  &  & 1.8 & 2.2 & 2.5 & 2.7 & 3.2 & 3.4 & 3.5 & 3.8 & 4.6 & 4.3 & 4.8 &  &  &  \cr
0.70 &  &  &  &  &  &  &  &  & 1.6 & 2.2 & 2.6 & 2.9 & 3.2 & 3.4 & 3.8 & 4.0 & 4.3 & 4.5 & 5.5 & 5.0 & 5.1 &  &  \cr
0.65 &  &  &  &  &  &  &  & 1.7 & 2.2 & 2.7 & 2.9 & 3.1 & 4.0 & 3.6 & 4.1 & 4.7 & 4.8 & 5.2 & 5.2 & 5.6 & 5.3 &  &  \cr
0.60 &  &  &  &  &  &  & 1.4 & 2.4 & 2.4 & 3.3 & 3.7 & 4.1 & 4.2 & 4.5 & 5.2 & 5.4 & 5.2 & 5.0 & 5.9 & 6.0 & 5.6 & 6.3 &  \cr
0.55 &  &  &  &  &  & 1.3 & 1.9 & 2.4 & 3.0 & 3.8 & 4.1 & 4.6 & 4.7 & 5.2 & 5.1 & 5.7 & 5.6 & 6.1 & 5.8 & 7.0 & 7.4 & 7.6 & 7.4 \cr
0.50 &  &  &  &  &  & 1.8 & 2.4 & 3.2 & 3.6 & 4.4 & 4.8 & 5.4 & 5.6 & 5.8 & 5.8 & 6.7 & 6.6 & 6.6 & 7.0 & 6.2 & 7.5 & 7.3 & 7.1 \cr
0.45 &  &  &  &  & 1.5 & 2.3 & 3.1 & 4.0 & 4.7 & 5.2 & 5.7 & 5.5 & 6.1 & 6.5 & 6.4 & 6.6 & 6.9 & 7.7 & 7.1 & 7.4 & 6.6 & 7.9 & 8.1 \cr
0.40 &  &  &  & 1.3 & 2.5 & 3.2 & 4.2 & 4.8 & 5.1 & 6.1 & 6.5 & 6.7 & 6.7 & 7.1 & 7.1 & 7.3 & 7.5 & 7.1 & 8.1 & 7.9 & 8.6 & 8.4 & 9.3 \cr
0.35 &  &  & 1.1 & 2.0 & 3.1 & 4.0 & 4.8 & 5.7 & 6.4 & 6.8 & 6.8 & 7.0 & 7.4 & 7.8 & 7.6 & 8.2 & 8.1 & 8.1 & 8.7 & 9.1 & 8.7 & 8.9 & 8.8 \cr
0.30 &  &  & 1.9 & 3.1 & 4.1 & 5.1 & 6.0 & 6.1 & 7.4 & 7.4 & 7.4 & 8.0 & 8.0 & 8.1 & 8.6 & 9.1 & 9.2 & 9.0 & 8.7 & 8.7 & 8.9 & 8.7 & 9.8 \cr
0.25 &  & 1.8 & 3.0 & 4.1 & 5.3 & 5.9 & 6.7 & 6.8 & 7.5 & 7.7 & 8.2 & 8.2 & 8.1 & 8.5 & 9.0 & 9.0 & 9.8 & 9.4 & 9.4 & 8.8 & 9.5 & 9.7 & 10.1 \cr
0.20 & 1.0 & 2.9 & 4.2 & 5.2 & 6.0 & 6.7 & 6.9 & 7.6 & 8.0 & 8.8 & 8.5 & 8.8 & 8.7 & 9.3 & 9.2 & 9.5 & 9.6 & 9.5 & 9.9 & 10.3 & 10.2 & 11.5 & 11.2 \cr
0.15 & 1.9 & 4.1 & 5.4 & 6.3 & 7.1 & 7.4 & 7.5 & 8.3 & 8.7 & 9.1 & 9.0 & 9.3 & 9.2 & 9.6 & 9.4 & 9.6 & 9.6 & 9.9 & 10.6 & 11.6 & 10.8 & 11.2 & 10.3 \cr
0.10 & 3.1 & 5.2 & 6.6 & 7.0 & 7.4 & 7.6 & 7.9 & 8.7 & 8.2 & 9.2 & 9.3 & 9.4 & 9.5 & 9.1 & 9.0 & 9.5 & 9.2 & 9.5 & 10.4 & 9.4 & 10.5 & 10.2 & 9.8 \cr
0.05 & 4.3 & 6.0 & 6.5 & 7.0 & 7.3 & 7.4 & 7.6 & 8.0 & 7.9 & 8.0 & 8.0 & 8.3 & 8.6 & 8.2 & 8.6 & 8.2 & 8.4 & 7.8 & 8.6 & 9.3 & 9.3 & 9.6 & 10.0 \cr
0.00 & 1.9 & 2.7 & 3.3 & 3.6 & 3.9 & 4.3 & 4.6 & 4.8 & 5.2 & 5.4 & 5.4 & 5.2 & 5.8 & 5.9 & 6.0 & 6.4 & 6.6 & 6.1 & 6.3 & 6.7 & 7.2 &  &  \cr
\end{tabular}
\end{center}
\end{table*}

\begin{table*}[h!]
\begin{center}
\caption{The percent uncertainty of the $\partial\sigma(E_{\nu})/\partial(Q^2)$ results from Fig.~\ref{fig:qsqvenuxsec0} is shown.  Each bin is labeled according to its low edge.  The columns are bins of neutrino energy (MeV) and the rows are bins of $Q^2$ (GeV$^2$/c$^4$).  Empty bins indicate regions where no measurement has been made.\label{tab:uqsqvenu}}
\begin{tabular}{l|ccccccccccccccccccccccc}\hline\hline
Bin &  500 &  600 &  650 &  700 &  750 &  800 &  850 &  900 &  950 & 1000 & 1050 & 1100 & 1150 & 1200 & 1250 & 1300 & 1350 & 1400 & 1450 & 1500 & 1550 & 1600 & 1650 \cr\hline
1.15 &  &  &  &  &  &  &  &  &  &  &  &  &  &  &  &  & 25.7 & 26.5 & 34.4 &  &  &  &  \cr
1.05 &  &  &  &  &  &  &  &  &  &  &  &  &  & 27.9 & 33.9 & 26.0 & 21.6 & 22.1 & 33.5 &  &  &  &  \cr
0.95 &  &  &  &  &  &  &  &  &  &  &  &  & 22.7 & 23.5 & 20.8 & 21.9 & 21.5 & 21.5 & 20.6 & 25.1 & 22.0 & 21.9 &  \cr
0.90 &  &  &  &  &  &  &  &  &  &  &  &  & 21.8 & 25.5 & 22.6 & 19.8 & 21.7 & 21.1 & 22.8 &  &  &  &  \cr
0.85 &  &  &  &  &  &  &  &  &  &  & 24.1 & 20.0 & 22.1 & 18.8 & 20.0 & 22.1 & 19.6 & 21.3 & 20.6 &  &  &  &  \cr
0.80 &  &  &  &  &  &  &  &  &  & 26.4 & 22.7 & 20.2 & 18.9 & 17.7 & 18.5 & 19.5 & 17.9 & 21.8 & 21.8 &  &  &  &  \cr
0.75 &  &  &  &  &  &  &  &  &  & 22.2 & 22.5 & 17.3 & 16.1 & 16.8 & 16.4 & 18.9 & 18.9 & 20.5 & 20.0 & 23.0 &  &  &  \cr
0.70 &  &  &  &  &  &  &  &  & 21.0 & 15.9 & 19.4 & 17.2 & 15.7 & 15.1 & 15.6 & 18.3 & 18.7 & 22.6 & 20.9 & 25.0 & 23.7 &  &  \cr
0.65 &  &  &  &  &  &  &  & 23.6 & 17.9 & 21.8 & 15.0 & 14.8 & 14.9 & 15.7 & 17.6 & 19.1 & 19.0 & 20.7 & 19.8 & 18.9 & 23.5 &  &  \cr
0.60 &  &  &  &  &  &  & 25.6 & 24.0 & 15.3 & 16.3 & 14.7 & 13.6 & 16.4 & 15.5 & 17.0 & 18.2 & 17.7 & 18.0 & 19.9 & 22.2 & 25.7 & 23.2 &  \cr
0.55 &  &  &  &  &  & 17.3 & 18.4 & 18.0 & 14.2 & 13.8 & 12.9 & 14.3 & 15.7 & 16.2 & 16.3 & 16.3 & 17.4 & 18.6 & 18.0 & 22.6 & 22.2 & 22.6 & 25.5 \cr
0.50 &  &  &  &  &  & 22.0 & 14.5 & 14.3 & 13.3 & 12.8 & 14.2 & 14.8 & 15.3 & 14.3 & 15.5 & 16.7 & 17.7 & 19.7 & 19.1 & 21.4 & 22.7 & 25.2 & 29.9 \cr
0.45 &  &  &  &  & 19.7 & 13.9 & 13.4 & 13.1 & 13.3 & 13.5 & 13.9 & 13.9 & 14.4 & 16.7 & 15.3 & 16.4 & 16.9 & 18.2 & 19.0 & 20.8 & 22.8 & 27.1 & 29.2 \cr
0.40 &  &  &  & 19.7 & 16.3 & 12.1 & 12.4 & 13.4 & 13.4 & 13.4 & 14.1 & 14.5 & 15.9 & 13.9 & 15.4 & 14.6 & 16.8 & 18.6 & 18.8 & 19.5 & 21.2 & 22.4 & 23.5 \cr
0.35 &  &  & 19.8 & 14.8 & 12.0 & 12.8 & 11.7 & 13.1 & 12.3 & 13.5 & 13.9 & 13.8 & 14.0 & 14.8 & 15.7 & 15.3 & 17.5 & 16.5 & 18.9 & 20.7 & 21.9 & 21.1 & 25.2 \cr
0.30 &  &  & 13.0 & 11.0 & 12.7 & 12.1 & 12.5 & 12.6 & 12.6 & 12.8 & 12.3 & 12.8 & 13.1 & 14.5 & 15.3 & 15.2 & 16.6 & 17.9 & 18.0 & 19.8 & 19.8 & 21.5 & 23.2 \cr
0.25 &  & 14.1 & 12.3 & 12.8 & 11.9 & 12.0 & 11.9 & 11.7 & 11.6 & 11.9 & 11.9 & 12.7 & 13.7 & 13.9 & 14.4 & 15.4 & 16.8 & 17.0 & 18.7 & 18.6 & 19.3 & 23.2 & 22.4 \cr
0.20 & 14.8 & 13.3 & 11.2 & 11.5 & 11.6 & 11.2 & 11.1 & 11.1 & 11.5 & 12.3 & 12.8 & 12.7 & 13.4 & 14.3 & 14.8 & 15.2 & 15.0 & 17.0 & 18.2 & 18.2 & 21.7 & 22.8 & 27.1 \cr
0.15 & 10.9 & 10.6 & 11.2 & 11.0 & 10.9 & 11.1 & 10.9 & 12.0 & 11.8 & 11.9 & 12.4 & 12.3 & 13.7 & 14.0 & 15.0 & 14.4 & 16.4 & 16.2 & 19.4 & 19.0 & 20.6 & 21.7 & 22.6 \cr
0.10 & 10.3 & 11.2 & 10.8 & 10.9 & 11.2 & 12.1 & 11.9 & 11.5 & 12.1 & 11.9 & 12.0 & 12.1 & 13.6 & 14.5 & 14.8 & 15.7 & 16.1 & 16.7 & 17.6 & 18.4 & 19.0 & 21.5 & 23.1 \cr
0.05 & 12.2 & 12.8 & 12.2 & 13.9 & 13.8 & 13.3 & 12.8 & 11.8 & 11.9 & 11.7 & 12.8 & 13.0 & 13.7 & 14.8 & 15.8 & 17.5 & 18.4 & 21.1 & 20.4 & 19.1 & 21.4 & 20.9 & 23.9 \cr
0.00 & 20.0 & 18.6 & 19.1 & 18.9 & 17.7 & 17.2 & 16.2 & 14.9 & 14.8 & 13.6 & 15.8 & 15.3 & 14.9 & 14.7 & 15.3 & 17.3 & 20.9 & 20.7 & 21.9 & 24.0 & 21.2 &  &  \cr
\end{tabular}
\end{center}
\end{table*}

\begin{table*}[h!]
\begin{center}
\caption{The $\partial\sigma(E_{\nu})/\partial(KE_{\mu})$ results from Fig.~\ref{fig:mukevenuxsec0} are shown ($10^{-41}\cm^2/\mev$).  Each bin is labeled according to its low edge.  The columns are bins of neutrino energy (MeV) and the rows are bins of muon kinetic energy (MeV).  Empty bins indicate regions where no measurement has been made.\label{tab:mukevenu}}
\begin{tabular}{l|ccccccccccccccccccccccc}\hline\hline
Bin &  500 &  600 &  650 &  700 &  750 &  800 &  850 &  900 &  950 & 1000 & 1050 & 1100 & 1150 & 1200 & 1250 & 1300 & 1350 & 1400 & 1450 & 1500 & 1550 & 1600 & 1650 \cr\hline
1200 &  &  &  &  &  &  &  &  &  &  &  &  &  &  &  &  &  &  &  &  &  &  & 1.9 \cr
1100 &  &  &  &  &  &  &  &  &  &  &  &  &  &  &  &  &  &  &  &  & 4.1 & 6.3 & 8.0 \cr
1000 &  &  &  &  &  &  &  &  &  &  &  &  &  &  &  &  &  & 1.8 & 3.6 & 5.9 & 8.0 & 8.9 & 9.0 \cr
 950 &  &  &  &  &  &  &  &  &  &  &  &  &  &  &  &  & 2.8 & 4.6 & 6.9 & 8.6 & 9.4 & 9.5 & 9.6 \cr
 900 &  &  &  &  &  &  &  &  &  &  &  &  &  &  &  & 2.5 & 4.9 & 6.9 & 8.8 & 9.4 & 9.8 & 9.1 & 9.5 \cr
 850 &  &  &  &  &  &  &  &  &  &  &  &  &  &  & 2.5 & 4.2 & 6.8 & 8.4 & 9.1 & 10.3 & 9.7 & 9.5 & 9.1 \cr
 800 &  &  &  &  &  &  &  &  &  &  &  &  & 1.2 & 2.4 & 4.7 & 7.1 & 7.9 & 9.5 & 9.4 & 9.6 & 9.5 & 8.5 & 9.2 \cr
 750 &  &  &  &  &  &  &  &  &  &  &  & 1.3 & 2.7 & 4.8 & 7.0 & 8.5 & 9.5 & 9.4 & 9.5 & 8.3 & 8.3 & 9.1 & 9.5 \cr
 700 &  &  &  &  &  &  &  &  &  &  & 1.1 & 2.4 & 4.8 & 7.0 & 8.1 & 8.9 & 8.8 & 9.1 & 9.1 & 8.5 & 8.8 & 8.6 & 7.5 \cr
 650 &  &  &  &  &  &  &  &  &  & 1.2 & 2.5 & 4.6 & 6.9 & 8.4 & 9.7 & 9.7 & 9.9 & 9.2 & 8.9 & 8.0 & 8.1 & 7.3 & 7.4 \cr
 600 &  &  &  &  &  &  &  &  & 1.2 & 2.5 & 4.8 & 7.0 & 8.7 & 9.1 & 9.0 & 9.1 & 9.2 & 9.3 & 8.4 & 8.4 & 7.8 & 7.9 &  \cr
 550 &  &  &  &  &  &  &  & 1.2 & 2.4 & 4.7 & 6.7 & 8.3 & 9.1 & 9.4 & 8.9 & 9.8 & 8.8 & 7.6 & 8.9 & 7.9 & 8.1 & 7.1 &  \cr
 500 &  &  &  &  &  &  & 1.2 & 2.6 & 4.5 & 7.1 & 8.4 & 9.4 & 9.2 & 9.3 & 9.1 & 8.9 & 8.9 & 8.0 & 8.2 & 8.0 & 7.1 & 7.4 &  \cr
 450 &  &  &  &  &  & 1.1 & 2.4 & 4.5 & 7.1 & 8.3 & 9.1 & 9.8 & 9.7 & 9.0 & 8.7 & 8.6 & 8.2 & 8.9 & 7.4 & 7.7 & 7.6 & 6.3 &  \cr
 400 &  &  &  &  & 1.0 & 2.2 & 4.3 & 7.0 & 8.7 & 9.6 & 9.9 & 9.4 & 8.9 & 9.1 & 9.1 & 8.4 & 7.9 & 7.5 & 7.9 & 7.0 & 6.5 & 6.2 &  \cr
 350 &  &  &  & 1.0 & 2.1 & 4.3 & 6.8 & 8.3 & 9.8 & 10.4 & 9.9 & 9.5 & 9.1 & 8.4 & 8.9 & 8.4 & 7.6 & 7.1 & 7.0 & 7.5 & 6.1 &  &  \cr
 300 &  & 0.3 & 1.0 & 2.1 & 4.3 & 6.2 & 7.9 & 9.3 & 9.5 & 10.2 & 10.2 & 9.1 & 9.4 & 8.4 & 8.4 & 8.3 & 7.3 & 7.2 & 7.3 & 6.7 & 6.7 &  &  \cr
 250 & 0.2 & 0.9 & 2.0 & 3.9 & 6.2 & 7.8 & 8.5 & 9.5 & 9.8 & 9.4 & 9.4 & 9.0 & 8.5 & 8.3 & 7.6 & 7.7 & 6.7 & 6.8 & 6.6 & 6.0 &  &  &  \cr
 200 & 0.5 & 1.8 & 4.0 & 6.0 & 7.7 & 8.5 & 9.3 & 9.5 & 9.6 & 9.3 & 8.9 & 8.3 & 8.0 & 7.5 & 6.8 & 6.4 & 6.8 & 6.0 & 5.7 & 5.8 &  &  &  \cr
 150 & 1.3 & 3.7 & 5.7 & 7.6 & 8.5 & 9.3 & 9.3 & 8.9 & 8.5 & 8.2 & 7.5 & 7.4 & 6.7 & 6.7 & 6.2 & 5.7 & 5.4 & 4.7 & 4.9 &  &  &  &  \cr
 100 & 2.2 & 5.4 & 7.1 & 8.0 & 8.5 & 8.5 & 7.9 & 7.5 & 6.5 & 6.6 & 6.1 & 5.6 & 5.4 & 4.8 & 4.5 & 4.0 & 4.1 &  &  &  &  &  &  \cr
  50 & 4.0 & 6.7 & 7.5 & 7.2 & 7.0 & 6.2 & 5.6 & 5.1 & 4.5 & 4.2 & 3.9 & 3.6 & 3.4 & 3.2 &  &  &  &  &  &  &  &  &  \cr
   0 & 4.5 & 4.8 & 4.6 & 3.8 & 3.5 & 3.1 & 2.7 & 2.6 & 2.4 & 2.4 & 2.2 &  &  &  &  &  &  &  &  &  &  &  &  \cr
\end{tabular}
\end{center}
\end{table*}

\begin{table*}[h!]
\begin{center}
\caption{The percent uncertainty of the $\partial\sigma(E_{\nu})/\partial(KE_{\mu})$ results from Fig.~\ref{fig:mukevenuxsec0} is shown.  Each bin is labeled according to its low edge.  The columns are bins of neutrino energy (MeV) and the rows are bins of muon kinetic energy (MeV).  Empty bins indicate regions where no measurement has been made.\label{tab:umukevenu}}
\begin{tabular}{l|ccccccccccccccccccccccc}\hline\hline
Bin &  500 &  600 &  650 &  700 &  750 &  800 &  850 &  900 &  950 & 1000 & 1050 & 1100 & 1150 & 1200 & 1250 & 1300 & 1350 & 1400 & 1450 & 1500 & 1550 & 1600 & 1650 \cr\hline
1200 &  &  &  &  &  &  &  &  &  &  &  &  &  &  &  &  &  &  &  &  &  &  & 28.6 \cr
1100 &  &  &  &  &  &  &  &  &  &  &  &  &  &  &  &  &  &  &  &  & 22.9 & 22.9 & 24.8 \cr
1000 &  &  &  &  &  &  &  &  &  &  &  &  &  &  &  &  &  & 21.9 & 22.0 & 19.8 & 21.2 & 23.3 & 23.8 \cr
 950 &  &  &  &  &  &  &  &  &  &  &  &  &  &  &  &  & 22.2 & 19.6 & 19.7 & 20.3 & 20.9 & 22.1 & 26.8 \cr
 900 &  &  &  &  &  &  &  &  &  &  &  &  &  &  &  & 19.2 & 19.5 & 20.1 & 19.5 & 20.1 & 22.4 & 23.3 & 25.2 \cr
 850 &  &  &  &  &  &  &  &  &  &  &  &  &  &  & 20.7 & 19.0 & 17.9 & 17.9 & 18.8 & 19.8 & 23.3 & 23.7 & 25.9 \cr
 800 &  &  &  &  &  &  &  &  &  &  &  &  & 21.2 & 20.6 & 19.9 & 16.8 & 17.4 & 17.7 & 19.2 & 21.5 & 22.0 & 24.5 & 25.8 \cr
 750 &  &  &  &  &  &  &  &  &  &  &  & 18.3 & 16.4 & 15.1 & 15.6 & 15.9 & 16.1 & 18.1 & 19.1 & 21.1 & 20.5 & 26.1 & 30.5 \cr
 700 &  &  &  &  &  &  &  &  &  &  & 20.4 & 18.0 & 15.6 & 15.3 & 15.8 & 16.0 & 16.5 & 17.9 & 18.5 & 20.4 & 21.3 & 22.3 & 28.9 \cr
 650 &  &  &  &  &  &  &  &  &  & 20.0 & 18.4 & 15.6 & 15.1 & 14.3 & 15.3 & 16.1 & 18.0 & 17.9 & 19.6 & 21.0 & 21.9 & 22.0 & 25.8 \cr
 600 &  &  &  &  &  &  &  &  & 15.8 & 14.9 & 13.6 & 12.9 & 13.9 & 15.6 & 14.9 & 15.5 & 17.2 & 19.2 & 20.1 & 20.8 & 22.6 & 25.6 &  \cr
 550 &  &  &  &  &  &  &  & 15.5 & 14.4 & 14.1 & 13.5 & 13.7 & 13.7 & 15.1 & 15.2 & 16.4 & 17.0 & 19.1 & 20.9 & 21.3 & 24.7 & 22.9 &  \cr
 500 &  &  &  &  &  &  & 15.0 & 12.3 & 12.5 & 12.7 & 11.9 & 13.4 & 13.8 & 14.4 & 15.6 & 17.0 & 17.8 & 18.2 & 20.8 & 25.2 & 24.9 & 26.4 &  \cr
 450 &  &  &  &  &  & 17.0 & 13.7 & 12.3 & 11.8 & 10.8 & 11.7 & 12.3 & 13.5 & 14.2 & 14.8 & 15.5 & 16.4 & 19.2 & 19.2 & 21.7 & 23.0 & 22.8 &  \cr
 400 &  &  &  &  & 18.8 & 16.3 & 13.7 & 11.8 & 11.5 & 11.5 & 11.8 & 12.3 & 13.8 & 13.7 & 16.5 & 16.8 & 17.0 & 18.7 & 20.9 & 20.5 & 23.5 & 23.9 &  \cr
 350 &  &  &  & 20.3 & 17.4 & 14.5 & 12.0 & 10.9 & 11.0 & 12.0 & 12.5 & 13.2 & 13.8 & 14.8 & 15.8 & 16.8 & 18.8 & 19.6 & 18.9 & 20.4 & 21.6 &  &  \cr
 300 &  & 21.5 & 16.9 & 15.4 & 14.0 & 12.3 & 11.4 & 10.9 & 11.1 & 12.3 & 13.9 & 15.3 & 15.4 & 15.5 & 16.1 & 20.0 & 19.1 & 21.2 & 21.3 & 21.3 & 24.1 &  &  \cr
 250 & 22.5 & 16.5 & 13.6 & 12.6 & 12.0 & 11.9 & 10.4 & 11.0 & 11.7 & 12.4 & 13.3 & 14.7 & 15.6 & 16.3 & 18.4 & 17.4 & 19.4 & 22.9 & 22.8 & 23.0 &  &  &  \cr
 200 & 18.2 & 14.1 & 11.7 & 10.4 & 10.7 & 10.8 & 10.9 & 12.3 & 12.2 & 12.7 & 14.4 & 15.0 & 16.2 & 17.7 & 17.9 & 20.0 & 19.3 & 19.4 & 21.4 & 24.2 &  &  &  \cr
 150 & 14.6 & 12.6 & 10.8 & 10.7 & 11.0 & 11.5 & 11.3 & 12.0 & 13.4 & 13.7 & 15.4 & 14.8 & 15.7 & 17.0 & 18.9 & 18.7 & 20.3 & 21.7 & 20.6 &  &  &  &  \cr
 100 & 14.2 & 11.6 & 10.8 & 10.7 & 10.6 & 11.4 & 11.9 & 12.1 & 13.0 & 13.7 & 14.3 & 15.2 & 16.4 & 18.1 & 17.7 & 19.9 & 20.5 &  &  &  &  &  &  \cr
  50 & 13.3 & 12.9 & 11.2 & 12.0 & 11.9 & 11.7 & 13.8 & 14.9 & 13.9 & 15.4 & 16.7 & 17.5 & 17.9 & 18.5 &  &  &  &  &  &  &  &  &  \cr
   0 & 12.8 & 11.9 & 13.5 & 12.7 & 14.3 & 14.6 & 15.6 & 16.2 & 16.9 & 18.2 & 18.4 &  &  &  &  &  &  &  &  &  &  &  &  \cr
\end{tabular}
\end{center}
\end{table*}

\begin{table*}[h!]
\begin{center}
\caption{The $\partial\sigma(E_{\nu})/\partial(KE_{\pi})$ results from Fig.~\ref{fig:pikevenuxsec0} are shown ($10^{-41}\cm^2/\mev$).  Each bin is labeled according to its low edge.  The columns are bins of neutrino energy (MeV) and the rows are bins of pion kinetic energy (MeV).  Empty bins indicate regions where no measurement has been made.\label{tab:pikevenu}}
\begin{tabular}{l|ccccccccccccccccccccccc}\hline\hline
Bin &  500 &  600 &  650 &  700 &  750 &  800 &  850 &  900 &  950 & 1000 & 1050 & 1100 & 1150 & 1200 & 1250 & 1300 & 1350 & 1400 & 1450 & 1500 & 1550 & 1600 & 1650 \cr\hline
 375 &  &  &  &  &  & 1.6 & 1.8 & 2.3 & 2.7 & 3.5 & 4.3 & 4.8 & 5.1 & 5.2 & 6.0 & 6.8 & 7.0 & 7.5 & 8.0 &  &  &  &  \cr
 350 &  &  &  &  & 1.3 & 1.8 & 2.2 & 3.0 & 3.5 & 4.4 & 5.2 & 5.4 & 6.0 & 6.9 & 6.7 & 7.2 & 8.2 & 8.0 & 8.9 & 9.5 &  &  &  \cr
 325 &  &  &  & 1.1 & 1.9 & 2.6 & 3.2 & 4.0 & 5.0 & 5.2 & 6.0 & 7.0 & 6.8 & 7.5 & 7.9 & 8.6 & 9.2 & 10.1 & 10.9 & 10.9 & 11.5 &  &  \cr
 300 &  &  & 1.1 & 1.6 & 2.6 & 3.7 & 4.1 & 4.8 & 5.7 & 6.6 & 7.2 & 8.0 & 8.2 & 8.7 & 9.9 & 10.9 & 10.8 & 11.0 & 11.5 & 12.3 & 13.7 & 12.8 &  \cr
 275 &  & 0.9 & 1.6 & 2.7 & 3.6 & 4.4 & 5.2 & 6.4 & 7.3 & 7.8 & 8.4 & 8.7 & 9.4 & 10.4 & 10.6 & 11.7 & 10.9 & 12.0 & 14.4 & 13.5 & 14.2 & 15.8 & 13.6 \cr
 250 &  & 1.4 & 2.4 & 3.4 & 4.6 & 5.4 & 6.3 & 7.1 & 8.2 & 9.5 & 10.0 & 11.1 & 11.6 & 11.3 & 11.8 & 12.8 & 12.8 & 13.5 & 14.3 & 14.8 & 16.5 & 16.0 & 16.7 \cr
 225 & 0.7 & 2.4 & 3.3 & 4.4 & 5.5 & 6.4 & 7.5 & 8.3 & 9.5 & 11.0 & 11.2 & 12.1 & 12.8 & 12.8 & 13.4 & 14.3 & 15.4 & 15.7 & 15.3 & 16.2 & 17.9 & 17.5 & 17.2 \cr
 200 & 1.4 & 3.1 & 4.2 & 5.6 & 6.9 & 8.3 & 9.2 & 10.0 & 11.3 & 12.3 & 12.0 & 13.1 & 14.9 & 14.2 & 15.0 & 15.8 & 16.5 & 16.8 & 17.8 & 18.0 & 17.7 & 20.3 & 18.8 \cr
 175 & 1.9 & 3.8 & 5.2 & 6.3 & 8.1 & 9.5 & 10.0 & 11.3 & 12.3 & 13.2 & 13.1 & 14.9 & 15.5 & 15.6 & 16.0 & 16.9 & 18.2 & 17.2 & 19.9 & 20.6 & 19.1 & 19.7 & 22.0 \cr
 150 & 2.6 & 5.2 & 6.9 & 7.9 & 8.8 & 9.9 & 11.4 & 12.8 & 13.6 & 15.2 & 16.3 & 15.8 & 16.5 & 16.1 & 17.7 & 18.2 & 18.8 & 20.2 & 18.9 & 20.4 & 22.1 & 22.4 & 19.8 \cr
 125 & 3.4 & 5.7 & 7.2 & 8.6 & 10.5 & 11.2 & 12.4 & 14.7 & 15.9 & 17.5 & 17.6 & 18.0 & 18.4 & 19.5 & 19.7 & 21.2 & 20.6 & 22.0 & 21.2 & 22.2 & 24.0 & 24.1 & 24.4 \cr
 100 & 3.8 & 6.5 & 8.9 & 10.6 & 11.7 & 12.7 & 14.5 & 15.8 & 17.1 & 18.5 & 19.0 & 20.3 & 20.2 & 21.1 & 21.9 & 22.1 & 22.6 & 22.7 & 25.0 & 25.3 & 25.3 & 26.5 & 26.7 \cr
  75 & 4.2 & 7.0 & 9.0 & 10.5 & 12.3 & 13.7 & 14.9 & 16.2 & 16.8 & 18.3 & 19.7 & 19.3 & 18.9 & 20.7 & 20.8 & 22.0 & 22.9 & 21.7 & 23.3 & 25.0 & 24.1 & 21.6 & 24.8 \cr
  50 & 3.7 & 5.9 & 7.4 & 8.6 & 9.9 & 11.2 & 12.4 & 13.7 & 14.1 & 15.4 & 14.7 & 15.5 & 16.3 & 16.8 & 18.0 & 17.0 & 17.8 & 17.8 & 18.8 & 19.0 & 18.9 & 19.2 &  \cr
   0 &  &  & 2.9 & 3.4 & 3.8 & 4.5 & 5.1 & 5.4 & 5.8 & 6.1 & 6.6 & 6.6 & 7.0 & 6.8 & 7.0 & 7.7 & 8.1 & 7.7 &  &  &  &  &  \cr
\end{tabular}
\end{center}
\end{table*}

\begin{table*}[h!]
\begin{center}
\caption{The percent uncertainty of the $\partial\sigma(E_{\nu})/\partial(KE_{\pi})$ results from Fig.~\ref{fig:pikevenuxsec0} is shown.  Each bin is labeled according to its low edge.  The columns are bins of neutrino energy (MeV) and the rows are bins of pion kinetic energy (MeV).  Empty bins indicate regions where no measurement has been made.\label{tab:upikevenu}}
\begin{tabular}{l|ccccccccccccccccccccccc}\hline\hline
Bin &  500 &  600 &  650 &  700 &  750 &  800 &  850 &  900 &  950 & 1000 & 1050 & 1100 & 1150 & 1200 & 1250 & 1300 & 1350 & 1400 & 1450 & 1500 & 1550 & 1600 & 1650 \cr\hline
 375 &  &  &  &  &  & 22.5 & 20.1 & 20.9 & 19.5 & 20.8 & 20.5 & 20.8 & 20.8 & 20.6 & 21.2 & 20.3 & 23.3 & 23.2 & 24.2 &  &  &  &  \cr
 350 &  &  &  &  & 23.2 & 20.2 & 18.9 & 19.3 & 18.2 & 19.0 & 20.3 & 19.3 & 20.7 & 19.9 & 19.2 & 21.1 & 23.3 & 23.0 & 26.8 & 25.8 &  &  &  \cr
 325 &  &  &  & 20.6 & 19.4 & 18.8 & 19.3 & 18.8 & 18.2 & 18.4 & 17.2 & 18.6 & 19.3 & 19.9 & 20.1 & 20.2 & 22.3 & 21.3 & 22.5 & 24.3 & 23.5 &  &  \cr
 300 &  &  & 22.6 & 19.0 & 17.0 & 18.2 & 17.9 & 17.6 & 16.6 & 16.8 & 17.7 & 17.6 & 19.7 & 20.0 & 19.0 & 19.9 & 20.3 & 21.9 & 27.2 & 21.9 & 22.8 & 23.2 &  \cr
 275 &  & 23.0 & 19.9 & 17.6 & 17.3 & 17.3 & 16.4 & 17.2 & 16.0 & 17.0 & 17.1 & 17.5 & 18.4 & 18.4 & 17.9 & 18.9 & 19.9 & 19.7 & 19.6 & 24.3 & 21.6 & 23.0 & 25.5 \cr
 250 &  & 19.8 & 18.1 & 17.1 & 15.6 & 15.5 & 15.1 & 16.2 & 16.0 & 16.0 & 16.2 & 15.8 & 17.5 & 16.7 & 17.9 & 18.2 & 18.7 & 18.7 & 21.7 & 21.3 & 24.6 & 22.7 & 26.3 \cr
 225 & 19.4 & 17.3 & 16.7 & 16.0 & 15.3 & 15.1 & 15.0 & 14.7 & 14.9 & 14.6 & 14.3 & 14.4 & 16.7 & 15.9 & 18.4 & 16.7 & 17.2 & 19.4 & 22.8 & 24.8 & 19.4 & 22.5 & 24.9 \cr
 200 & 17.0 & 16.3 & 15.6 & 15.3 & 13.7 & 14.1 & 14.1 & 13.5 & 13.7 & 13.0 & 13.5 & 13.4 & 15.7 & 15.2 & 15.9 & 16.2 & 18.5 & 18.4 & 18.8 & 20.0 & 21.2 & 21.9 & 23.4 \cr
 175 & 16.6 & 17.3 & 15.9 & 14.1 & 12.9 & 12.8 & 12.5 & 12.2 & 12.2 & 12.7 & 13.3 & 13.5 & 13.9 & 13.8 & 14.8 & 16.8 & 16.2 & 17.4 & 18.6 & 19.4 & 19.9 & 22.5 & 24.4 \cr
 150 & 15.6 & 14.9 & 13.5 & 12.8 & 12.5 & 12.1 & 11.5 & 11.6 & 11.9 & 11.3 & 12.5 & 12.4 & 13.3 & 14.9 & 14.7 & 14.8 & 16.2 & 18.1 & 19.5 & 20.2 & 20.6 & 22.6 & 21.3 \cr
 125 & 16.5 & 14.6 & 11.6 & 10.8 & 11.5 & 11.1 & 10.5 & 10.5 & 10.2 & 10.5 & 11.4 & 12.1 & 13.1 & 13.0 & 14.2 & 14.7 & 16.1 & 17.3 & 17.7 & 17.4 & 19.9 & 20.0 & 24.4 \cr
 100 & 14.4 & 11.9 & 10.9 & 9.8 & 10.0 & 10.1 & 9.2 & 9.8 & 9.7 & 11.5 & 10.8 & 11.1 & 11.9 & 12.9 & 13.2 & 14.3 & 15.4 & 15.7 & 17.2 & 17.4 & 18.3 & 20.1 & 21.2 \cr
  75 & 12.0 & 12.8 & 10.4 & 11.3 & 11.6 & 13.0 & 12.0 & 12.8 & 12.7 & 12.7 & 14.0 & 13.7 & 13.2 & 13.8 & 14.4 & 14.2 & 15.3 & 16.6 & 15.9 & 18.7 & 18.6 & 18.8 & 21.5 \cr
  50 & 14.5 & 13.9 & 11.9 & 13.9 & 14.3 & 13.9 & 15.4 & 14.3 & 13.3 & 14.1 & 14.4 & 14.7 & 16.7 & 14.5 & 13.9 & 15.0 & 17.1 & 16.6 & 15.9 & 18.0 & 18.5 & 18.3 &  \cr
   0 &  &  & 15.3 & 16.0 & 15.9 & 15.9 & 16.1 & 16.5 & 12.9 & 14.0 & 14.3 & 14.0 & 14.9 & 13.6 & 13.8 & 14.7 & 14.2 & 15.5 &  &  &  &  &  \cr
\end{tabular}
\end{center}
\end{table*}

\begin{table*}[h!]
\begin{center}
\caption{The $\partial^2\sigma/\partial(KE_{\mu})\partial(\cos(\theta_{\mu,\nu}))$ results from Fig.~\ref{fig:muctvkexsec0} are shown ($10^{-42}\cm^2/\mev$).  Each bin is labeled according to its low edge.  The columns are bins of muon kinetic energy (MeV) and the rows are bins of cos(muon, neutrino angle).  Empty bins indicate regions where no measurement has been made.\label{tab:muctvke}}
\begin{tabular}{l|cccccccccccccccccccc}\hline\hline
Bin &  150 &  200 &  250 &  300 &  350 &  400 &  450 &  500 &  550 &  600 &  650 &  700 &  750 &  800 &  850 &  900 &  950 & 1000 & 1100 & 1200 \cr\hline
0.95 & 23.0 & 26.1 & 32.1 & 38.5 & 42.8 & 47.9 & 55.8 & 58.2 & 59.8 & 63.7 & 65.8 & 69.3 & 71.5 & 65.9 & 58.5 & 59.3 & 56.0 & 48.4 & 43.3 & 27.7 \cr
0.90 & 30.5 & 35.8 & 41.3 & 46.7 & 55.0 & 59.5 & 64.3 & 70.8 & 70.3 & 72.6 & 73.3 & 65.0 & 65.7 & 59.4 & 58.0 & 50.8 & 46.4 & 38.3 & 28.1 & 12.6 \cr
0.85 & 31.7 & 37.0 & 44.8 & 50.9 & 58.7 & 70.2 & 67.6 & 67.5 & 70.4 & 71.4 & 69.5 & 56.0 & 52.7 & 49.3 & 42.3 & 36.1 & 27.3 & 19.8 & 12.9 & 5.7 \cr
0.80 & 31.7 & 39.1 & 43.4 & 53.8 & 63.6 & 62.5 & 64.7 & 67.3 & 61.7 & 57.2 & 55.3 & 45.4 & 37.1 & 33.6 & 25.1 & 19.9 & 15.5 & 9.0 &  &  \cr
0.75 & 33.7 & 40.0 & 45.5 & 53.3 & 57.3 & 58.6 & 58.5 & 60.4 & 53.0 & 44.9 & 37.0 & 33.1 & 25.8 & 18.7 & 13.7 &  &  &  &  &  \cr
0.70 & 37.8 & 40.5 & 48.6 & 58.5 & 57.3 & 58.2 & 53.8 & 52.4 & 42.4 & 35.5 & 28.8 & 22.0 & 15.4 & 12.3 &  &  &  &  &  &  \cr
0.65 & 35.9 & 41.7 & 46.8 & 50.6 & 56.1 & 49.2 & 44.8 & 38.4 & 29.6 & 23.5 & 16.9 & 14.1 & 10.0 &  &  &  &  &  &  &  \cr
0.60 & 33.5 & 43.3 & 47.0 & 47.1 & 51.6 & 51.1 & 40.6 & 31.0 & 25.0 & 16.7 & 11.5 & 8.5 &  &  &  &  &  &  &  &  \cr
0.55 & 35.0 & 41.7 & 46.1 & 49.4 & 42.3 & 37.3 & 33.2 & 24.1 & 17.1 & 11.1 &  &  &  &  &  &  &  &  &  &  \cr
0.50 & 32.8 & 38.3 & 44.2 & 44.2 & 39.5 & 33.6 & 26.1 & 15.9 & 11.8 &  &  &  &  &  &  &  &  &  &  &  \cr
0.45 & 32.8 & 37.8 & 39.0 & 39.1 & 32.8 & 25.8 & 17.9 & 13.3 & 8.8 &  &  &  &  &  &  &  &  &  &  &  \cr
0.40 & 32.5 & 38.1 & 42.3 & 37.7 & 28.9 & 22.4 & 14.9 & 9.2 &  &  &  &  &  &  &  &  &  &  &  &  \cr
0.35 & 36.9 & 40.5 & 36.9 & 31.6 & 27.1 & 17.0 & 12.1 &  &  &  &  &  &  &  &  &  &  &  &  &  \cr
0.30 & 31.1 & 34.7 & 29.9 & 28.8 & 20.9 & 12.3 &  &  &  &  &  &  &  &  &  &  &  &  &  &  \cr
0.25 & 30.2 & 31.8 & 30.2 & 22.9 & 17.9 & 11.2 &  &  &  &  &  &  &  &  &  &  &  &  &  &  \cr
0.20 & 27.2 & 31.6 & 27.1 & 22.3 & 14.2 & 10.6 &  &  &  &  &  &  &  &  &  &  &  &  &  &  \cr
0.15 & 26.2 & 26.8 & 23.1 & 18.4 & 9.4 &  &  &  &  &  &  &  &  &  &  &  &  &  &  &  \cr
0.10 & 25.3 & 25.3 & 21.3 & 15.6 &  &  &  &  &  &  &  &  &  &  &  &  &  &  &  &  \cr
0.05 & 24.7 & 25.4 & 20.6 & 13.3 &  &  &  &  &  &  &  &  &  &  &  &  &  &  &  &  \cr
0.00 & 24.5 & 23.1 & 17.6 & 10.3 &  &  &  &  &  &  &  &  &  &  &  &  &  &  &  &  \cr
-0.05 & 24.6 & 18.6 & 13.1 &  &  &  &  &  &  &  &  &  &  &  &  &  &  &  &  &  \cr
-0.10 & 21.3 & 17.9 & 12.4 &  &  &  &  &  &  &  &  &  &  &  &  &  &  &  &  &  \cr
-0.20 & 21.0 & 16.7 & 10.2 & 5.4 &  &  &  &  &  &  &  &  &  &  &  &  &  &  &  &  \cr
-0.30 & 17.5 & 12.3 & 5.7 &  &  &  &  &  &  &  &  &  &  &  &  &  &  &  &  &  \cr
-0.40 & 14.7 & 10.2 & 6.0 &  &  &  &  &  &  &  &  &  &  &  &  &  &  &  &  &  \cr
-0.50 & 13.1 & 8.3 &  &  &  &  &  &  &  &  &  &  &  &  &  &  &  &  &  &  \cr
-0.60 & 11.6 & 5.7 &  &  &  &  &  &  &  &  &  &  &  &  &  &  &  &  &  &  \cr
-0.80 & 8.5 & 3.0 &  &  &  &  &  &  &  &  &  &  &  &  &  &  &  &  &  &  \cr
-1.00 & 3.9 &  &  &  &  &  &  &  &  &  &  &  &  &  &  &  &  &  &  &  \cr
\end{tabular}
\end{center}
\end{table*}

\begin{table*}[h!]
\begin{center}
\caption{The percent uncertainty of the $\partial^2\sigma/\partial(KE_{\mu})\partial(\cos(\theta_{\mu,\nu}))$ results from Fig.~\ref{fig:muctvkexsec0} is shown.  Each bin is labeled according to its low edge.  The columns are bins of muon kinetic energy (MeV) and the rows are bins of cos(muon, neutrino angle).  Empty bins indicate regions where no measurement has been made.\label{tab:umuctvke}}
\begin{tabular}{l|cccccccccccccccccccc}\hline\hline
Bin &  150 &  200 &  250 &  300 &  350 &  400 &  450 &  500 &  550 &  600 &  650 &  700 &  750 &  800 &  850 &  900 &  950 & 1000 & 1100 & 1200 \cr\hline
0.95 & 27.0 & 25.9 & 20.7 & 19.0 & 17.5 & 17.1 & 14.7 & 16.2 & 16.3 & 14.5 & 15.0 & 13.9 & 13.9 & 15.2 & 16.3 & 15.9 & 15.8 & 15.8 & 15.6 & 19.7 \cr
0.90 & 18.9 & 18.1 & 17.6 & 16.7 & 16.0 & 14.6 & 13.9 & 15.0 & 14.3 & 14.5 & 14.3 & 15.4 & 14.6 & 13.8 & 14.4 & 14.0 & 16.8 & 16.8 & 20.6 & 25.0 \cr
0.85 & 20.3 & 16.9 & 16.0 & 17.0 & 16.4 & 16.0 & 13.8 & 14.7 & 13.6 & 14.8 & 14.5 & 13.8 & 14.3 & 15.4 & 17.6 & 18.1 & 19.4 & 20.3 & 24.6 & 36.6 \cr
0.80 & 18.5 & 16.6 & 16.7 & 15.1 & 15.8 & 15.9 & 14.2 & 14.5 & 13.2 & 14.0 & 14.3 & 14.7 & 15.1 & 15.7 & 18.2 & 22.4 & 27.1 & 33.1 &  &  \cr
0.75 & 17.0 & 16.3 & 15.9 & 14.9 & 14.0 & 14.5 & 14.5 & 14.1 & 14.2 & 15.6 & 16.4 & 15.9 & 16.5 & 17.3 & 21.8 &  &  &  &  &  \cr
0.70 & 15.9 & 15.6 & 15.6 & 15.8 & 16.5 & 15.9 & 14.7 & 15.0 & 14.8 & 15.0 & 15.7 & 20.2 & 20.5 & 21.0 &  &  &  &  &  &  \cr
0.65 & 16.4 & 15.4 & 14.5 & 15.7 & 15.4 & 14.8 & 15.4 & 15.0 & 16.2 & 16.5 & 16.1 & 20.1 & 26.5 &  &  &  &  &  &  &  \cr
0.60 & 15.9 & 15.1 & 15.3 & 16.1 & 16.0 & 16.2 & 15.1 & 15.4 & 16.1 & 20.7 & 21.0 & 23.9 &  &  &  &  &  &  &  &  \cr
0.55 & 16.4 & 16.8 & 16.4 & 15.1 & 14.7 & 15.1 & 17.2 & 18.9 & 17.8 & 21.8 &  &  &  &  &  &  &  &  &  &  \cr
0.50 & 16.4 & 17.4 & 15.1 & 17.1 & 16.4 & 17.1 & 16.8 & 19.1 & 25.5 &  &  &  &  &  &  &  &  &  &  &  \cr
0.45 & 15.5 & 17.8 & 17.0 & 16.3 & 15.6 & 17.8 & 19.0 & 17.6 & 26.0 &  &  &  &  &  &  &  &  &  &  &  \cr
0.40 & 16.4 & 16.8 & 17.1 & 17.0 & 16.0 & 17.2 & 18.7 & 26.3 &  &  &  &  &  &  &  &  &  &  &  &  \cr
0.35 & 19.1 & 20.3 & 17.7 & 16.2 & 17.3 & 22.5 & 25.2 &  &  &  &  &  &  &  &  &  &  &  &  &  \cr
0.30 & 18.1 & 20.6 & 17.2 & 15.5 & 18.1 & 20.8 &  &  &  &  &  &  &  &  &  &  &  &  &  &  \cr
0.25 & 17.1 & 16.9 & 16.2 & 16.8 & 18.5 & 24.5 &  &  &  &  &  &  &  &  &  &  &  &  &  &  \cr
0.20 & 17.1 & 17.8 & 18.0 & 20.2 & 19.2 & 35.3 &  &  &  &  &  &  &  &  &  &  &  &  &  &  \cr
0.15 & 17.9 & 16.6 & 16.9 & 21.1 & 22.2 &  &  &  &  &  &  &  &  &  &  &  &  &  &  &  \cr
0.10 & 16.4 & 17.1 & 17.7 & 21.1 &  &  &  &  &  &  &  &  &  &  &  &  &  &  &  &  \cr
0.05 & 18.7 & 18.5 & 21.0 & 24.0 &  &  &  &  &  &  &  &  &  &  &  &  &  &  &  &  \cr
0.00 & 18.4 & 21.1 & 22.5 & 27.8 &  &  &  &  &  &  &  &  &  &  &  &  &  &  &  &  \cr
-0.05 & 20.1 & 20.0 & 20.8 &  &  &  &  &  &  &  &  &  &  &  &  &  &  &  &  &  \cr
-0.10 & 21.8 & 17.7 & 25.6 &  &  &  &  &  &  &  &  &  &  &  &  &  &  &  &  &  \cr
-0.20 & 19.1 & 18.9 & 26.1 & 36.8 &  &  &  &  &  &  &  &  &  &  &  &  &  &  &  &  \cr
-0.30 & 17.7 & 21.6 & 28.3 &  &  &  &  &  &  &  &  &  &  &  &  &  &  &  &  &  \cr
-0.40 & 18.7 & 20.4 & 35.2 &  &  &  &  &  &  &  &  &  &  &  &  &  &  &  &  &  \cr
-0.50 & 21.7 & 27.5 &  &  &  &  &  &  &  &  &  &  &  &  &  &  &  &  &  &  \cr
-0.60 & 24.0 & 32.9 &  &  &  &  &  &  &  &  &  &  &  &  &  &  &  &  &  &  \cr
-0.80 & 24.2 & 29.5 &  &  &  &  &  &  &  &  &  &  &  &  &  &  &  &  &  &  \cr
-1.00 & 25.4 &  &  &  &  &  &  &  &  &  &  &  &  &  &  &  &  &  &  &  \cr
\end{tabular}
\end{center}
\end{table*}

\begin{table*}[h!]
\begin{center}
\caption{The $\partial^2\sigma/\partial(KE_{\pi})\partial(\cos(\theta_{\pi,\nu}))$ results from Fig.~\ref{fig:pictvkexsec0} are shown ($10^{-41}\cm^2/\mev$).  Each bin is labeled according to its low edge.  The columns are bins of pion kinetic energy (MeV) and the rows are bins of cos(pion, neutrino angle).  Empty bins indicate regions where no measurement has been made.\label{tab:pictvke}}
\begin{tabular}{l|cccccccccc}\hline\hline
Bin &  150 &  175 &  200 &  225 &  250 &  275 &  300 &  325 &  350 &  375 \cr\hline
0.95 & 7.9 & 8.9 & 10.3 & 10.7 & 10.0 & 10.3 & 9.4 & 8.0 & 6.8 & 6.0 \cr
0.90 & 8.5 & 9.6 & 10.1 & 9.9 & 9.9 & 9.1 & 8.8 & 7.5 & 6.3 & 5.4 \cr
0.85 & 9.7 & 10.0 & 10.8 & 10.0 & 9.2 & 8.5 & 7.5 & 6.7 & 5.5 & 4.7 \cr
0.80 & 9.2 & 9.8 & 9.6 & 9.2 & 8.4 & 7.5 & 6.9 & 5.9 & 5.3 & 4.4 \cr
0.75 & 9.6 & 9.0 & 9.1 & 8.6 & 7.8 & 7.2 & 6.3 & 5.3 & 4.2 & 4.1 \cr
0.70 & 9.2 & 9.7 & 8.9 & 8.2 & 7.4 & 6.3 & 5.4 & 4.3 & 3.9 & 3.4 \cr
0.65 & 9.4 & 8.7 & 8.2 & 7.6 & 6.5 & 5.8 & 4.9 & 4.1 & 3.2 & 2.9 \cr
0.60 & 9.3 & 8.4 & 8.4 & 7.0 & 5.9 & 5.0 & 3.9 & 3.4 & 3.0 & 2.5 \cr
0.55 & 9.0 & 7.7 & 7.2 & 6.1 & 5.5 & 4.6 & 3.5 & 3.1 & 2.4 & 2.2 \cr
0.50 & 8.4 & 7.5 & 6.4 & 5.6 & 5.0 & 3.8 & 3.2 & 2.6 & 2.1 &  \cr
0.45 & 7.5 & 6.7 & 5.8 & 4.9 & 4.2 & 3.3 & 2.7 &  &  &  \cr
0.40 & 7.5 & 6.2 & 5.4 & 4.4 & 3.5 & 2.6 & 2.2 &  &  &  \cr
0.35 & 7.0 & 5.8 & 4.9 & 3.6 & 3.1 & 2.5 &  &  &  &  \cr
0.30 & 6.8 & 5.5 & 4.3 & 3.3 & 2.5 &  &  &  &  &  \cr
0.25 & 6.1 & 5.1 & 3.6 & 2.6 & 2.3 &  &  &  &  &  \cr
0.20 & 5.7 & 4.2 & 3.4 & 2.7 &  &  &  &  &  &  \cr
0.15 & 5.1 & 3.7 & 2.8 & 1.9 &  &  &  &  &  &  \cr
0.10 & 4.5 & 3.5 & 2.3 & 1.7 &  &  &  &  &  &  \cr
0.05 & 4.3 & 2.9 & 2.4 &  &  &  &  &  &  &  \cr
0.00 & 3.7 & 2.4 & 2.0 &  &  &  &  &  &  &  \cr
-0.05 & 3.1 & 2.1 &  &  &  &  &  &  &  &  \cr
-0.10 & 2.8 & 1.8 &  &  &  &  &  &  &  &  \cr
-0.15 & 2.3 &  &  &  &  &  &  &  &  &  \cr
-0.20 & 2.1 &  &  &  &  &  &  &  &  &  \cr
-0.25 & 2.0 &  &  &  &  &  &  &  &  &  \cr
\end{tabular}
\end{center}
\end{table*}

\begin{table*}[h!]
\begin{center}
\caption{The percent uncertainty of the $\partial^2\sigma/\partial(KE_{\pi})\partial(\cos(\theta_{\pi,\nu}))$ results from Fig.~\ref{fig:pictvkexsec0} is shown.  Each bin is labeled according to its low edge.  The columns are bins of pion kinetic energy (MeV) and the rows are bins of cos(pion, neutrino angle).  Empty bins indicate regions where no measurement has been made.\label{tab:upictvke}}
\begin{tabular}{l|cccccccccc}\hline\hline
Bin &  150 &  175 &  200 &  225 &  250 &  275 &  300 &  325 &  350 &  375 \cr\hline
0.95 & 16.5 & 16.7 & 17.3 & 17.8 & 19.7 & 20.8 & 21.3 & 22.6 & 22.1 & 22.8 \cr
0.90 & 14.4 & 16.1 & 15.9 & 17.4 & 18.1 & 19.2 & 19.8 & 20.1 & 22.0 & 21.4 \cr
0.85 & 12.8 & 13.5 & 14.8 & 16.0 & 17.4 & 18.1 & 19.0 & 18.3 & 19.1 & 20.0 \cr
0.80 & 13.8 & 14.3 & 14.8 & 15.6 & 17.5 & 19.8 & 19.2 & 19.5 & 20.1 & 19.8 \cr
0.75 & 13.5 & 15.3 & 15.9 & 15.8 & 17.3 & 17.7 & 19.6 & 18.7 & 21.5 & 20.3 \cr
0.70 & 14.3 & 16.0 & 15.3 & 17.1 & 17.6 & 17.8 & 19.9 & 18.0 & 19.0 & 19.8 \cr
0.65 & 14.0 & 16.0 & 15.5 & 16.8 & 16.8 & 17.7 & 17.1 & 22.4 & 19.2 & 21.8 \cr
0.60 & 14.0 & 16.9 & 16.4 & 17.6 & 19.6 & 19.7 & 22.2 & 20.3 & 21.7 & 23.3 \cr
0.55 & 15.7 & 17.4 & 20.4 & 19.4 & 18.6 & 19.7 & 19.8 & 21.2 & 32.3 & 21.1 \cr
0.50 & 16.2 & 17.9 & 18.8 & 18.1 & 19.2 & 19.0 & 22.6 & 23.5 & 20.0 &  \cr
0.45 & 17.0 & 18.2 & 19.3 & 19.4 & 19.4 & 23.1 & 23.0 &  &  &  \cr
0.40 & 17.1 & 16.7 & 19.5 & 23.2 & 22.5 & 20.8 & 25.2 &  &  &  \cr
0.35 & 16.8 & 16.3 & 18.4 & 21.7 & 21.9 & 23.8 &  &  &  &  \cr
0.30 & 17.8 & 19.0 & 19.7 & 20.8 & 22.2 &  &  &  &  &  \cr
0.25 & 16.9 & 18.7 & 21.3 & 22.9 & 22.0 &  &  &  &  &  \cr
0.20 & 18.3 & 18.8 & 21.9 & 20.8 &  &  &  &  &  &  \cr
0.15 & 18.5 & 19.7 & 18.5 & 20.9 &  &  &  &  &  &  \cr
0.10 & 17.3 & 16.6 & 17.8 & 20.5 &  &  &  &  &  &  \cr
0.05 & 17.1 & 16.4 & 17.5 &  &  &  &  &  &  &  \cr
0.00 & 17.4 & 20.4 & 18.4 &  &  &  &  &  &  &  \cr
-0.05 & 16.8 & 17.6 &  &  &  &  &  &  &  &  \cr
-0.10 & 18.4 & 20.1 &  &  &  &  &  &  &  &  \cr
-0.15 & 21.5 &  &  &  &  &  &  &  &  &  \cr
-0.20 & 18.9 &  &  &  &  &  &  &  &  &  \cr
-0.25 & 20.4 &  &  &  &  &  &  &  &  &  \cr
\end{tabular}
\end{center}
\end{table*}

\begin{table*}[h!]
\begin{tabular}{cccc|cccc|cccc}
\hline
\hline
& $E_\nu$ bin && $\nu_\mu$ flux && $E_\nu$ bin && $\nu_\mu$ flux && $E_\nu$ bin && $\nu_\mu$ flux \\
& (GeV) && ($\numu$/POT/GeV/cm$^2$) && (GeV) && ($\numu$/POT/GeV/cm$^2$) && (GeV) && ($\numu$/POT/GeV/cm$^2$) \\
\hline
&0.00-0.05&&$4.54\times 10^{-11}$&&1.00-1.05&&$3.35\times 10^{-10}$&&2.00-2.05&&$1.92\times 10^{-11}$\\
&0.05-0.10&&$1.71\times 10^{-10}$&&1.05-1.10&&$3.12\times 10^{-10}$&&2.05-2.10&&$1.63\times 10^{-11}$\\
&0.10-0.15&&$2.22\times 10^{-10}$&&1.10-1.15&&$2.88\times 10^{-10}$&&2.10-2.15&&$1.39\times 10^{-11}$\\
&0.15-0.20&&$2.67\times 10^{-10}$&&1.15-1.20&&$2.64\times 10^{-10}$&&2.15-2.20&&$1.19\times 10^{-11}$\\
&0.20-0.25&&$3.32\times 10^{-10}$&&1.20-1.25&&$2.39\times 10^{-10}$&&2.20-2.25&&$1.03\times 10^{-11}$\\
&0.25-0.30&&$3.64\times 10^{-10}$&&1.25-1.30&&$2.14\times 10^{-10}$&&2.25-2.30&&$8.96\times 10^{-12}$\\
&0.30-0.35&&$3.89\times 10^{-10}$&&1.30-1.35&&$1.90\times 10^{-10}$&&2.30-2.35&&$7.87\times 10^{-12}$\\
&0.35-0.40&&$4.09\times 10^{-10}$&&1.35-1.40&&$1.67\times 10^{-10}$&&2.35-2.40&&$7.00\times 10^{-12}$\\
&0.40-0.45&&$4.32\times 10^{-10}$&&1.40-1.45&&$1.46\times 10^{-10}$&&2.40-2.45&&$6.30\times 10^{-12}$\\
&0.45-0.50&&$4.48\times 10^{-10}$&&1.45-1.50&&$1.26\times 10^{-10}$&&2.45-2.50&&$5.73\times 10^{-12}$\\
&0.50-0.55&&$4.56\times 10^{-10}$&&1.50-1.55&&$1.08\times 10^{-10}$&&2.50-2.55&&$5.23\times 10^{-12}$\\
&0.55-0.60&&$4.58\times 10^{-10}$&&1.55-1.60&&$9.20\times 10^{-11}$&&2.55-2.60&&$4.82\times 10^{-12}$\\
&0.60-0.65&&$4.55\times 10^{-10}$&&1.60-1.65&&$7.80\times 10^{-11}$&&2.60-2.65&&$4.55\times 10^{-12}$\\
&0.65-0.70&&$4.51\times 10^{-10}$&&1.65-1.70&&$6.57\times 10^{-11}$&&2.65-2.70&&$4.22\times 10^{-12}$\\
&0.70-0.75&&$4.43\times 10^{-10}$&&1.70-1.75&&$5.52\times 10^{-11}$&&2.70-2.75&&$3.99\times 10^{-12}$\\
&0.75-0.80&&$4.31\times 10^{-10}$&&1.75-1.80&&$4.62\times 10^{-11}$&&2.75-2.80&&$3.84\times 10^{-12}$\\
&0.80-0.85&&$4.16\times 10^{-10}$&&1.80-1.85&&$3.86\times 10^{-11}$&&2.80-2.85&&$3.63\times 10^{-12}$\\
&0.85-0.90&&$3.98\times 10^{-10}$&&1.85-1.90&&$3.23\times 10^{-11}$&&2.85-2.90&&$3.45\times 10^{-12}$\\
&0.90-0.95&&$3.79\times 10^{-10}$&&1.90-1.95&&$2.71\times 10^{-11}$&&2.90-2.95&&$3.33\times 10^{-12}$\\
&0.95-1.00&&$3.58\times 10^{-10}$&&1.95-2.00&&$2.28\times 10^{-11}$&&2.95-3.00&&$3.20\times 10^{-12}$\\
\hline
\hline
\end{tabular}
\caption{
Predicted $\nu_\mu$ flux at the MiniBooNE detector.\label{tab:numuflux}}
\end{table*}

\clearpage

\bibliography{ccpip}

\begin{thebibliography}{36}
\expandafter\ifx\csname natexlab\endcsname\relax\def\natexlab#1{#1}\fi
\expandafter\ifx\csname bibnamefont\endcsname\relax
  \def\bibnamefont#1{#1}\fi
\expandafter\ifx\csname bibfnamefont\endcsname\relax
  \def\bibfnamefont#1{#1}\fi
\expandafter\ifx\csname citenamefont\endcsname\relax
  \def\citenamefont#1{#1}\fi
\expandafter\ifx\csname url\endcsname\relax
  \def\url#1{\texttt{#1}}\fi
\expandafter\ifx\csname urlprefix\endcsname\relax\def\urlprefix{URL }\fi
\providecommand{\bibinfo}[2]{#2}
\providecommand{\eprint}[2][]{\url{#2}}

\bibitem[{\citenamefont{Athar et~al.}(2008)\citenamefont{Athar, Chauhan, and
  Singh}}]{th1}
\bibinfo{author}{\bibfnamefont{M.~S.} \bibnamefont{Athar}},
  \bibinfo{author}{\bibfnamefont{S.}~\bibnamefont{Chauhan}}, \bibnamefont{and}
  \bibinfo{author}{\bibfnamefont{S.~K.} \bibnamefont{Singh}},
  \bibinfo{journal}{arXiv:0808.2103v1}  (\bibinfo{year}{2008}).

\bibitem[{\citenamefont{Hern\'andez et~al.}(2008)\citenamefont{Hern\'andez,
  Nieves, and Valverde}}]{th2}
\bibinfo{author}{\bibfnamefont{E.}~\bibnamefont{Hern\'andez}},
  \bibinfo{author}{\bibfnamefont{J.}~\bibnamefont{Nieves}}, \bibnamefont{and}
  \bibinfo{author}{\bibfnamefont{M.}~\bibnamefont{Valverde}},
  \bibinfo{journal}{Mod. Phys. Lett.} \textbf{\bibinfo{volume}{A23}},
  \bibinfo{pages}{2317} (\bibinfo{year}{2008}).

\bibitem[{\citenamefont{Leitner et~al.}(2009)\citenamefont{Leitner, Buss,
  Alvarez-Ruso, and Mosel}}]{th3}
\bibinfo{author}{\bibfnamefont{T.}~\bibnamefont{Leitner}},
  \bibinfo{author}{\bibfnamefont{O.}~\bibnamefont{Buss}},
  \bibinfo{author}{\bibfnamefont{L.}~\bibnamefont{Alvarez-Ruso}},
  \bibnamefont{and} \bibinfo{author}{\bibfnamefont{U.}~\bibnamefont{Mosel}},
  \bibinfo{journal}{Phys. Rev.} \textbf{\bibinfo{volume}{C79}},
  \bibinfo{pages}{034601} (\bibinfo{year}{2009}).

\bibitem[{\citenamefont{Praet et~al.}(2009)\citenamefont{Praet, Lalakulich,
  Jackowicz, and Ryckebusch}}]{th4}
\bibinfo{author}{\bibfnamefont{C.}~\bibnamefont{Praet}},
  \bibinfo{author}{\bibfnamefont{O.}~\bibnamefont{Lalakulich}},
  \bibinfo{author}{\bibfnamefont{N.}~\bibnamefont{Jackowicz}},
  \bibnamefont{and}
  \bibinfo{author}{\bibfnamefont{J.}~\bibnamefont{Ryckebusch}},
  \bibinfo{journal}{Phys. Rev.} \textbf{\bibinfo{volume}{C79}},
  \bibinfo{pages}{044603} (\bibinfo{year}{2009}).

\bibitem[{\citenamefont{Paschos and Rakshit}(2008)}]{th5}
\bibinfo{author}{\bibfnamefont{E.~A.} \bibnamefont{Paschos}} \bibnamefont{and}
  \bibinfo{author}{\bibfnamefont{S.}~\bibnamefont{Rakshit}},
  \bibinfo{journal}{arXiv:0812.4234v1}  (\bibinfo{year}{2008}).

\bibitem[{\citenamefont{Graczyk et~al.}(2009)\citenamefont{Graczyk,
  Kielczewska, and Sobczyk}}]{th6}
\bibinfo{author}{\bibfnamefont{K.~M.} \bibnamefont{Graczyk}},
  \bibinfo{author}{\bibfnamefont{D.}~\bibnamefont{Kielczewska}},
  \bibnamefont{and} \bibinfo{author}{\bibfnamefont{J.~T.}
  \bibnamefont{Sobczyk}}, \bibinfo{journal}{arXiv:0907.1886v1}
  (\bibinfo{year}{2009}).

\bibitem[{\citenamefont{Martini et~al.}(2009)\citenamefont{Martini, Ericson,
  Chanfray, and Marteau}}]{th7}
\bibinfo{author}{\bibfnamefont{M.}~\bibnamefont{Martini}},
  \bibinfo{author}{\bibfnamefont{M.}~\bibnamefont{Ericson}},
  \bibinfo{author}{\bibfnamefont{G.}~\bibnamefont{Chanfray}}, \bibnamefont{and}
  \bibinfo{author}{\bibfnamefont{J.}~\bibnamefont{Marteau}},
  \bibinfo{journal}{Phys. Rev.} \textbf{\bibinfo{volume}{C80}},
  \bibinfo{pages}{065501} (\bibinfo{year}{2009}).

\bibitem[{\citenamefont{Gershtein et~al.}(1980)\citenamefont{Gershtein,
  Komachenko, and Khlopov}}]{th8}
\bibinfo{author}{\bibfnamefont{S.~S.} \bibnamefont{Gershtein}},
  \bibinfo{author}{\bibfnamefont{Y.~Y.} \bibnamefont{Komachenko}},
  \bibnamefont{and} \bibinfo{author}{\bibfnamefont{M.~Y.}
  \bibnamefont{Khlopov}}, \bibinfo{journal}{Sov. J. Nucl. Phys.}
  \textbf{\bibinfo{volume}{32}}, \bibinfo{pages}{861} (\bibinfo{year}{1980}).

\bibitem[{\citenamefont{Aguilar-Arevalo
  et~al.}(2009{\natexlab{a}})}]{minibooneratio}
\bibinfo{author}{\bibfnamefont{A.~A.} \bibnamefont{Aguilar-Arevalo}}
  \bibnamefont{et~al.}, \bibinfo{journal}{Phys. Rev. Lett.}
  \textbf{\bibinfo{volume}{103}}, \bibinfo{pages}{081801}
  (\bibinfo{year}{2009}{\natexlab{a}}).

\bibitem[{\citenamefont{Rodriguez et~al.}(2008)}]{k2kratio}
\bibinfo{author}{\bibfnamefont{A.}~\bibnamefont{Rodriguez}}
  \bibnamefont{et~al.}, \bibinfo{journal}{Phys. Rev.}
  \textbf{\bibinfo{volume}{D78}}, \bibinfo{pages}{032003}
  (\bibinfo{year}{2008}).

\bibitem[{\citenamefont{Radecky et~al.}(1982)}]{radecky}
\bibinfo{author}{\bibfnamefont{G.~M.} \bibnamefont{Radecky}}
  \bibnamefont{et~al.}, \bibinfo{journal}{Phys. Rev.}
  \textbf{\bibinfo{volume}{D25}}, \bibinfo{pages}{1161} (\bibinfo{year}{1982}).

\bibitem[{\citenamefont{Campbell et~al.}(1973)}]{campbell}
\bibinfo{author}{\bibfnamefont{J.}~\bibnamefont{Campbell}}
  \bibnamefont{et~al.}, \bibinfo{journal}{Phys. Rev. Lett.}
  \textbf{\bibinfo{volume}{30}}, \bibinfo{pages}{335} (\bibinfo{year}{1973}).

\bibitem[{\citenamefont{Barish et~al.}(1979)}]{barish}
\bibinfo{author}{\bibfnamefont{S.~J.} \bibnamefont{Barish}}
  \bibnamefont{et~al.}, \bibinfo{journal}{Phys. Rev}
  \textbf{\bibinfo{volume}{D19}}, \bibinfo{pages}{2521} (\bibinfo{year}{1979}).

\bibitem[{\citenamefont{Kitagaki et~al.}(1986)}]{kitagaki}
\bibinfo{author}{\bibfnamefont{T.}~\bibnamefont{Kitagaki}}
  \bibnamefont{et~al.}, \bibinfo{journal}{Phys. Rev.}
  \textbf{\bibinfo{volume}{D34}}, \bibinfo{pages}{2554} (\bibinfo{year}{1986}).

\bibitem[{\citenamefont{Aguilar-Arevalo
  et~al.}(2009{\natexlab{b}})}]{fluxpaper}
\bibinfo{author}{\bibfnamefont{A.~A.} \bibnamefont{Aguilar-Arevalo}}
  \bibnamefont{et~al.}, \bibinfo{journal}{Phys. Rev.}
  \textbf{\bibinfo{volume}{D79}}, \bibinfo{pages}{072002}
  (\bibinfo{year}{2009}{\natexlab{b}}).

\bibitem[{\citenamefont{Aguilar-Arevalo
  et~al.}(2009{\natexlab{c}})}]{detectorpaper}
\bibinfo{author}{\bibfnamefont{A.~A.} \bibnamefont{Aguilar-Arevalo}}
  \bibnamefont{et~al.}, \bibinfo{journal}{Nucl. Instr. Meth.}
  \textbf{\bibinfo{volume}{A599}}, \bibinfo{pages}{28}
  (\bibinfo{year}{2009}{\natexlab{c}}).

\bibitem[{\citenamefont{Casper}(2001)}]{NUANCE}
\bibinfo{author}{\bibfnamefont{D.}~\bibnamefont{Casper}},
  \bibinfo{journal}{arXiv:hep-ph/0208030v1}  (\bibinfo{year}{2001}).

\bibitem[{\citenamefont{Aguilar-Arevalo
  et~al.}(2010{\natexlab{a}})}]{minibooneccqe2}
\bibinfo{author}{\bibfnamefont{A.~A.} \bibnamefont{Aguilar-Arevalo}}
  \bibnamefont{et~al.}, \bibinfo{journal}{Phys. Rev.}
  \textbf{\bibinfo{volume}{D81}}, \bibinfo{pages}{092005}
  (\bibinfo{year}{2010}{\natexlab{a}}).

\bibitem[{\citenamefont{Moniz}(1971)}]{moniz}
\bibinfo{author}{\bibfnamefont{E.~J.} \bibnamefont{Moniz}},
  \bibinfo{journal}{Phys. Rev. Lett.} \textbf{\bibinfo{volume}{26}},
  \bibinfo{pages}{445} (\bibinfo{year}{1971}).

\bibitem[{\citenamefont{Rein and Sehgal}(1981)}]{reinsehgal}
\bibinfo{author}{\bibfnamefont{D.}~\bibnamefont{Rein}} \bibnamefont{and}
  \bibinfo{author}{\bibfnamefont{L.~H.} \bibnamefont{Sehgal}},
  \bibinfo{journal}{Annals Physics} \textbf{\bibinfo{volume}{133}},
  \bibinfo{pages}{79} (\bibinfo{year}{1981}).

\bibitem[{\citenamefont{Rein and Sehgal}(1983)}]{reinsehgal2}
\bibinfo{author}{\bibfnamefont{D.}~\bibnamefont{Rein}} \bibnamefont{and}
  \bibinfo{author}{\bibfnamefont{L.~M.} \bibnamefont{Sehgal}},
  \bibinfo{journal}{Nucl. Phys.} \textbf{\bibinfo{volume}{B223}},
  \bibinfo{pages}{29} (\bibinfo{year}{1983}).

\bibitem[{\citenamefont{Aguilar-Arevalo
  et~al.}(2010{\natexlab{b}})}]{miniboonencpi0}
\bibinfo{author}{\bibfnamefont{A.~A.} \bibnamefont{Aguilar-Arevalo}}
  \bibnamefont{et~al.}, \bibinfo{journal}{Phys. Rev.}
  \textbf{\bibinfo{volume}{D81}}, \bibinfo{pages}{013005}
  (\bibinfo{year}{2010}{\natexlab{b}}).

\bibitem[{\citenamefont{Budd et~al.}(2003)\citenamefont{Budd, Bodek, and
  Arrington}}]{bba}
\bibinfo{author}{\bibfnamefont{A.}~\bibnamefont{Budd}},
  \bibinfo{author}{\bibfnamefont{A.}~\bibnamefont{Bodek}}, \bibnamefont{and}
  \bibinfo{author}{\bibfnamefont{J.}~\bibnamefont{Arrington}},
  \bibinfo{journal}{arXiv:hep-ex/0308005}  (\bibinfo{year}{2003}).

\bibitem[{\citenamefont{Liu et~al.}(1995)}]{liu}
\bibinfo{author}{\bibfnamefont{K.~F.} \bibnamefont{Liu}} \bibnamefont{et~al.},
  \bibinfo{journal}{Phys. Rev. Lett.} \textbf{\bibinfo{volume}{74}},
  \bibinfo{pages}{2172} (\bibinfo{year}{1995}).

\bibitem[{\citenamefont{Aguilar-Arevalo et~al.}(2008)}]{minibooneccqe1}
\bibinfo{author}{\bibfnamefont{A.~A.} \bibnamefont{Aguilar-Arevalo}}
  \bibnamefont{et~al.}, \bibinfo{journal}{Phys. Rev. Lett.}
  \textbf{\bibinfo{volume}{100}}, \bibinfo{pages}{032301}
  (\bibinfo{year}{2008}).

\bibitem[{\citenamefont{Ashery et~al.}(1981)}]{ashery}
\bibinfo{author}{\bibfnamefont{D.}~\bibnamefont{Ashery}} \bibnamefont{et~al.},
  \bibinfo{journal}{Phys. Rev.} \textbf{\bibinfo{volume}{C23}},
  \bibinfo{pages}{2173} (\bibinfo{year}{1981}).

\bibitem[{\citenamefont{Jones et~al.}(1993)}]{jones}
\bibinfo{author}{\bibfnamefont{M.~K.} \bibnamefont{Jones}}
  \bibnamefont{et~al.}, \bibinfo{journal}{Phys. Rev.}
  \textbf{\bibinfo{volume}{C48}}, \bibinfo{pages}{2800} (\bibinfo{year}{1993}).

\bibitem[{\citenamefont{Ransome et~al.}(1992)}]{ransome}
\bibinfo{author}{\bibfnamefont{R.~D.} \bibnamefont{Ransome}}
  \bibnamefont{et~al.}, \bibinfo{journal}{Phys. Rev.}
  \textbf{\bibinfo{volume}{C45}}, \bibinfo{pages}{R509} (\bibinfo{year}{1992}).

\bibitem[{\citenamefont{Patterson et~al.}(2009)}]{reconstructionnim}
\bibinfo{author}{\bibfnamefont{R.~B.} \bibnamefont{Patterson}}
  \bibnamefont{et~al.}, \bibinfo{journal}{Nucl. Instr. Meth.}
  \textbf{\bibinfo{volume}{A608}}, \bibinfo{pages}{206} (\bibinfo{year}{2009}).

\bibitem[{\citenamefont{Wascko}(2006)}]{morgan}
\bibinfo{author}{\bibfnamefont{M.~O.} \bibnamefont{Wascko}},
  \bibinfo{journal}{Nucl. Phys. B, Proc. Suppl.}
  \textbf{\bibinfo{volume}{159}}, \bibinfo{pages}{50} (\bibinfo{year}{2006}).

\bibitem[{\citenamefont{Wilking}(2009)}]{wilkingthesis}
\bibinfo{author}{\bibfnamefont{M.~J.} \bibnamefont{Wilking}}, Ph.D. thesis,
  \bibinfo{school}{University of Colorado} (\bibinfo{year}{2009}).

\bibitem[{\citenamefont{D'Agostini}(1995)}]{unfoldmeth}
\bibinfo{author}{\bibfnamefont{G.}~\bibnamefont{D'Agostini}},
  \bibinfo{journal}{Nucl. Instrum. Meth.} \textbf{\bibinfo{volume}{A362}},
  \bibinfo{pages}{487} (\bibinfo{year}{1995}).

\bibitem[{\citenamefont{Sanford and Wang}(1967)}]{sw}
\bibinfo{author}{\bibfnamefont{J.~R.} \bibnamefont{Sanford}} \bibnamefont{and}
  \bibinfo{author}{\bibfnamefont{C.~L.} \bibnamefont{Wang}},
  \bibinfo{journal}{BNL 11299}  (\bibinfo{year}{1967}).

\bibitem[{\citenamefont{Catanesi et~al.}(2007)}]{harp}
\bibinfo{author}{\bibfnamefont{M.~G.} \bibnamefont{Catanesi}}
  \bibnamefont{et~al.}, \bibinfo{journal}{Eur. Phys. J.}
  \textbf{\bibinfo{volume}{C52}}, \bibinfo{pages}{29} (\bibinfo{year}{2007}).

\bibitem[{\citenamefont{Chemakin et~al.}(2008)}]{e910}
\bibinfo{author}{\bibfnamefont{I.}~\bibnamefont{Chemakin}}
  \bibnamefont{et~al.}, \bibinfo{journal}{(in preparation)}
  (\bibinfo{year}{2008}).

\bibitem[{\citenamefont{Zeitnitz and Gabriel}(1994)}]{gcalor}
\bibinfo{author}{\bibfnamefont{C.}~\bibnamefont{Zeitnitz}} \bibnamefont{and}
  \bibinfo{author}{\bibfnamefont{T.~A.} \bibnamefont{Gabriel}},
  \bibinfo{journal}{Nucl. Instrum. Meth.} \textbf{\bibinfo{volume}{A349}},
  \bibinfo{pages}{106} (\bibinfo{year}{1994}).

\end{thebibliography}

\end{document}